\documentclass{article}
\usepackage{graphicx}
\usepackage[letterpaper, total={7in, 9in}]{geometry}
\usepackage{wrapfig}
\usepackage{enumitem}
\usepackage{amsmath,stackengine}
\usepackage{amssymb}
\usepackage{marginnote}
\usepackage{hyperref}
\usepackage{url}
\usepackage{esint}
\usepackage{multirow}
\usepackage{comment}
\usepackage{caption}
\usepackage{subcaption}
\usepackage{tikz}
\usetikzlibrary{arrows.meta}

\usepackage{listings}
\usepackage{xcolor}
\usepackage{siunitx}
\usepackage{pgfplots}
\pgfplotsset{compat=1.9, , every axis/.append style={font=\footnotesize}}
\usepackage{xspace}
\usepackage{tensor}
\usepackage{authblk}

\definecolor{codegray}{rgb}{0.5,0.5,0.5}
\definecolor{codered}{rgb}{0.75,0,0}
\definecolor{backcolour}{rgb}{0.95,0.95,0.92}
\lstdefinestyle{mystyle}{
    backgroundcolor=\color{backcolour},
    keywordstyle=\color{blue},
    numberstyle=\tiny\color{codegray},
    stringstyle=\color{codered},
    basicstyle=\ttfamily\footnotesize,
    breakatwhitespace=false,         
    captionpos=b,                    
    keepspaces=true,                 
    numbers=left,                    
    numbersep=5pt,                  
    showspaces=false,                
    showstringspaces=false,
    showtabs=false,                  
    tabsize=2
}
\newcommand{\mb}{\mathbf}

\newcommand{\X}{\mathcal{X}}
\newcommand{\J}{\mathcal{J}}
\newcommand{\F}{\mathcal{F}}
\renewcommand{\i}{\mb{i}}
\renewcommand{\j}{\mb{j}}

\newcommand{\R}{\mathbb{R}}
\newcommand{\C}{\mathfrak{C}}
\newcommand{\N}{\mathcal{N}}
\newcommand{\D}{\mathfrak{D}}
\newcommand{\I}{\mathcal{I}}

\newcommand{\quotesaround}[1]{``#1"}

\title{Hyperchaos and complex dynamical regimes \\ in $N$-dimensional neuron lattices}

\author{Brandon B. Le\thanks{e-mail: \href{mailto:sxh3qf@virginia.edu}{sxh3qf@virginia.edu} (corresponding author)}}
\author{Dima Watkins\thanks{e-mail: \href{mailto:bem8mq@virginia.edu}{bem8mq@virginia.edu}}}

\affil[1]{Department of Physics, University of Virginia, Charlottesville, Virginia 22904-4714, USA}

\date{}

\begin{document}

\maketitle

\begin{abstract}
    We study the dynamics of $N$-dimensional lattices of nonchaotic Rulkov neurons coupled with a flow of electrical current. We consider both nearest-neighbor and next-nearest-neighbor couplings, homogeneous and heterogeneous neurons, and small and large lattices over a wide range of electrical coupling strengths. As the coupling strength is varied, the neurons exhibit a number of complex dynamical regimes, including unsynchronized chaotic spiking, local quasi-bursting, synchronized chaotic bursting, and synchronized hyperchaos. For lattices in higher spatial dimensions, we discover dynamical effects arising from the ``destructive interference'' of many connected neurons and miniature ``phase transitions'' from coordinated spiking threshold crossings. In large two- and three-dimensional neuron lattices, we observe emergent dynamics such as local synchronization, quasi-synchronization, and lag synchronization. These results illustrate the rich dynamics that emerge from coupled neurons in multiple spatial dimensions, highlighting how dimensionality, connectivity, and heterogeneity critically shape the collective behavior of neuronal systems.
\end{abstract}

\vspace{0.5cm}

\tableofcontents

\newpage

\section{Introduction}

The mathematical modeling of nervous system dynamics has a rich and evolving history. Typically, it involves constructing a set of continuous-time dynamical equations that capture the behavior of neurons --- the primary computational units of an animal's nervous system \cite{neuron-review}. A landmark in the field was the Hodgkin-Huxley model, which introduced a set of differential equations describing a neuron's membrane potential and transmembrane ion channels \cite{hh}. Since then, numerous models have been developed \cite{neuronal-bursting-model, 1d-comp-efficient-simulation, izh-cont, chay, buchholtz}, incorporating finer layers of biological complexity through frameworks such as transition state theory \cite{thermo-neurons} and mean field theory \cite{mft-neurons}.

Although these models can capture increasingly subtle aspects of neuronal behavior, they also tend to be mathematically intricate. This complexity can hinder both analytical and computational approaches to their study. For example, the Hodgkin-Huxley model comprises four coupled nonlinear differential equations to describe the state of a single neuron \cite{hh}. In response to such mathematical intractability and computational prohibitivity, discrete-time models have emerged as a promising alternative \cite{izh-map, chialvo, nagumo, discrete-memristors, ibarz}. Among them are the Rulkov maps \cite{rulkov, rulkov2}, which offer several notable advantages and have seen wide application in the literature, including in stability analysis \cite{wang2}, control of chaos \cite{lopez}, symbolic analysis \cite{budzinski}, final state sensitivity \cite{le}, machine learning \cite{ge}, information patterns \cite{njitacke}, and digital watermarking \cite{ding}. Its utility in studying synchronization phenomena is well demonstrated in Refs. \cite{hu, complete-synched-rulkov, scale-free-rulkov}, where coupled Rulkov neurons are shown to exhibit robust synchronous behavior across various network topologies. Its adaptability is further illustrated in Ref. \cite{rulkov-memristor}, which reports increased dynamical complexity when the map is embedded in discrete memristive models to simulate neuromorphic behavior. The rigorous proof of chaos in Ref. \cite{chaos-marotto-rulkov} reinforces the Rulkov model’s credibility as a foundation for nonlinear analysis. Finally, Ref. \cite{small-work-rulkov} highlights the map’s ability to capture complex collective dynamics, such as noise-induced pattern formation in small-world networks. The Rulkov map also retains key features of biological neuronal dynamics, including chaotic and non-chaotic spiking, bursting, and quiescence \cite{izhikevich-article}, while dramatically simplifying the mathematics involved \cite{izh-time}. Two principal types of Rulkov dynamics are known, often called the chaotic Rulkov model \cite{rulkov2} and the non-chaotic Rulkov model \cite{rulkov}. While the collective behavior of chaotic Rulkov neurons has been extensively studied \cite{ibarz}, we turn our attention here to the comparatively less-explored territory of non-chaotic Rulkov neurons coupled in lattice geometries.

A common topology used when studying coupled dynamical systems is a ring topology \cite{banerjee, jampa, chen}, which has been applied in the context of Rulkov neuron lattices \cite{omelchenko, ring, three-ring, lag-sync-model}. In this paper, we generalize these studies by analyzing the dynamics of Rulkov neuron lattices in $N$ spatial dimensions. Specifically, we are interested in $N$-dimensional cubic lattices of nonchaotic Rulkov neurons that are electrically coupled with one another by a flow of current. One motivation for considering $N$-dimensional lattices stems from the desire to better understand how complex dynamical systems behave across different topological and dimensional contexts. While two- and three-dimensional systems are most familiar and often the easiest to visualize, higher-dimensional lattices provide a powerful theoretical framework for exploring the influence of connectivity, neighborhood structure, and spatial embedding on emergent dynamics. In particular, studying Rulkov maps on $N$-dimensional lattices allows for the investigation of how dimensionality affects synchronization, wave propagation, and the formation of spatiotemporal patterns. Such generalizations are not only mathematically interesting but also potentially relevant to high-dimensional data modeling in neuroscience, where interactions among many neurons or brain regions can be viewed through the lens of high-dimensional coupling.

The organization of this paper is as follows. In Sec. \ref{sec:the-model}, we introduce the dynamics of the Rulkov neuron, the electrical coupling scheme, and the mathematical framework of our $N$-dimensional lattice model. Section \ref{sec:2d-lattices} begins our analysis of the hyperchaotic dynamics and the emergence of complex dynamical regimes in small and large two-dimensional Rulkov lattices across a variety of individual neuron behaviors and a wide range of electrical coupling strengths. Naturally extending this analysis, Sec. \ref{sec:nd-lattices} examines the general case of $N$-dimensional lattice systems and offers insight into the behavior of highly connected neuron lattices. Finally, we summarize our findings and offer some future directions in Sec. \ref{sec:conclusions}.

\section{The model}
\label{sec:the-model}

The (nonchaotic) Rulkov map is a two-dimensional slow-fast map that accurately models the various dynamics of neurons, including regular spiking, irregular spiking, nonchaotic bursting, and chaotic spiking-bursting patterns \cite{rulkov}. The Rulkov map offers a simplified approach to modeling neuronal dynamics, in contrast to traditional neuron models that rely on complex, highly nonlinear differential equations. Rather than serving as a fully accurate physical representation, the Rulkov map captures the essential behaviors of biological neurons to facilitate the exploration of more intricate collective phenomena that can be obscured by the mathematical intricacies of more detailed models. 

For this reason, our study utilizes the Rulkov neuron to build lattices of neurons arranged within an $N$-dimensional physical space. This approach allows us to investigate how local interactions between simplified neurons can give rise to complex global behaviors, providing insight into the collective dynamics of neuronal systems in a phenomenological framework. In this section, we first provide a brief overview of the individual dynamics of the Rulkov neuron and its response from an injection of direct current (Sec. \ref{subsec:rulkov-neuron}). Then, we describe the $N$-dimensional lattice model and the electrical coupling scheme in Sec. \ref{subsec:nd-model}.

\subsection{The Rulkov neuron}
\label{subsec:rulkov-neuron}

In order to facilitate an easy transition to the $N$-dimensional model, we denote the state of an individual Rulkov neuron at timestep $t=k$ as
\begin{equation}
    X(k) = X^a(k)\hat{\theta}_a = \begin{pmatrix}
        X^0(k) \\ X^1(k)
    \end{pmatrix} = \begin{pmatrix}
        x(k) \\ y(k)
    \end{pmatrix},
\end{equation}
where $\hat{\theta}_a$ are two-dimensional basis vectors, and we use the Einstein summation convention. Here, $x$ is the fast variable representing the voltage of the neuron, and $y$ is the slow variable that controls slow oscillations of the neuron. The discrete-time iteration function $F$ that the Rulkov neuron obeys is
\begin{equation}
    X(k+1) = \begin{pmatrix}
        x(k+1) \\ y(k+1)
    \end{pmatrix} = F(X(k)) = \begin{pmatrix}
        f(x(k), y(k); \alpha) \\
        y(k) - \mu[x(k) - \sigma]
    \end{pmatrix},
    \label{eq:rulkov_iteration}
\end{equation}
where $f$ is the piecewise function
\begin{equation}
    f(x,y;\alpha) = 
    \begin{cases}
        \alpha/(1-x) + y, & x\leq 0 \\
        \alpha + y, & 0 < x < \alpha + y \\
        -1, & x\geq \alpha + y
    \end{cases}
    \label{eq:rulkov_fast_equation}
\end{equation}
and $\alpha$, $\sigma$, and $\mu$ are parameters. To make $y$ a slow variable, we require $0<\mu\ll 1$, so we choose the standard value of $\mu=0.001$. 

\begin{figure}[t!]
    \centering
    \begin{subfigure}{0.475\textwidth}
        \centering
        \begin{tikzpicture}[scale=0.9]
            \begin{axis}[
                    axis lines = left,
                    x=1cm,
                    y=1cm,
                    xlabel =\large \(x(k)\),
                    xtick align=outside,
                    xtick pos=left,
                    xmin=-3, xmax=3.5,
                    xtick={-3,-2,-1,0,1,2,3},
                    minor xtick={-2.5,-1.5,-0.5,0.5,1.5,2.5},
                    ylabel = \large \(x(k+1)\),
                    ytick align=outside,
                    ytick pos=left,
                    ymin=-3, ymax=3.5,
                    ytick={-3,-2,-1,0,1,2,3},
                    minor ytick={-2.5,-1.5,-0.5,0.5,1.5,2.5},
                ]
            \addplot [
                domain=-3:0, 
                samples=100,
                color=blue!80!black,
                very thick
            ]
            {6/(1-x)-3.93};
            \addplot [
                domain=0:6-3.93, 
                color=blue!80!black,
                very thick
            ]
            {6-3.93};
            \addplot [
                domain=6-3.93:3.5, 
                color=blue!80!black,
                very thick
            ]
            {-1};
            \addplot [
                domain=-3:3.5, 
                color=black,
                thick,
                dashed
            ]
            {x};
            \addplot[color=green!80!black, mark=*] coordinates {(-1.741,-1.741)};
            \addplot[color=red, mark=*] coordinates {(-1.189,-1.189)};
            \addplot[color=blue, mark=o] coordinates {(2.07,2.07)};
            \addplot[color=blue, mark=*] coordinates {(2.07,-1)};
            \addplot+[
                color=green!80!black,
                mark=none,
                densely dotted,
                very thick
            ]
            coordinates
            {(2.07,-1) (-1,-1) (-1,-0.93) (-0.93,-0.93) (-0.93,-0.821) (-0.821,-0.821) (-0.821,-0.635) (-0.635,-0.635) (-0.635,-0.26) (-0.26,-0.26) (-0.26,0.832) (0.832,0.832) (0.832,2.07) (2.07,2.07) (2.07,-1)};
            \end{axis}
        \end{tikzpicture}
        \caption{The function $x(k+1)=f(x(k),y;\alpha)$ graphed in blue for $y=-3.93$ and $\alpha=6$. The stable fixed point is shown in green and the unstable fixed point is shown in red at the intersection between the function and the dashed black line $x(k+1) = x(k)$. The periodic spiking orbit is shown with a dotted green line.}
    \label{fig:rulkov_fast_iteration}
    \end{subfigure}
    \hfill
    \begin{minipage}[t]{0.475\textwidth}
        \vspace{-10cm}
        \begin{subfigure}{\textwidth}
            \centering
            \includegraphics[width=\textwidth]{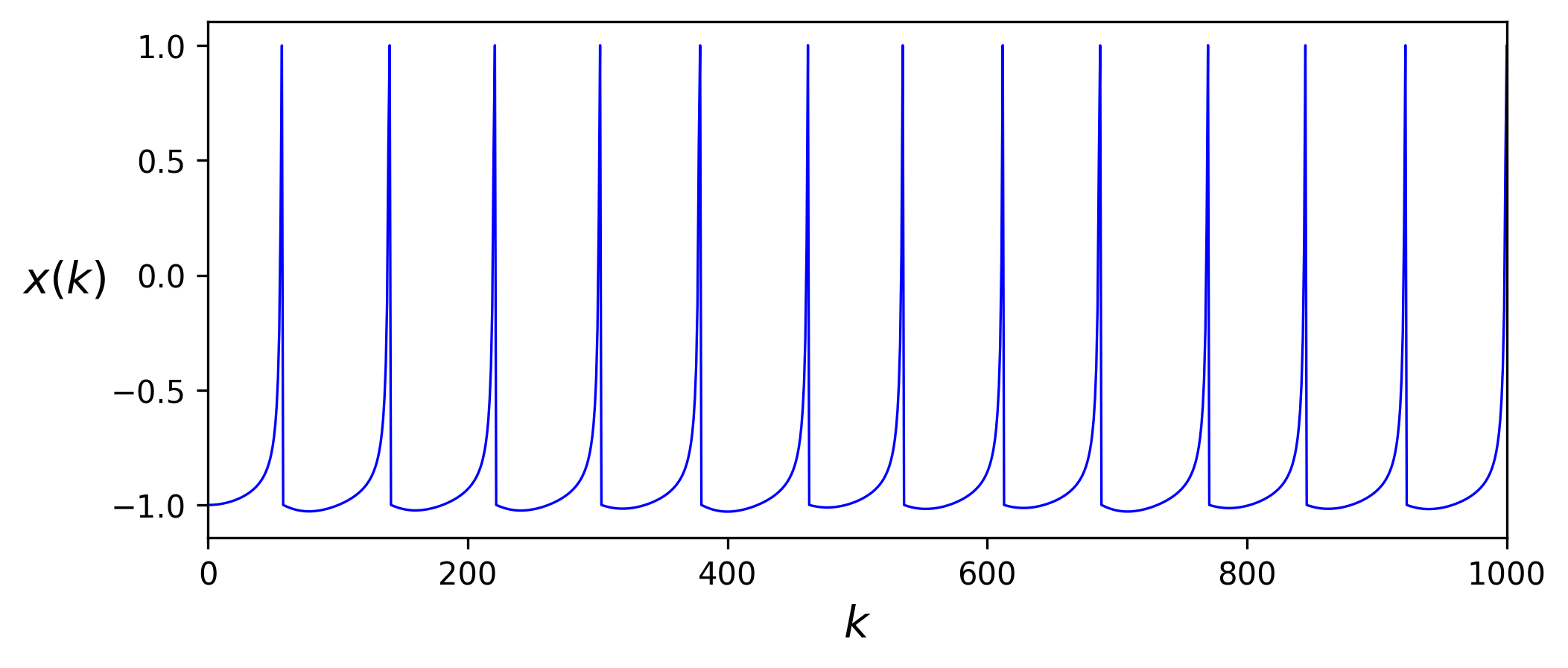}
            \caption{A graph of the voltage $x(k)$ of an individually nonchaotic spiking Rulkov neuron with $\alpha=4$, $\sigma=-0.9$, and $X(0) = (-1,-3)$.}
            \label{fig:rulkov_example_graph}
        \end{subfigure}
        \begin{subfigure}{\textwidth}
            \centering
            \includegraphics[width=\textwidth]{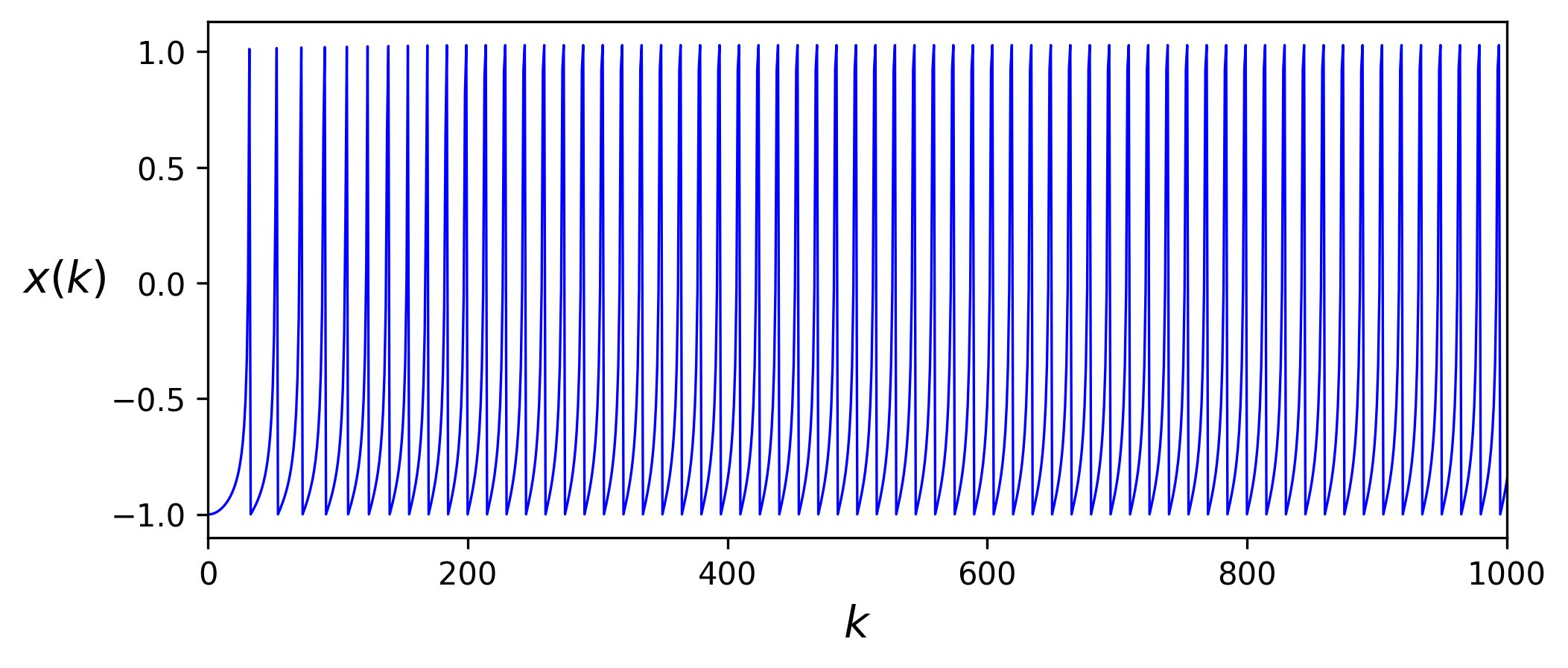}
            \caption{A graph of the voltage $x(k)$ of an individually nonchaotic spiking Rulkov neuron with $\alpha=4$, $\sigma=-0.5$, and $X(0) = (-1,-3)$.}
            \label{fig:rulkov_example_graph_2}
        \end{subfigure}
    \end{minipage}
    \begin{subfigure}{0.475\textwidth}
        \centering
        \includegraphics[width=\textwidth]{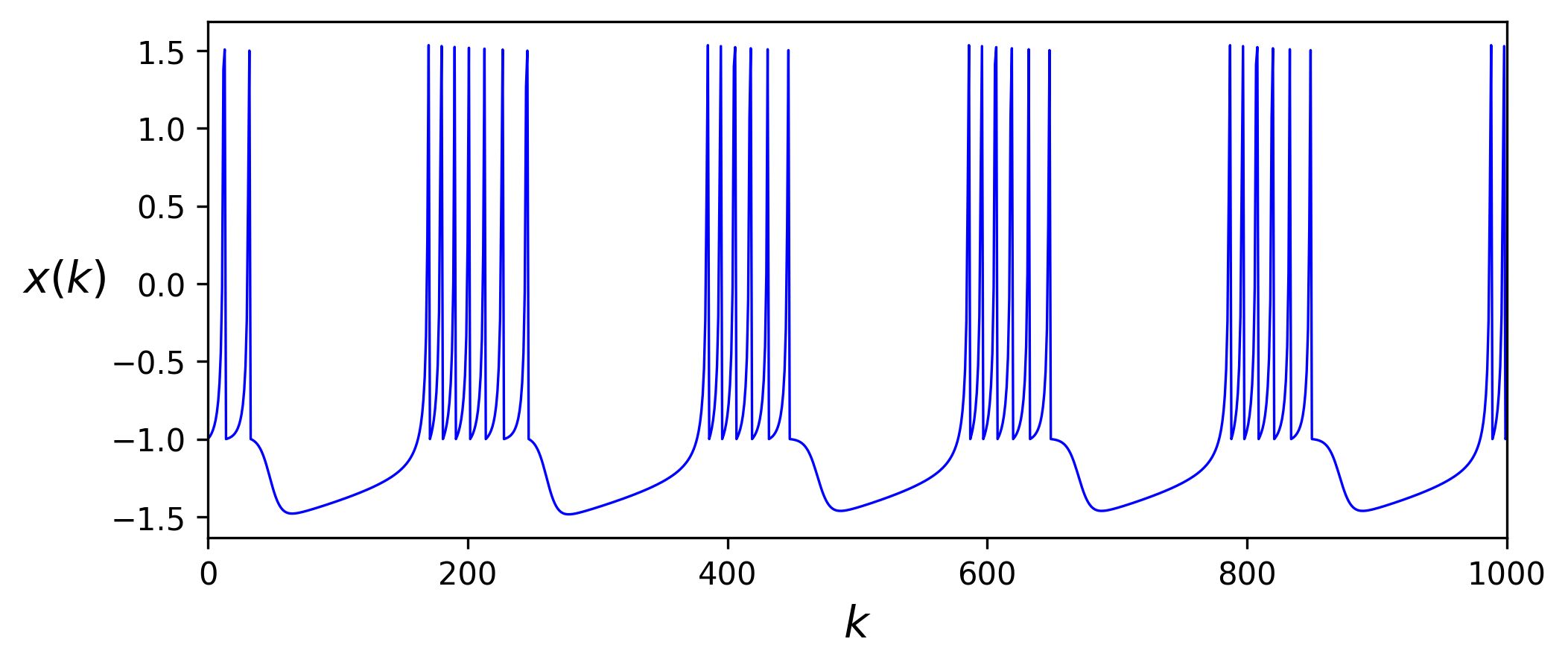}
        \caption{A graph of the voltage $x(k)$ of an individually nonchaotic bursting Rulkov neuron with $\alpha=5$, $\sigma=-1$, and $X(0) = (-1,-3.49)$.}
        \label{fig:rulkov_example_graph_3}
    \end{subfigure}
    \hfill
    \begin{subfigure}{0.475\textwidth}
        \centering
        \includegraphics[width=\textwidth]{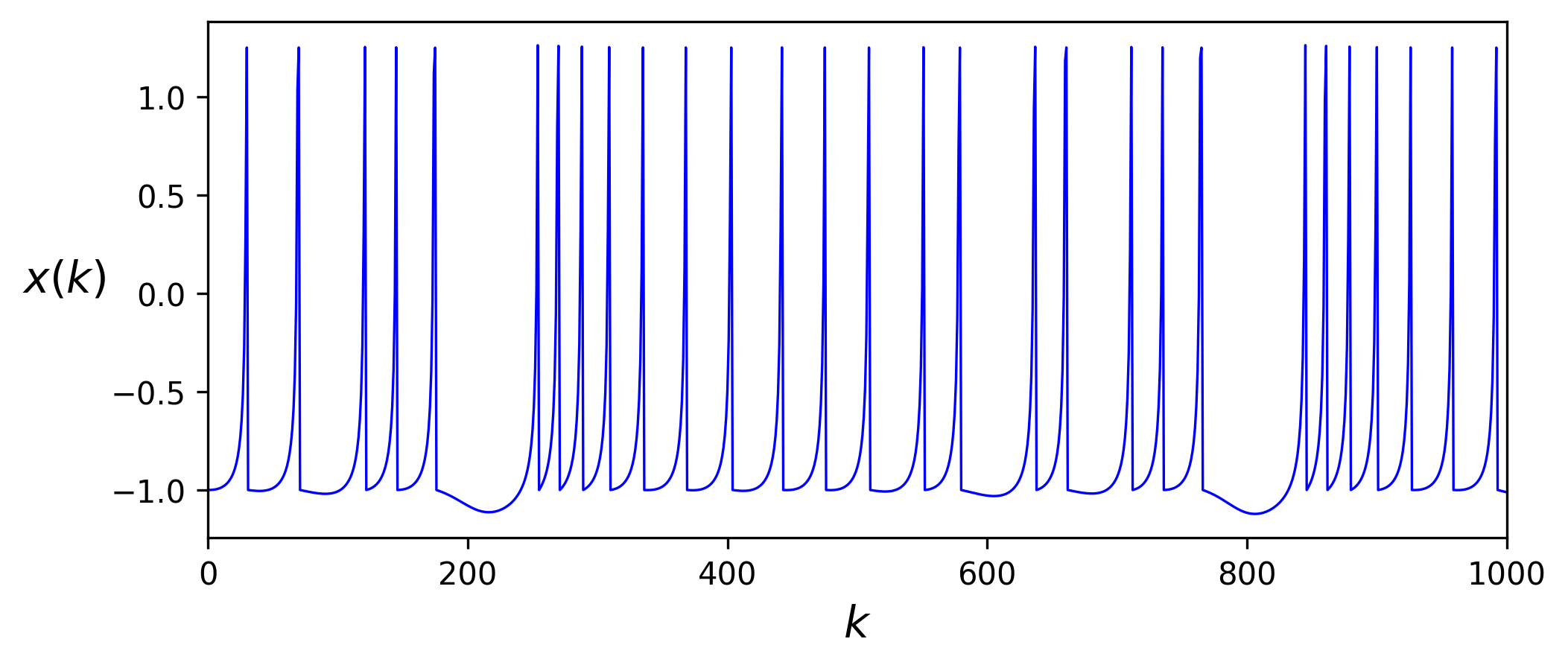}
        \caption{A graph of the voltage $x(k)$ of an individually chaotic spiking Rulkov neuron with $\alpha=4.5$, $\sigma=-0.82$, and $X(0) = (-1,-3.25)$.}
        \label{fig:rulkov_example_graph_4}
    \end{subfigure}
    \caption{Examples of the dynamics of individual Rulkov neurons.}
    \label{fig:rulkov_examples}
\end{figure}
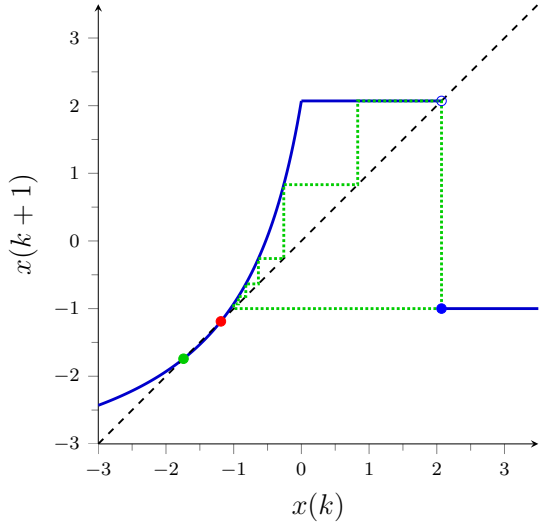
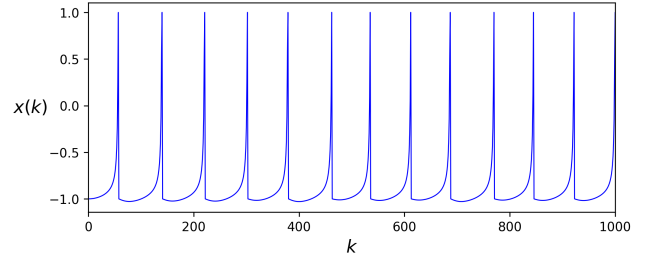
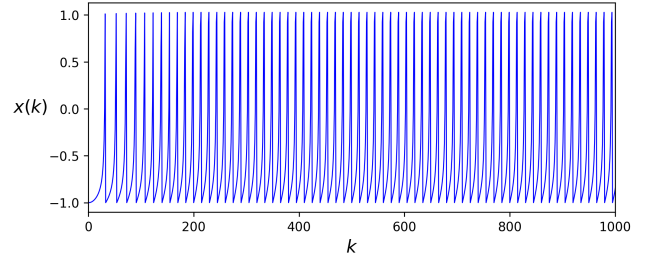
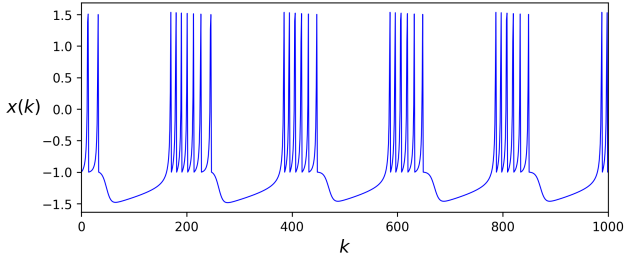
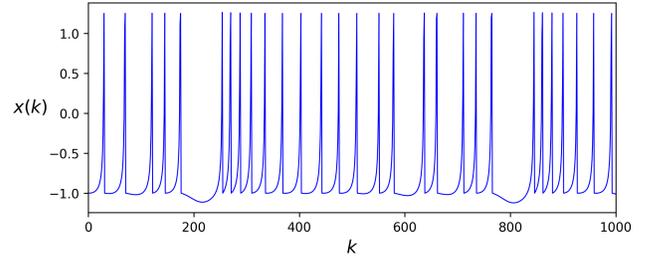

We now present a few examples that illustrate the range of dynamics the Rulkov neuron can model. In order to understand how these dynamics can manifest, we first graph the fast variable iteration function $f$ [Eq. \eqref{eq:rulkov_fast_equation}] in Fig. \ref{fig:rulkov_fast_iteration}. Here, we treat the slow variable $y$ as a fixed parameter and set it equal to $y=-3.93$. In Fig. \ref{fig:rulkov_fast_iteration}, we draw a cobweb plot with a dotted green line, which starts at $x=-1$, slowly rises, quickly reaches a maximum, then falls back to the starting point. This periodic orbit demonstrates how regular spiking manifests itself in the Rulkov map. We also observe that there are two fixed points at the intersection between the curve $f$ and the line $x(k+1) = x(k)$: one stable (green) and one unstable (red). Now, from the equation of $f$ [Eq. \eqref{eq:rulkov_fast_equation}], it is easy to see that $y$ controls the height of the curve to the left of the discontinuity. Increasing $y$ will cause the two fixed points to vanish, making quiescent orbits (voltage silence at the fixed point) impossible. A higher curve will also result in fewer iterations before the voltage is reset by the third piece of Eq. \eqref{eq:rulkov_fast_equation}, which indicates a shorter periodic orbit length, or faster spikes. If $y$ is lowered, the red unstable point will eventually rise higher than $x(k+1)=x(k)=-1$, which will cause all orbits to be attracted to the green stable point after reaching the resetting mechanism.

To get a better understanding of the dynamics of the full two-dimensional map, we first examine the role of the parameter $\sigma$. In Eq. \eqref{eq:rulkov_iteration}, the iteration function for $y$ indicates that $\sigma$ is the value of $x$ that will keep $y$ constant. If $x>\sigma$, $y$ will decrease to compensate, lowering the curve in Fig. \ref{fig:rulkov_fast_iteration} until the average value of $x$ has reached $\sigma$, and vice versa. We know that a higher value of $y$ results in faster spikes, so it is unsurprising that a higher value of $\sigma$ will result in faster spikes since a higher $\sigma$ means that $y$ will increase until the spikes are fast enough for the average value of $x$ to match $\sigma$. Therefore, $\sigma$ serves as an ``excitation parameter'' of the Rulkov map. For example, in Figs. \ref{fig:rulkov_example_graph} and \ref{fig:rulkov_example_graph_2}, we graph the fast variable $x(k)$ for a Rulkov neuron with $\alpha = 4$ and $X(0) = (-1,-3)$. Both neurons exhibit nonchaotic spiking dynamics, but since we increase $\sigma$ from $-0.9$ in Fig. \ref{fig:rulkov_example_graph} to $-0.5$ in Fig. \ref{fig:rulkov_example_graph_2}, the neuron is spiking with a much higher frequency in the latter graph.

The role of the parameter $\alpha$ is more subtle, but one of its roles is to control the existence of bursting. Specifically, for $\alpha>4$, certain values of $\sigma$ will cause the Rulkov neuron to exhibit bursting behavior, which is characterized by slow oscillations (controlled by the slow variable $y$) between bursts of spikes and periods of silence (e.g. Fig. \ref{fig:rulkov_example_graph_3}). We also note that by setting the parameters correctly, Rulkov neurons can also exhibit chaotic, irregular spiking (e.g. Fig. \ref{fig:rulkov_example_graph_4}). See Ref. \cite{thesis} for a detailed account of how bursting and chaotic dynamics arise in individual Rulkov neurons.

In order to electrically couple Rulkov neurons, we need a way to model the response of a Rulkov neuron from an injection of direct current $I(k)$. To do this, we modify the Rulkov iteration function in Eq. \eqref{eq:rulkov_iteration} slightly by adding two new time-varying parameters $\beta(k)$ and $\sigma(k)$ that incorporate the time-varying current $I(k)$:
\begin{equation}
    \beta(k) = \beta^c I(k),\quad \sigma(k) = \sigma + \sigma^c I(k).
    \label{eq:beta-and-sigma}
\end{equation}
Here, $\beta^c$ and $\sigma^c$ are coefficients selected to achieve the desired response behavior. Then, the modified Rulkov iteration function is
\begin{equation}
    \begin{pmatrix}
        x(k+1) \\ y(k+1)
    \end{pmatrix} = \begin{pmatrix}
        f(x(k), y(k) + \beta(k); \alpha) \\
        y(k) - \mu[x(k) - \sigma(k)]
    \end{pmatrix},
    \label{eq:rulkov_iteration_current}
\end{equation}
We will now briefly overview the effect of these new time-varying parameters $\sigma(k)$ and $\beta(k)$ in turn. 

\begin{figure}[t!]
    \centering
    \begin{subfigure}{0.475\textwidth}
        \centering
        \includegraphics[width=\textwidth]{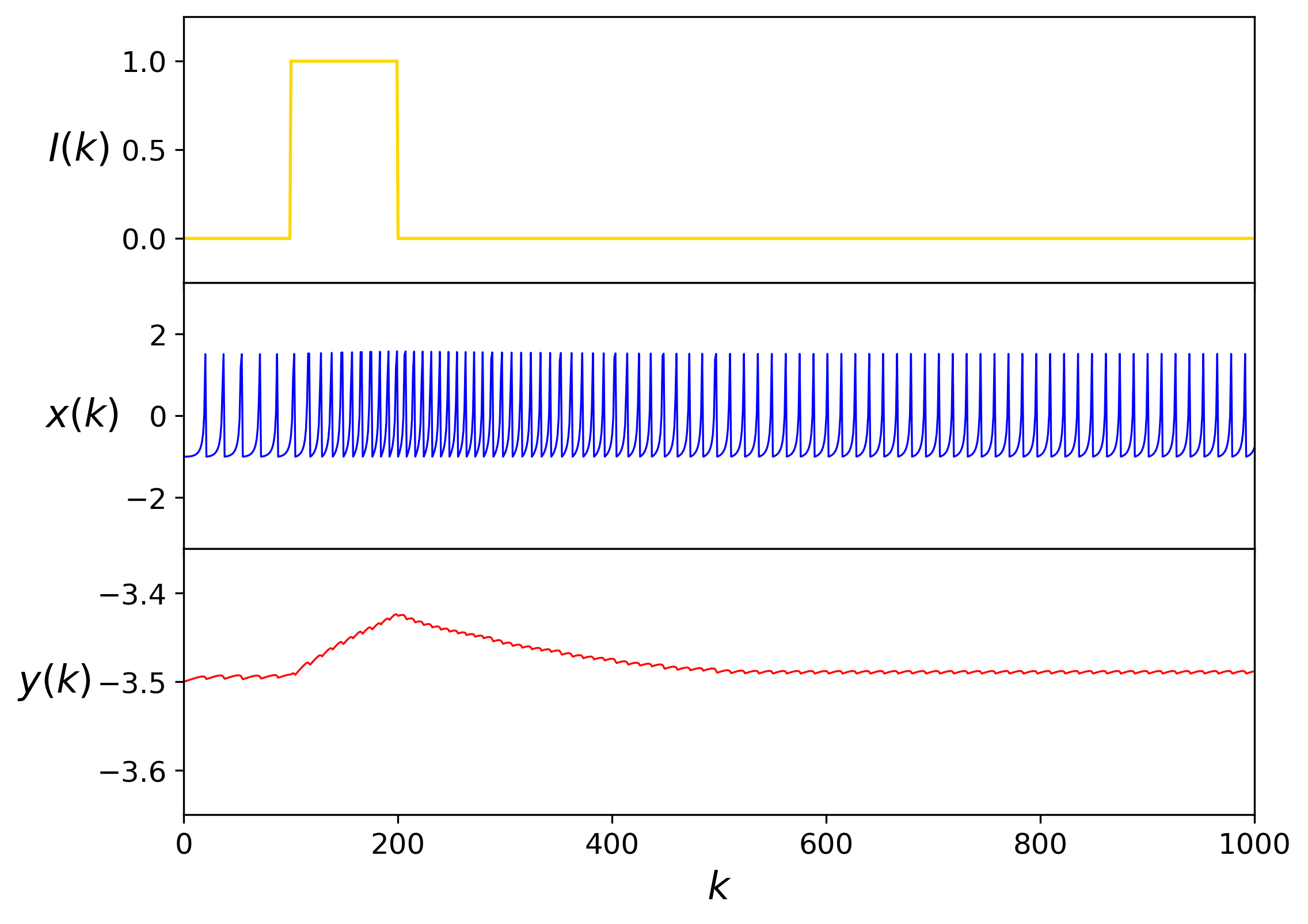}
        \caption{$\sigma^c=1$, $\beta^c=0$}
        \label{fig:current_response_sigma}
    \end{subfigure}
    \hfill
    \begin{subfigure}{0.475\textwidth}
        \centering
        \includegraphics[width=\textwidth]{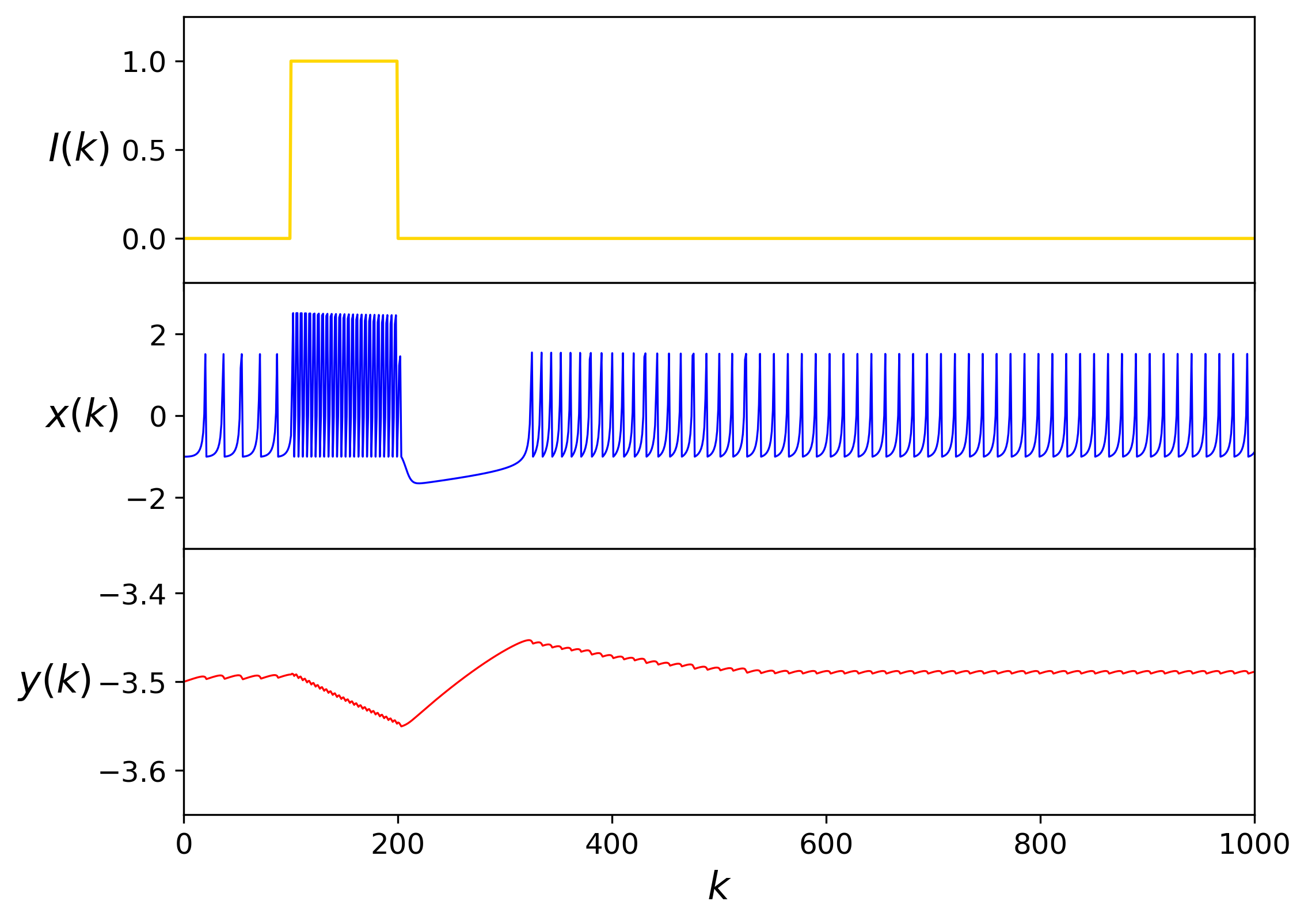}
        \caption{$\sigma^c=1$, $\beta^c=1$}
        \label{fig:current_response_beta}
    \end{subfigure}
    \caption{Graphs of the response of the fast variable $x(k)$ and the slow variable $y(k)$ of a regular spiking Rulkov neuron to a pulse of direct current $I(k)$. The neuron has parameters $\alpha=5$ and $\sigma=-0.6$, and the initial state of the neuron is $X(0)=(-1,-3.5)$.}
    \label{fig:current_response}
\end{figure}

Let us first set the coefficients $\sigma^c=1$ and $\beta^c = 0$. This configuration of parameters enables us to examine the effect of $\sigma(k)$ in isolation. From Eqs. \eqref{eq:beta-and-sigma} and \eqref{eq:rulkov_iteration_current}, it is clear that, under these conditions, $I(k)$ effectively changes the value of the parameter $\sigma$. To show how this effect manifests, we consider a pulse of current of magnitude $I=1$ that turns on at $k=100$ and turns off at $k=200$ on a regular spiking Rulkov neuron. The response of the fast variable $x(k)$ and the slow variable $y(k)$ are shown in Fig. \ref{fig:current_response_sigma}. When the current pulse turns on, the value of $\sigma(k)$ jumps up by 1, causing $y$ to slowly rise to compensate. As a result, the frequency of the spikes gradually starts to increase. Once the current pulse turns off, the value of $\sigma(k)$ goes back to $\sigma$, causing the value of $y$ and the frequency of spikes to gradually go back to their original values.

Now, let us set the coefficients $\sigma^c=1$ and $\beta^c=1$, incorporating the effects of $\beta(k)$. From Eqs. \eqref{eq:beta-and-sigma} and \eqref{eq:rulkov_iteration_current}, we can see that $\beta(k)$ changes the ``effective value'' of $y$ for the fast variable $x$. In other words, once the current pulse $I(k)$ is turned on, the fast variable $x$ will immediately ``think'' that $y$ is higher. This effect can be seen in Fig. \ref{fig:current_response_beta}, where as soon as the current pulse is turned on, the effective value of $y$ increases by 1, making the fast variable iteration function shoot up (Fig. \ref{fig:rulkov_fast_iteration}), causing extremely rapid spiking immediately. Now that the average value of $x$ is much larger, $y$ decreases to compensate, so once the current pulse turns off, the fast variable iteration function (Fig. \ref{fig:rulkov_fast_iteration}) is low, causing $x$ to be attracted to the stable fixed point. Now, the average value of $x$ is too low, so $y$ increases to compensate until the fast variable iteration function rises high enough for the fixed points to vanish. This causes $y$ to overshoot its original value, so the frequency of spikes slowly decreases until equilibrium is reached again. We also observe this high-frequency spiking followed by a period of silence due to a difference in current flow in Secs. \ref{sec:2d-lattices} and \ref{sec:nd-lattices} in the form of synchronized chaotic bursting as a result of the same phenomenon. For the rest of this paper, we set $\sigma^c=\beta^c=1$ in order to incorporate the effects of both time-varying parameters.

\subsection{The \texorpdfstring{$N$}{N}-dimensional lattice model}
\label{subsec:nd-model}

Now that we have established the dynamics of an individual Rulkov neuron, we introduce the $N$-dimensional lattice model. Consider an $N$-dimensional cubic lattice of Rulkov neurons with side length $\zeta$. Then, there are $\zeta^N$ total neurons, and since every neuron has one fast variable and one slow variable, the state space of the system is $n=2\zeta^N$ dimensional. It is natural to describe the state of the system as a type $(N+1, 0)$ ``tensor''
\begin{equation}
    \X = \X^{i_1i_2\cdots i_Na} \hat{e}_{i_1}\otimes \hat{e}_{i_2}\otimes \cdots\otimes\hat{e}_{i_N}\otimes \hat{\theta}_{a},
\end{equation}
where $i_1,i_2,\hdots,i_N$ range from $0$ to $\zeta-1$, $\hat{e}_{i_1}, \hat{e}_{i_2}, \hdots, \hat{e}_{i_N}$ are $\zeta$-dimensional basis vectors, and we use the Einstein summation convention. As a shorthand, we write $\X^{i_1i_2\cdots i_Na} = \X^{\mb{i}a}$, where $\mb{i} = \{i_1,i_2,\hdots,i_N\}$. This allows us to denote the state of the neuron at the $\mb{i}$th position in the lattice as $X = \X^{\mb{i}}$. 

We use the iteration function $\F:\underbrace{\R^\zeta\otimes\cdots\otimes\R^\zeta}_N\otimes\R^2\to \underbrace{\R^\zeta\otimes\cdots\otimes\R^\zeta}_N\otimes\R^2$ to describe how the state $\X$ evolves: 
\begin{equation}
    \X(k+1) = \F(\X(k)).
\end{equation}
To get the full iteration function $\F$, we can then define the iteration function of $\X^\i$ based on Eq. \eqref{eq:rulkov_iteration_current}:
\begin{equation}
    \begin{split}
        \X^\i(k+1) = \F^\i(\X(k)) &= \begin{pmatrix}
            f(\X^{\i0}(k), \X^{\i1}(k) + \C^{\i}(k); \alpha^{\i}) \\
            \X^{\i1}(k) - \mu \X^{\i0}(k) + \mu[\sigma^{\i} + \C^{\i}(k)]
        \end{pmatrix} \\
        &= \begin{pmatrix}
            f(x^{\i}(k), y^{\i}(k) + \C^{\i}(k); \alpha^{\i}) \\
            y^{\i}(k) - \mu x^{\i}(k) + \mu[\sigma^{\i} + \C^{\i}(k)]
        \end{pmatrix},
    \end{split}
    \label{eq:rulkov-iter-lattice}
\end{equation}
where we denote the fast variable of the $\i$th neuron as $x^{\i}$, the slow variable of the $\i$th neuron as $y^{\i}$, and the coupling parameter of the $\i$th neuron as $\C^\i(k)$. Note that we use the same coupling parameter $\C^\i(k)$ in place of both $\beta(k)$ and $\sigma(k)$ because we set $\sigma^c=\beta^c=1$ for all of the neurons. We will assume that the coupling parameter $\C^\i(k)$ of the $\i$th neuron takes the form
\begin{equation}
    \C^\i(k) = \frac{g}{|\N^\i|}\sum_{\X^\j\in\N^\i}(x^\j(k)-x^\i(k)),
    \label{eq:coup-param}
\end{equation}
where $\N^\i$ is the set of neurons with influence on $\X^\i$ and $g$ is the electrical coupling strength, or coupling conductance. Since current is equal to conductance multiplied by voltage, $\C^\i$ represents the average current flow into the neuron $\X^\i$ from its set of influence $\N^\i$. Here, we assume all the couplings are symmetric and all the coupling strengths are equal. In this paper, we examine electrical coupling strength values between 0 and 1, with $g=0$ being the minimum because it is associated with no coupling between the neurons and $g=1$ being the maximum because it is associated with the current flow equalling the voltage difference. 

One simple way to couple neurons in a lattice is through a nearest-neighbor (NN) coupling, where
\begin{equation}
    \N^{i_1i_2\cdots i_N} = \{\X^{(i_1+1)i_2\cdots i_N}, \X^{(i_1-1)i_2\cdots i_N}, \hdots,\X^{i_1i_2\cdots(i_N+1)},\X^{i_1i_2\cdots(i_N-1)}\}.
    \label{eq:nn-N}
\end{equation}
We will assume periodic boundary conditions, meaning there is an implied $\bmod\, N$ on every index $i$. In the case of NN coupling, $|\N^\i| = 2N$. We also consider a next-nearest-neighbor (NNN) coupling \cite{NNN1, NNN2}, where
\begin{equation}
    \N^{i_1i_2\cdots i_N} = \{\X^{(i_1+1)i_2\cdots i_N}, \X^{(i_1-1)i_2\cdots i_N}, \X^{(i_1+2)i_2\cdots i_N}, \X^{(i_1-2)i_2\cdots i_N}, \hdots, \X^{i_1i_2\cdots(i_N-2)}\}.
    \label{eq:nnn-N}
\end{equation}
In this case, $|\N^\i| = 4N$. We analyze both NN and NNN couplings in this paper. The most strongly coupled neuron lattice takes the form of an all-to-all coupling \cite{mirollo, rulkov2, chen, dePontes}, where
\begin{equation}
    \N^\i = \{\X^\j\mid\j\neq\i\}.
\end{equation}
In this case, $|\N^\i| = \zeta^N-1$. However, we do not examine all-to-all couplings in this paper because they are not affected by the spatial dimensionality of the lattice, which is this paper's focus. 

In addition to examining both NN and NNN couplings of the $N$-dimensional lattice system, we explore three different neuron parameter distributions in the following two sections:
\begin{enumerate}
    \item the homogeneous case, where all neurons have the same $\sigma^\i$ and $\alpha^\i$ values [$\sigma^\i = -0.5$, $\alpha^\i = 4.5$],
    \item the partially heterogeneous case, where each neuron has its own $\sigma^\i$ value but the same $\alpha^\i$ values [$\sigma^\i\in\mathcal{U}(-1.5,-0.5)$, $\alpha^\i = 4.5$],
    \item and the fully heterogeneous case, where each neuron has its own $\sigma^\i$ and $\alpha^\i$ values [$\sigma^\i\in\mathcal{U}(-1.5,-0.5)$, $\alpha^\i\in \mathcal{U}(4.25, 4.75)$].
\end{enumerate}
In all three cases, each neuron has a different $x^\i(0)$ value [$x^\i(0)\in\mathcal{U}(-1,1)$], but they all have the same $y^\i(0)$ value [$y^\i(0) = -3.25$]. For a given neuron in the lattice, the different parameter and initial state values are chosen randomly and uniformly from the specified intervals. These are standard, biologically relevant parameter values that result in regular (tonic) spiking dynamics in the homogeneous case, as well as chaotic spiking and chaotic bursting dynamics in the heterogeneous cases \cite{rulkov}.

Finally, it will be useful to create an analog of the Jacobian matrix for the state tensor $\X$ and iteration function $\F$ in order to compute Lyapunov exponents. In this case, the Jacobian $\J(\X)$ is a type $(N+1, N+1)$ tensor:
\begin{equation}
    \tensor{\J}{^{\i'a'}_{\i a}} = \tensor{\J}{^{i_1'i_2'\cdots i_N'a'}_{i_1i_2\cdots i_Na}} = \frac{\partial \F^{i_1'i_2'\cdots i_N'a'}}{\partial\X^{i_1i_2\cdots i_Na}} = \frac{\partial\F^{\i'a'}}{\partial\X^{\i a}}.
\end{equation}
In Appendix \ref{appdx:jacobian_deriv}, we derive the explicit form of the Jacobian tensor $\tensor{\J}{^{\i'a'}_{\i a}}$ in the NN coupling case:
\begin{equation}
    \tensor{\J}{^{\i'a'}_{\i a}} = 
    \begin{cases}
        \begin{cases}
            0, & \hspace{2.78cm}\text{if $a' = 0$ and $\X^{\i'0}\geq \D^{\i'}$,} \\
            1, & \hspace{2.78cm}\text{otherwise},
        \end{cases} & \text{for $\i = \i'$ and $a=1$} \\
        \\
        \begin{cases}
            -\mu(1+g), & \text{if $a'=1$,} \\
            \alpha^{\i'}(1-\X^{\i'0})^{-2} - g, & \text{if $a'=0$ and $\X^{\i'0}\leq 0$,} \\
            -g, & \text{if $a'=0$ and $0 < \X^{\i'0} < \D^{\i'}$,} \\
            0, & \text{otherwise,}
        \end{cases} & \text{for $\i = \i'$ and $a=0$} \\
        \\
        \begin{cases}
            \mu g/2N, & \hspace{1.92cm}\text{if $a'=1$,} \\
            0, & \hspace{1.92cm}\text{if $a' = 0$ and $\X^{\i'0}\geq \D^{\i'}$,} \\
            g/2N, & \hspace{1.92cm}\text{otherwise}
        \end{cases} & \text{for $\i = \i'^*_{m\pm}$ and $a=0$} \\
        \\
        0, & \text{otherwise}
    \end{cases}.
    \label{eq:NN-jacobian}
\end{equation}
Here, we use the notation $\D^{\i'} = \alpha^{\i'} + \X^{\i'1} + \C^{\i'}$ and $\i^*_{m\pm} = \{i_1,\hdots,i_{m-1},i_m\pm 1,i_{m+1},\hdots,i_N\}$ for some $1\leq m\leq N$. The Jacobian tensor in the NNN coupling case takes a similar form [Eq. \eqref{eq:NNN-jacobian}]. In Appendix \ref{appdx:tensor-matrix-scheme}, we discuss how to convert from this tensorial form of the state tensor $\X^{\i a}$ and Jacobian $\tensor{\J}{^{\i'a'}_{\i a}}$ to a standard state vector $\mb{X}^\nu$ and Jacobian matrix $\tensor{\mb{J}}{^{\nu'}_{\nu}}$ for the purposes of computational implementation. Now that we have introduced the $N$-dimensional lattice model and the relevant tools, we now present our analysis of the model's dynamics.

\begin{figure}[t!]
    \centering
    \begin{tikzpicture}[scale=1.25]
    \begin{axis}[
        view={60}{30},
        axis equal image,
        hide axis,
        width=14cm,
    ]
        \addplot3[
            surf,
            shader=flat,
            color=gray!30,
            samples=45,
            samples y=25,
            mesh/ordering=y varies,
            domain=0:360,
            y domain=0:360,
            opacity=0.35,
        ]
        ({(2 + 0.6 * cos(y)) * cos(x)},
         {(2 + 0.6 * cos(y)) * sin(x)},
         {0.6 * sin(y)});
        
        \def\N{8}
        \def\circlelinewidth{0.8pt}
        
        \foreach \m in {0,...,7} {
            \foreach \n in {0,...,7} {
                \pgfmathsetmacro{\theta}{\m * 360/\N}
                \pgfmathsetmacro{\phi}{\n * 360/\N}
        
                \pgfmathsetmacro{\X}{(2 + 0.6 * cos(\phi)) * cos(\theta)}
                \pgfmathsetmacro{\Y}{(2 + 0.6 * cos(\phi)) * sin(\theta)}
                \pgfmathsetmacro{\Z}{0.6 * sin(\phi)}
                \addplot3[only marks, mark=*, mark size=1.5pt, color=blue] coordinates {(\X,\Y,\Z)};
            }
        }
        
        \foreach \n in {0,...,7} {
            \pgfmathsetmacro{\phi}{\n * 360/\N}
            \addplot3[mark=none, color=yellow!95!black, very thick, samples=45, variable=t, domain=0:360]
            ({(2 + 0.6 * cos(\phi)) * cos(t)},
             {(2 + 0.6 * cos(\phi)) * sin(t)},
             {0.6 * sin(\phi)});
        }
        
        \foreach \m in {0,...,7} {
            \pgfmathsetmacro{\theta}{\m * 360/\N}
            \addplot3[mark=none, color=yellow!95!black, very thick, samples=25, variable=t, domain=0:360]
            ({(2 + 0.6 * cos(t)) * cos(\theta)},
             {(2 + 0.6 * cos(t)) * sin(\theta)},
             {0.6 * sin(t)});
        }
    \end{axis}
    \end{tikzpicture}
    \caption{Visualization of a two-dimensional electrically coupled neuron lattice with $\zeta=8$ neurons per side and periodic boundary conditions. The neurons are shown as blue points, and the electrical coupling connections are shown in gold.}
    \label{fig:torus}
\end{figure}
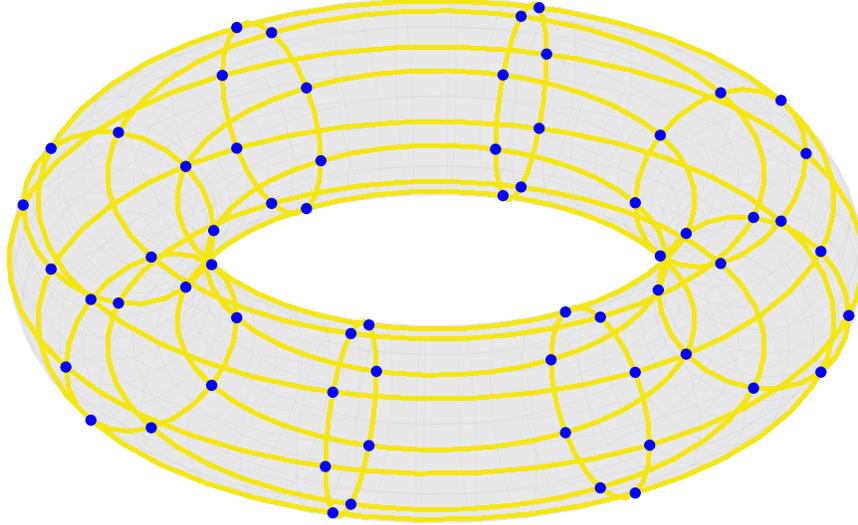

\section{Two-dimensional lattices}
\label{sec:2d-lattices}

We first consider the $N=2$ case, or two-dimensional lattices of NN electrically coupled Rulkov neurons with periodic boundary conditions. In Fig. \ref{fig:torus}, we present a visualization of one such lattice when $\zeta=8$ and the couplings are nearest-neighbor (NN), where each neuron is represented by a blue point and an electrical coupling connection is shown with a gold line. Because of the periodic boundary conditions, the resulting topology of the lattice is toroidal. In the figure, we can see that each neuron is electrically coupled with four other neurons in the lattice: two in the toroidal direction and two in the poloidal direction. We also note that there are $\zeta^2 = 64$ neurons in total, and the state space of the system is $n=2\zeta^2 = 128$-dimensional.

In this section, we begin by performing a detailed analysis of the system shown in Fig. \ref{fig:torus} in the homogeneous, partially heterogeneous, and fully heterogeneous cases. We first analyze the chaotic dynamics of the system by computing the system's maximal Lyapunov exponents across a wide range of electrical coupling strength values and analyzing the resulting curve with representative examples of the different dynamical regimes (Sec. \ref{subsec:2D-NN}). Then, we change the lattice coupling to next-nearest-neighbor (NNN) and examine the effect of the greater number of coupling connections on the dynamics (Sec. \ref{subsec:2D-NNN}). Finally, we qualitatively discuss the collective dynamics and complex dynamical regimes of a large two-dimensional neuron lattice with $\zeta = 300$, resulting in an $n=180000$-dimensional state space (Sec. \ref{subsec:2D-large}).

\begin{figure}[t!]
    \centering
    \begin{subfigure}{0.6\textwidth}
        \centering
        \includegraphics[width=\textwidth]{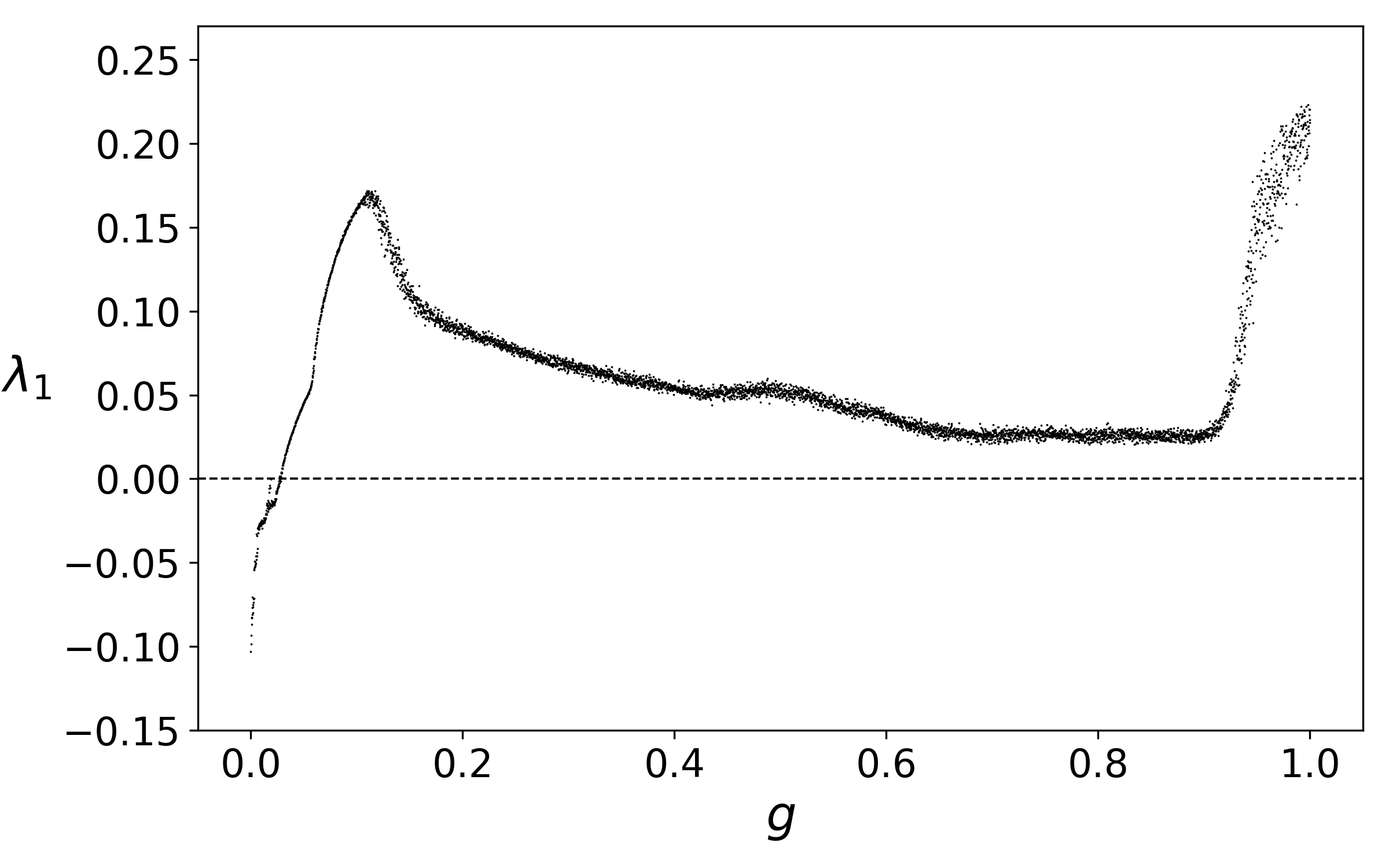}
        \caption{Homogeneous case}
        \label{fig:lambda1_vs_g_homo}
    \end{subfigure} \\[0.5cm]
    \begin{subfigure}{0.475\textwidth}
        \centering
        \includegraphics[width=\textwidth]{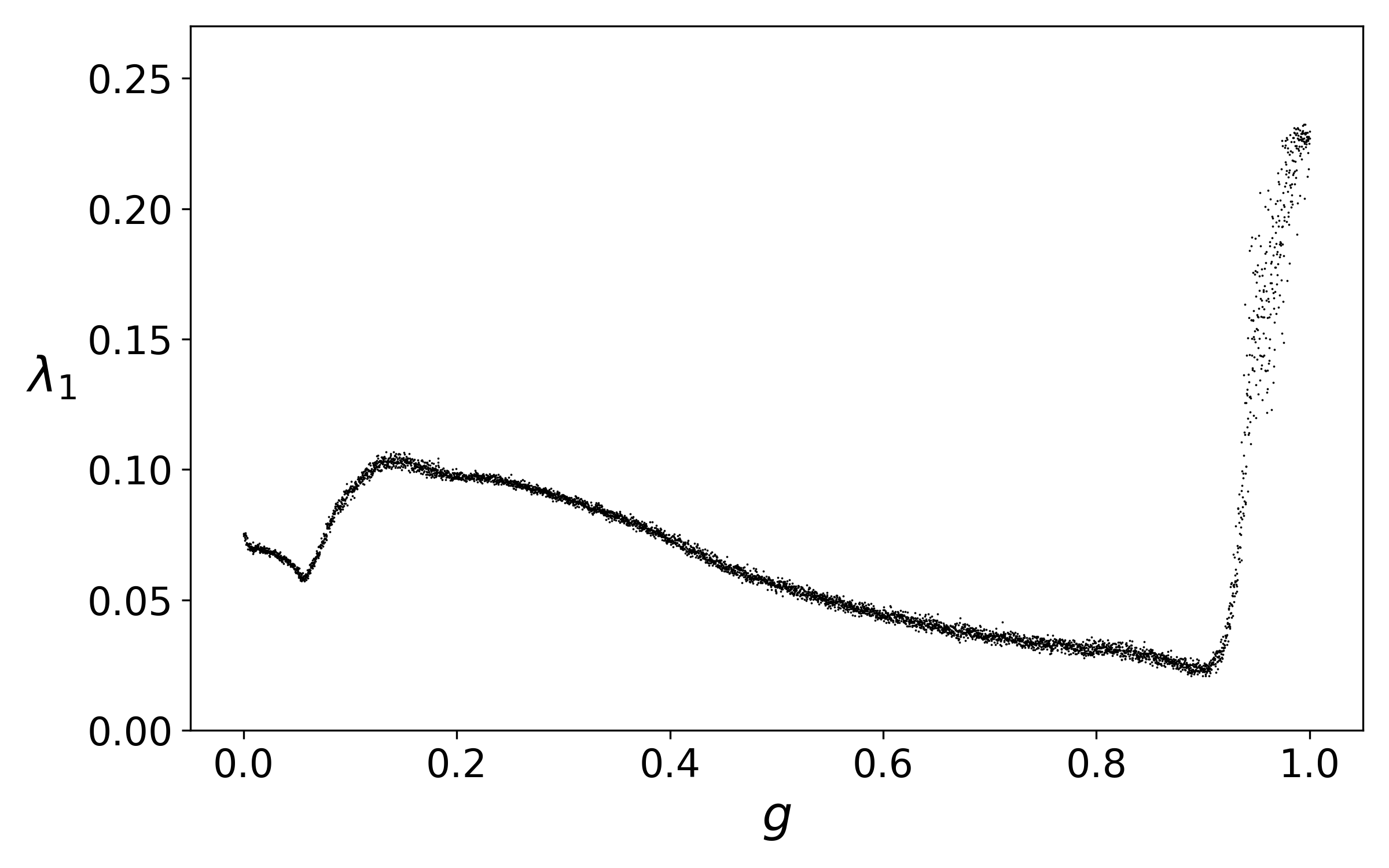}
        \caption{Partially heterogeneous case}
        \label{fig:lambda1_vs_g_phetero}
    \end{subfigure}
    \hfill
    \begin{subfigure}{0.475\textwidth}
        \centering
        \includegraphics[width=\textwidth]{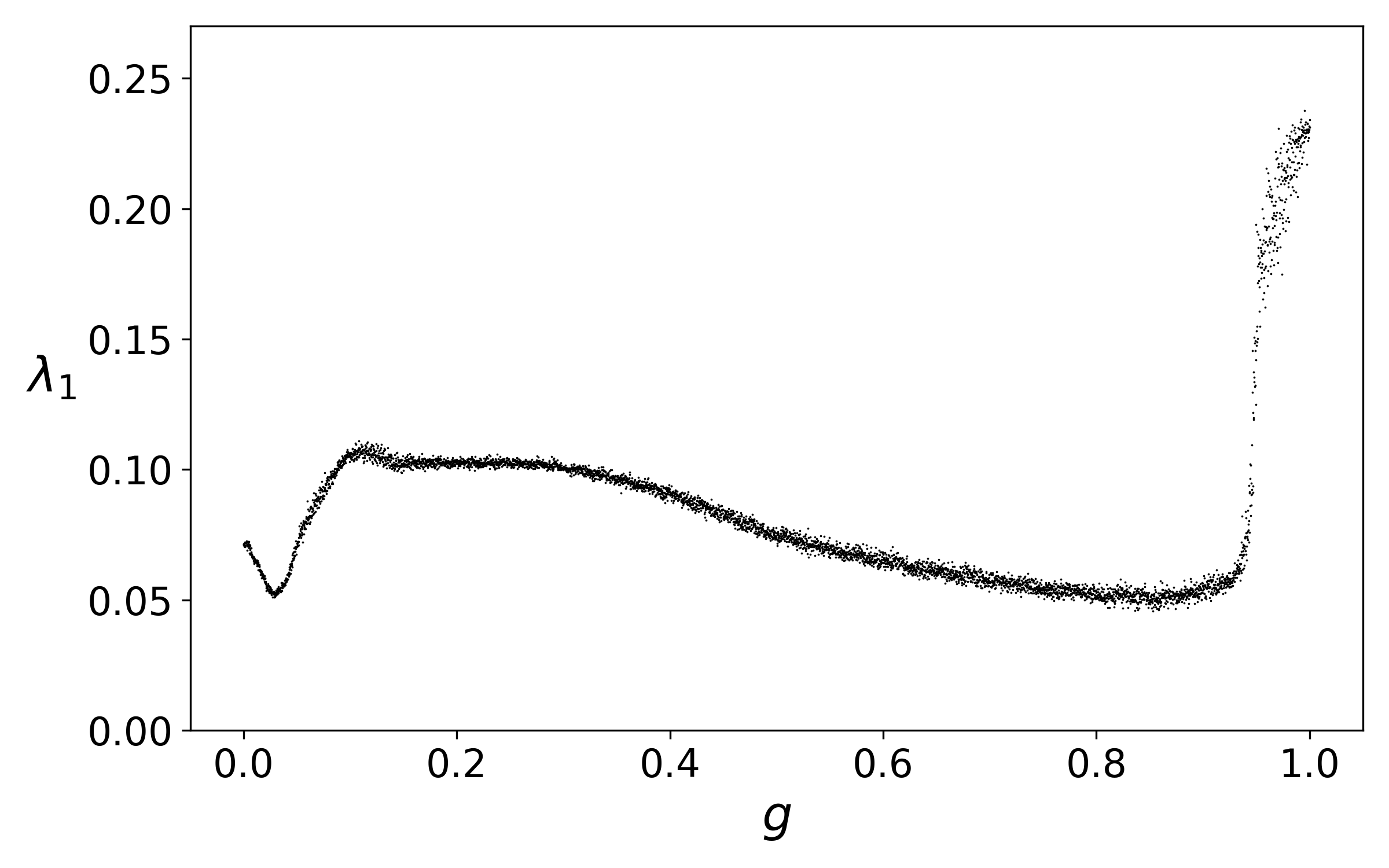}
        \caption{Fully heterogeneous case}
        \label{fig:lambda1_vs_g_fhetero}
    \end{subfigure}
    \caption{Graphs of the maximal Lyapunov exponent $\lambda_1$ for 5000 electrical coupling strength values $g$ between 0 and 1 in the homogeneous, partially heterogeneous, and fully heterogeneous cases of an $N=2$-dimensional lattice of NN electrically coupled Rulkov neurons with $\zeta=8$ neurons per side. The Lyapunov exponents are calculated using orbits of length $k=10000$.}
    \label{fig:lambda1_vs_g_graphs}
\end{figure}

\subsection{NN dynamics}
\label{subsec:2D-NN}

In this section, we analyze the dynamics of the system displayed in Fig. \ref{fig:torus}: a two-dimensional lattice of nearest-neighbor (NN) electrically coupled Rulkov neurons with $\zeta = 8$ neurons per side. We examine the system's dynamics over a range of 5000 electrical coupling strength values $g$ between 0 (no electrical interactions between neurons) and 1 (current flow equal to voltage difference) by computing the $2\zeta^2 = 128$ Lyapunov exponents of the system at each value of $g$. Specifically, we compute the system's Lyapunov exponents by determining the Jacobian matrices $\J(\X)$ [Eq. \eqref{eq:NN-jacobian}] at each $\X\in\{\X(0),\X(1),\hdots,\X(10000)\}$ and using the QR factorization method of computing Lyapunov spectra \cite{eckmann, thesis}. In Fig. \ref{fig:lambda1_vs_g_graphs}, we plot the maximal Lyapunov exponents $\lambda_1$ for the 5000 values of $g$ between 0 and 1 in the homogeneous case, partially heterogeneous case, and fully heterogeneous case. 

\begin{figure}[t!]
    \centering
    \begin{subfigure}{0.485\textwidth}
        \centering
        \includegraphics[width=0.8\textwidth]{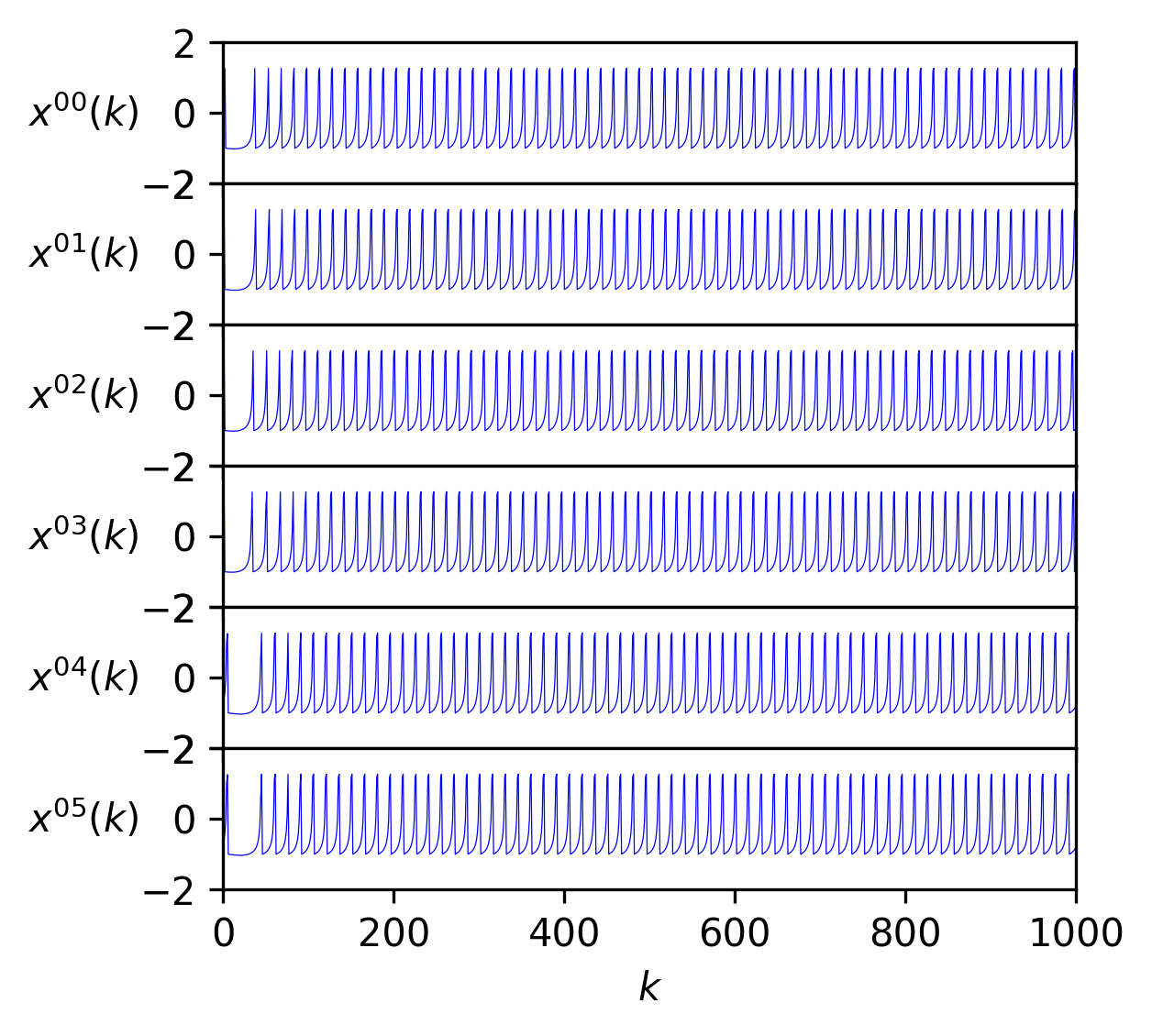}
        \caption{\centering $g=0$, $\lambda_1\approx -0.103$ (uncoupled nonchaotic spiking regime)}
        \label{fig:xvk_homo_g0}
    \end{subfigure}
    \hfill
    \begin{subfigure}{0.485\textwidth}
        \centering
        \includegraphics[width=0.8\textwidth]{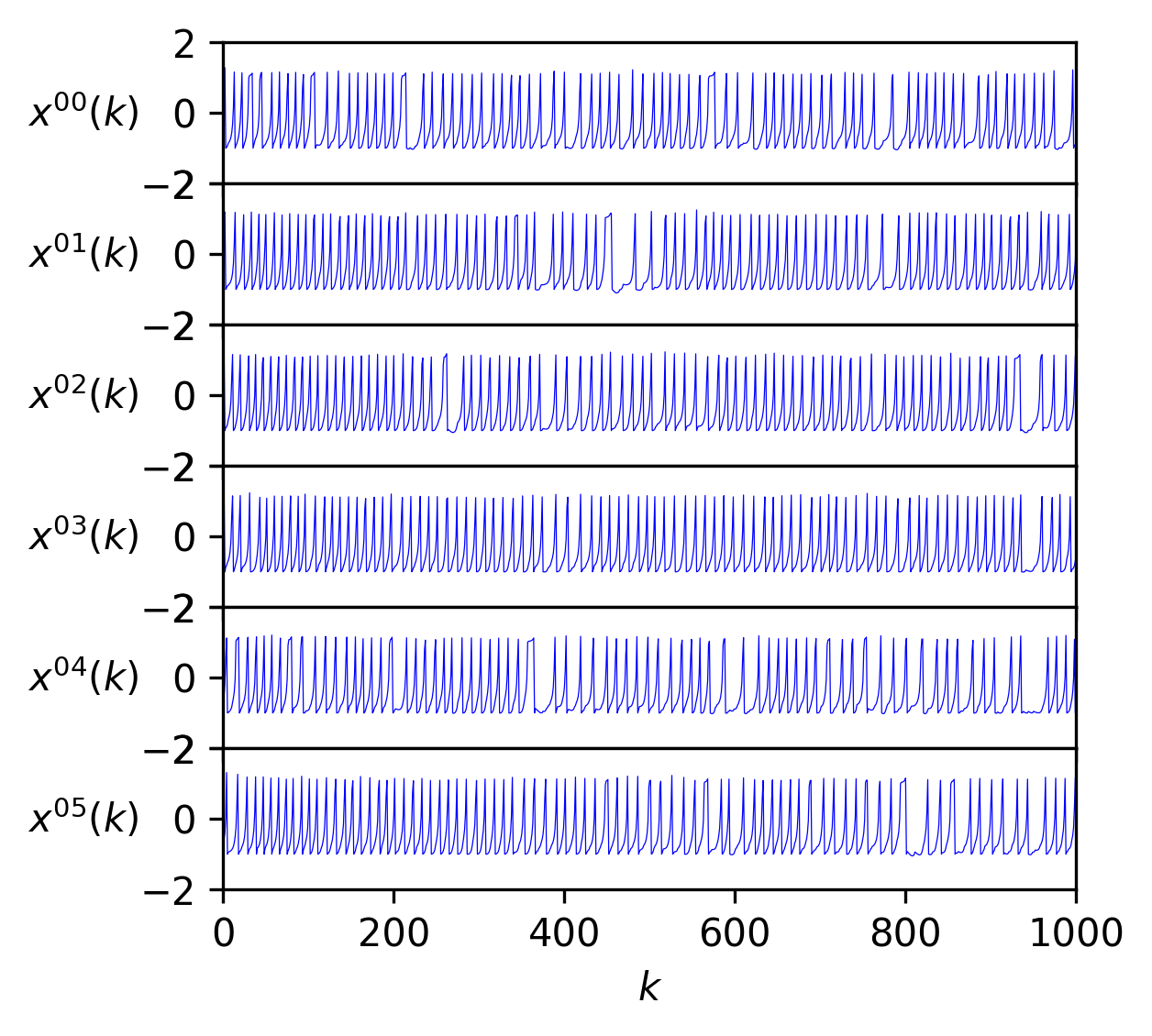}
        \caption{\centering $g=0.1$, $\lambda_1\approx 0.162$ (unsynchronized chaotic spiking regime)}
        \label{fig:xvk_homo_g0.1}
    \end{subfigure} \\[0.5cm]
    \begin{subfigure}{0.495\textwidth}
        \centering
        \includegraphics[width=0.8\textwidth]{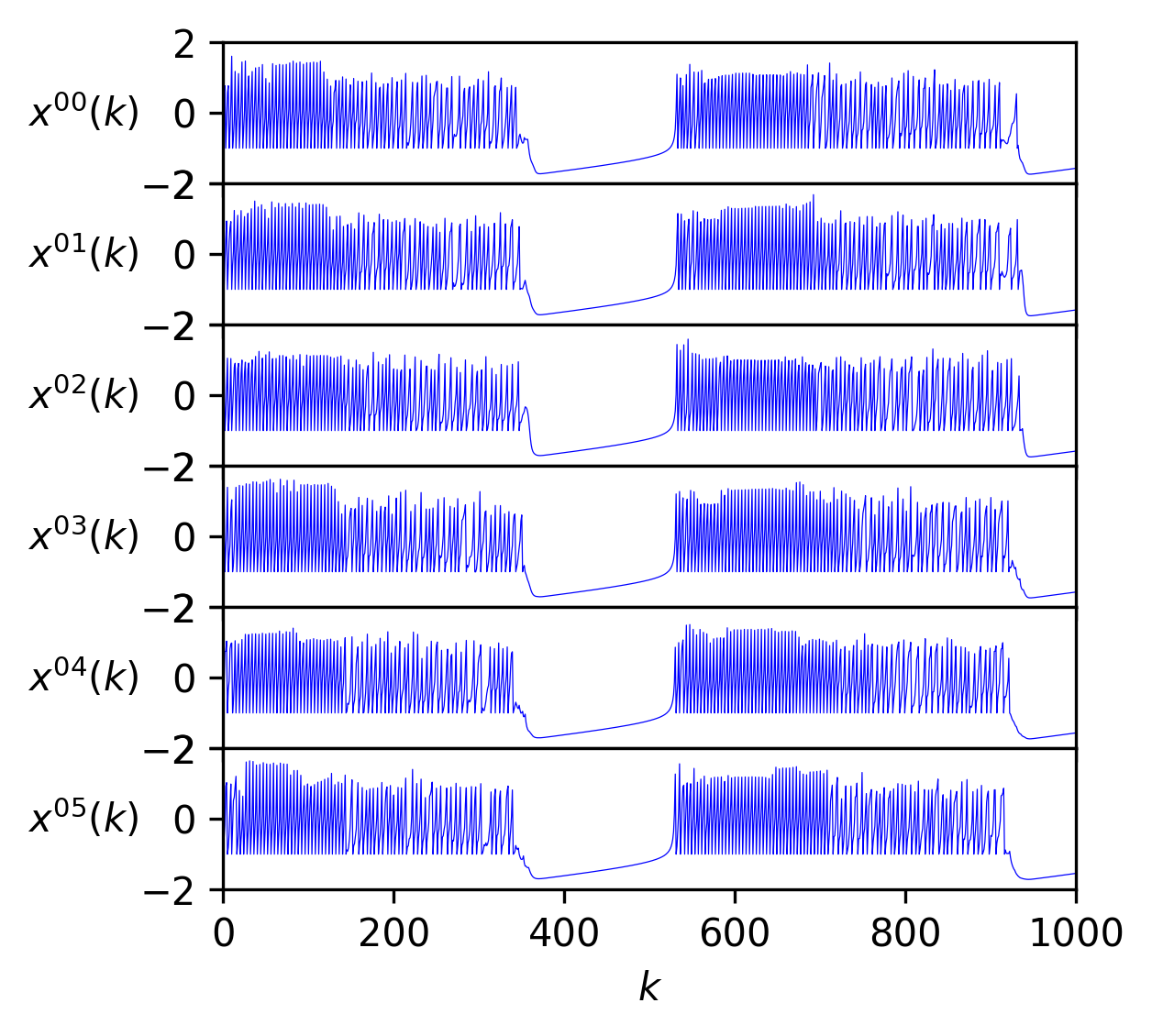}
        \caption{\centering $g=0.5$, $\lambda_1\approx 0.051$ (synchronized chaotic bursting regime)}
        \label{fig:xvk_homo_g0.5}
    \end{subfigure}
    \hfill
    \begin{subfigure}{0.495\textwidth}
        \centering
        \includegraphics[width=0.8\textwidth]{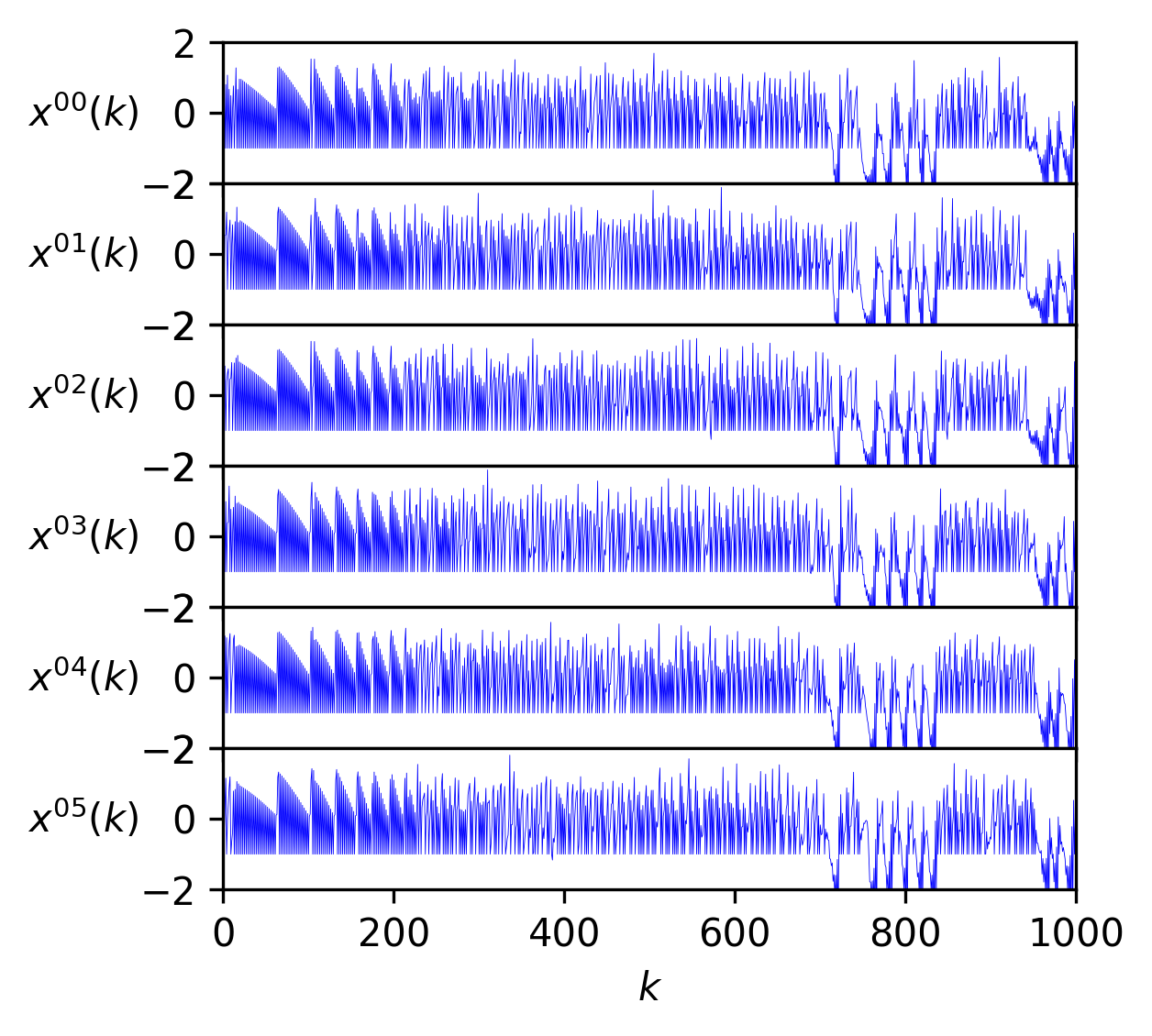}
        \caption{\centering $g=1$, $\lambda_1\approx 0.220$ (synchronized hyperchaotic regime)}
        \label{fig:xvk_homo_g1}
    \end{subfigure}
    \caption{Graphs of the voltages $x(k)$ of six adjacent homogeneous NN coupled neurons for four select values of $g$. The system is an $N=2$-dimensional lattice of NN electrically coupled homogeneous Rulkov neurons with $\zeta=8$ neurons per side.}
    \label{fig:xvk_homo_graphs}
\end{figure}

We begin by discussing the homogeneous case, where all of the neurons in the lattice have the same parameters as specified in Sec. \ref{sec:the-model}. We find that the dynamics of the two-dimensional lattice of homogeneous Rulkov neurons follows similar patterns to those of the 1-dimensional lattice in Ref. \cite{ring}, which we loosely categorize into four dynamical regimes of behavior: (a) the uncoupled nonchaotic spiking regime, (b) the unsynchronized chaotic spiking regime, (c) the synchronized chaotic bursting regime, and (d) the synchronized hyperchaotic regime. In Fig. \ref{fig:lambda1_vs_g_homo}, the shape of the curve of $\lambda_1$ vs. $g$ follows these dynamical regimes. $\lambda_1$ begins negative in the uncoupled nonchaotic spiking regime at $g=0$ (Fig. \ref{fig:xvk_homo_g0}) before following a steep increase through the unsynchronized chaotic spiking regime (Fig. \ref{fig:xvk_homo_g0.1}) and becomes positive when $0\lesssim g\lesssim 0.15$. Then, there is a sharp drop at the transition to the synchronized chaotic bursting regime (Fig. \ref{fig:xvk_homo_g0.5}) and a gradual slope down for $0.2\lesssim g\lesssim 0.8$ as the frequency of the slow oscillations between spiking and silence decreases, creating longer periods of silence that contribute to a lower $\lambda_1$. Finally, there is a sharp turn in the $\lambda_1$ curve at the transition to synchronized hyperchaos (Fig. \ref{fig:xvk_homo_g1}) and a steep increase in $\lambda_1$ up through $0.9\lesssim g\lesssim 1$ where each neuron has a complete influence over its neighbors. Having provided an overview of the chaotic behavior emerging from nearest-neighbor interactions in a two-dimensional lattice of $64$ neurons, we now discuss each of these dynamical regimes in more detail.

When the electrical coupling between neurons is $g=0$, each individual neuron is operating under its own parameters, insensitive to the topology of the system. Consider the dynamics in Fig. \ref{fig:xvk_homo_g0}: each individual homogeneous neuron is spiking with the same characteristic frequency, and the system is effectively operating as 64 identical copies of the same, lone, regular spiking neuron. As the coupling parameter $g$ slowly increases from $0$ to $0.15$, we observe irregularities resulting from the small interactions of neurons with their neighbors that introduce chaos into the lattice dynamics. Since $g$ is still small, these dynamics should be understood as a small perturbation to ``free neuron'' dynamics, the behavior of which we have just discussed and represented in Fig. \ref{fig:xvk_homo_g0}. An example of the dynamics in the unsynchronized chaotic spiking regime is shown in Fig. \ref{fig:xvk_homo_g0.1}, where irregularities occur when the voltage of one neuron happens to catch onto the voltage of one of its neighbors. Since this chaos manifests itself through the chance interactions between neighboring neurons, a perturbation in the initial condition of any neuron in the lattice will almost certainly be magnified with time. This sensitivity results in large maximal Lyapunov exponents ($\lambda_1\approx 0.162$ for $g=0.1$) and a large number of positive Lyapunov exponents (24 out of the 128 Lyapunov exponents are positive for $g=0.1$). More precisely, there are many orthogonal eigenvectors of the Jacobian matrix $\J(\X)$ with associated positive eigenvalues since there are many directions in $128$-dimensional state space that chaos can manifest.

When the coupling strength enters the region of $0.2 \lesssim g \lesssim 0.8$, the lattice undergoes synchronized chaotic bursting. Bursting is a dynamical phase of Rulkov neurons characterized by slow oscillations between bursts of spikes and quiescence \cite{rulkov}. In the synchronized chaotic bursting regime, the neurons exhibit synchronized slow oscillations with unsynchronized spikes within the bursts, as depicted in Fig. \ref{fig:xvk_homo_g0.5}. The emergence of synchronized bursts of spikes can be attributed to the same phenomenon that arises due to the injection of direct current discussed in Sec. \ref{subsec:rulkov-neuron} (see Fig. \ref{fig:current_response_beta}). The periods of silence between bursts of spikes contribute to a decrease in the maximal Lyapunov exponent of the system, since a perturbation to the system at any point during the quiescent phase will quickly vanish. As previously mentioned, as the electrical coupling is strengthened through the synchronized chaotic bursting regime, the stronger synchronization results in a decrease in the slow oscillations between spiking and silence. We comment that a primary reason why this dynamical phase is important is because it clarifies when ``something special'' is happening in the neuronal system --- if the dynamics of the neuronal system follows distinguishable periods of chaotic spiking and resting, then one may easily distinguish between a neuron (or groups of neurons) that is ``on'' or ``off,'' as is done in statistical modeling of neuronal systems \cite{on-off-prop-logic-model, hopfield-network}.

The last dynamical regime is realized by increasing $g$ until $g \simeq 1$. Since in this regime the current flow between adjacent neurons is essentially equal to their voltage difference [see Eq. \eqref{eq:coup-param}], the lattice undergoes synchronized hyperchaos, in which each neuron's dynamics effectively mirrors its neighbors'. The strong influence each neuron has on its neighbors manifests as the quick and jagged spikes displayed in Fig. \ref{fig:xvk_homo_g1}, and results in the largest maximal Lyapunov exponent value seen so far in this system ($\lambda_1\approx 0.220$ for $g=1$). However, it should be noted that the number of positive Lyapunov exponents in the synchronized hyperchaotic regime (11 out of 128 for $g=1$) is much lower than the number of positive Lyapunov exponents in the unsynchronized chaotic spiking regime (24 out of 128 for $g=0.1$), which are the two $\lambda_1$ peaks in Fig. \ref{fig:lambda1_vs_g_homo}. This is due to the strongly synchronized nature of this hyperchaotic regime: a perturbation that takes a neuron out of synchronization will quickly vanish as the neuron falls back into spiking with its neighbors. In other words, the strongly chaotic dynamics are confined to only a few directions in 128-dimensional state space, which results in fewer eigenvectors of the Jacobian $\J(\X)$ having positive eigenvalues. 

\begin{figure}[t!]
    \centering
    \begin{subfigure}{0.495\textwidth}
        \centering
        \includegraphics[width=0.8\textwidth]{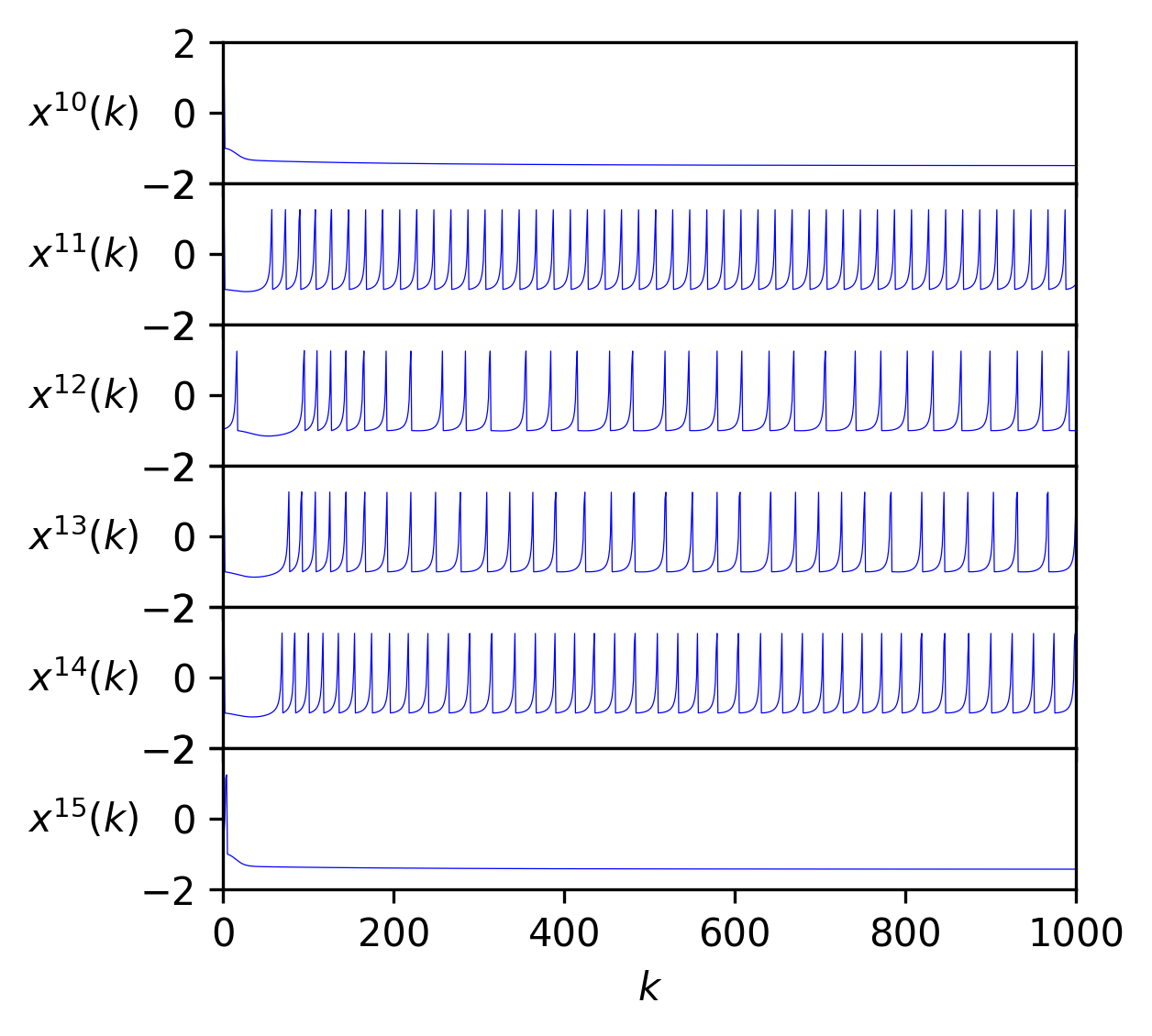}
        \caption{\centering $g=0$, $\lambda_1\approx 0.075$ (uncoupled regime)}
        \label{fig:xvk_phetero_g0}
    \end{subfigure}
    \hfill
    \begin{subfigure}{0.495\textwidth}
        \centering
        \includegraphics[width=0.8\textwidth]{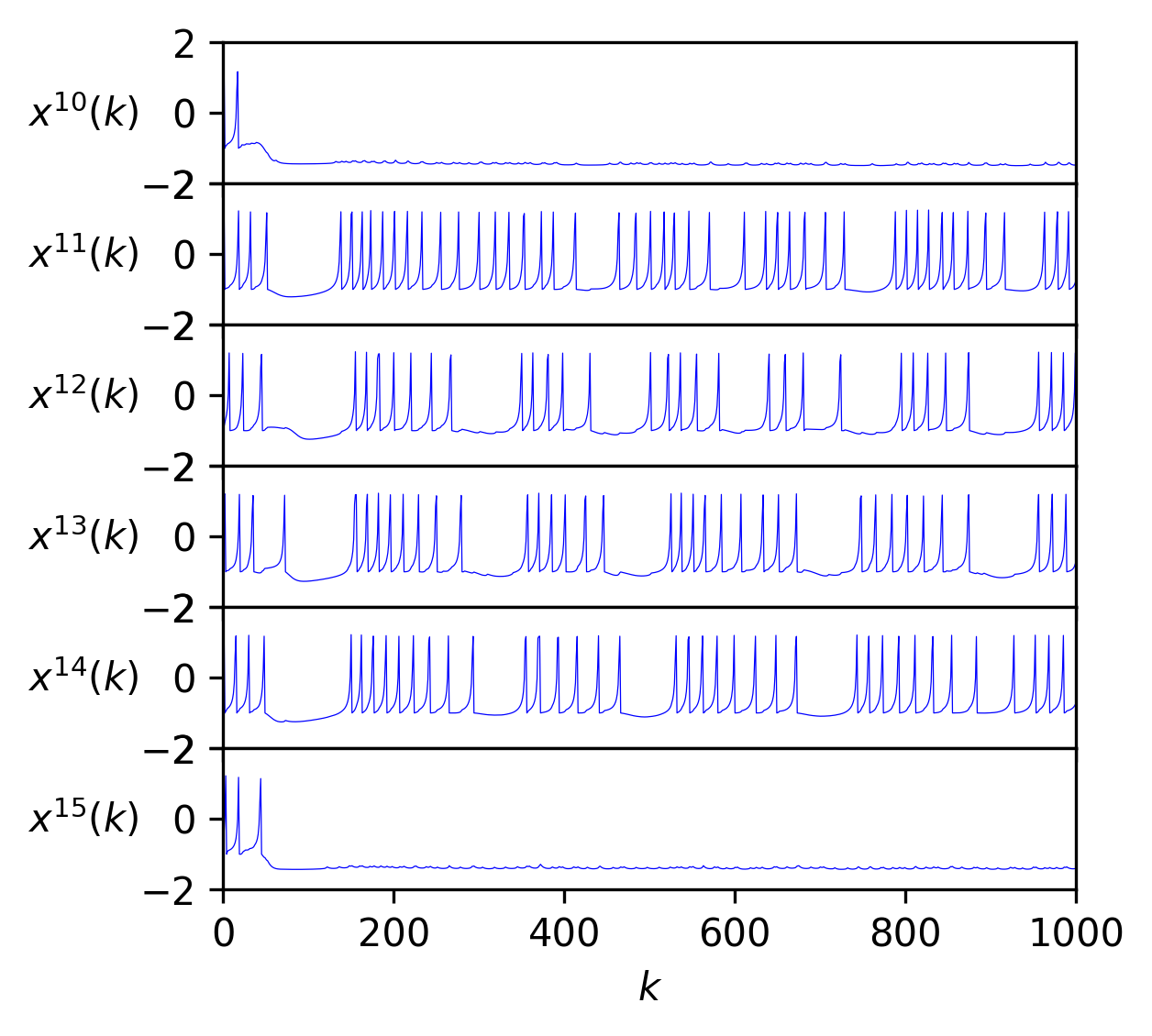}
        \caption{\centering $g=0.05$, $\lambda_1\approx 0.065$ (local quasi-bursting regime)}
        \label{fig:xvk_phetero_g0.05}
    \end{subfigure} \\[0.5cm]
    \begin{subfigure}{0.32\textwidth}
        \centering
        \includegraphics[width=\textwidth]{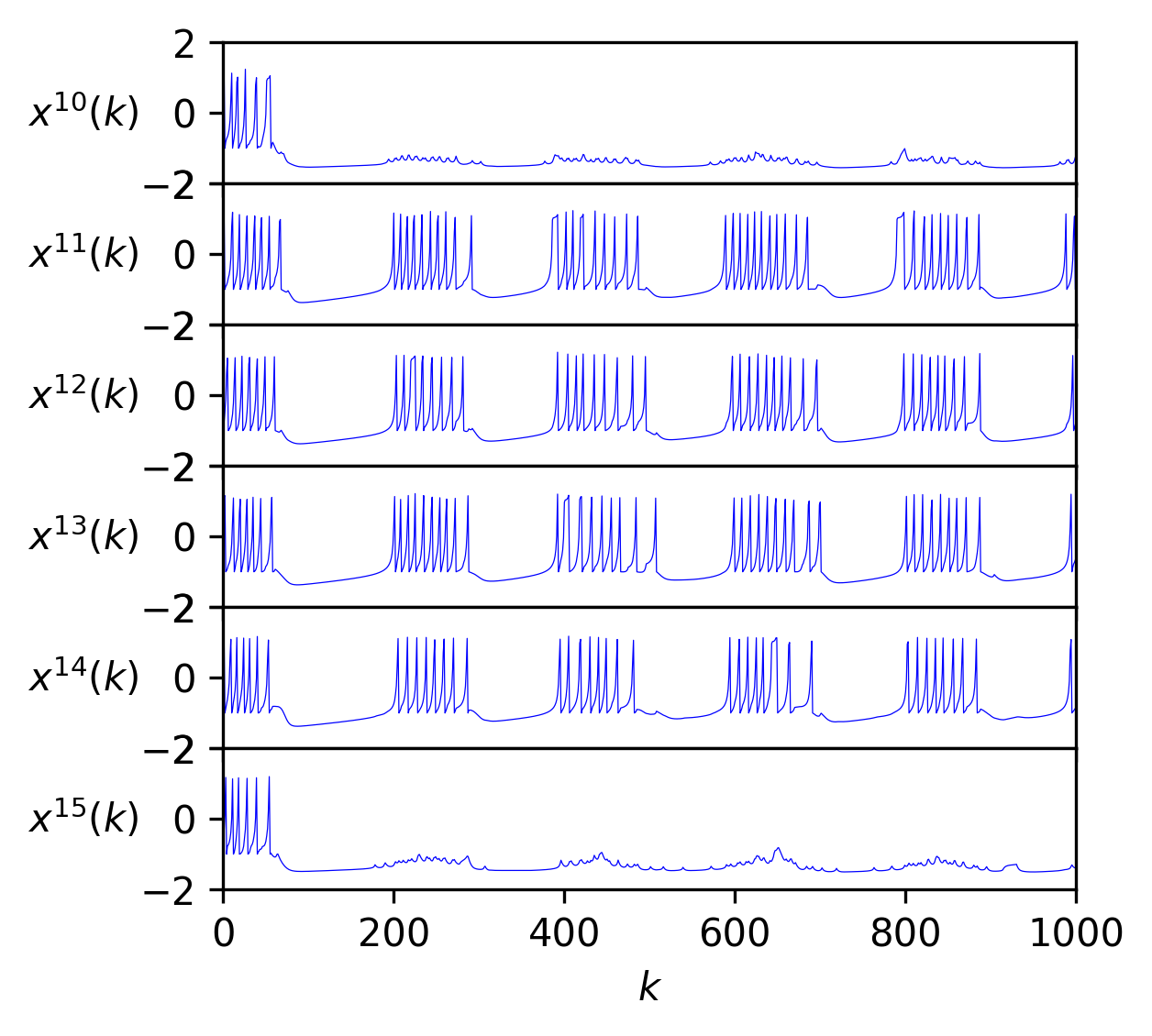}
        \caption{\centering $g=0.13$, $\lambda_1\approx 0.109$ (local chaotic high-frequency bursting regime)}
        \label{fig:xvk_phetero_g0.13}
    \end{subfigure}
    \hfill
    \begin{subfigure}{0.32\textwidth}
        \centering
        \includegraphics[width=\textwidth]{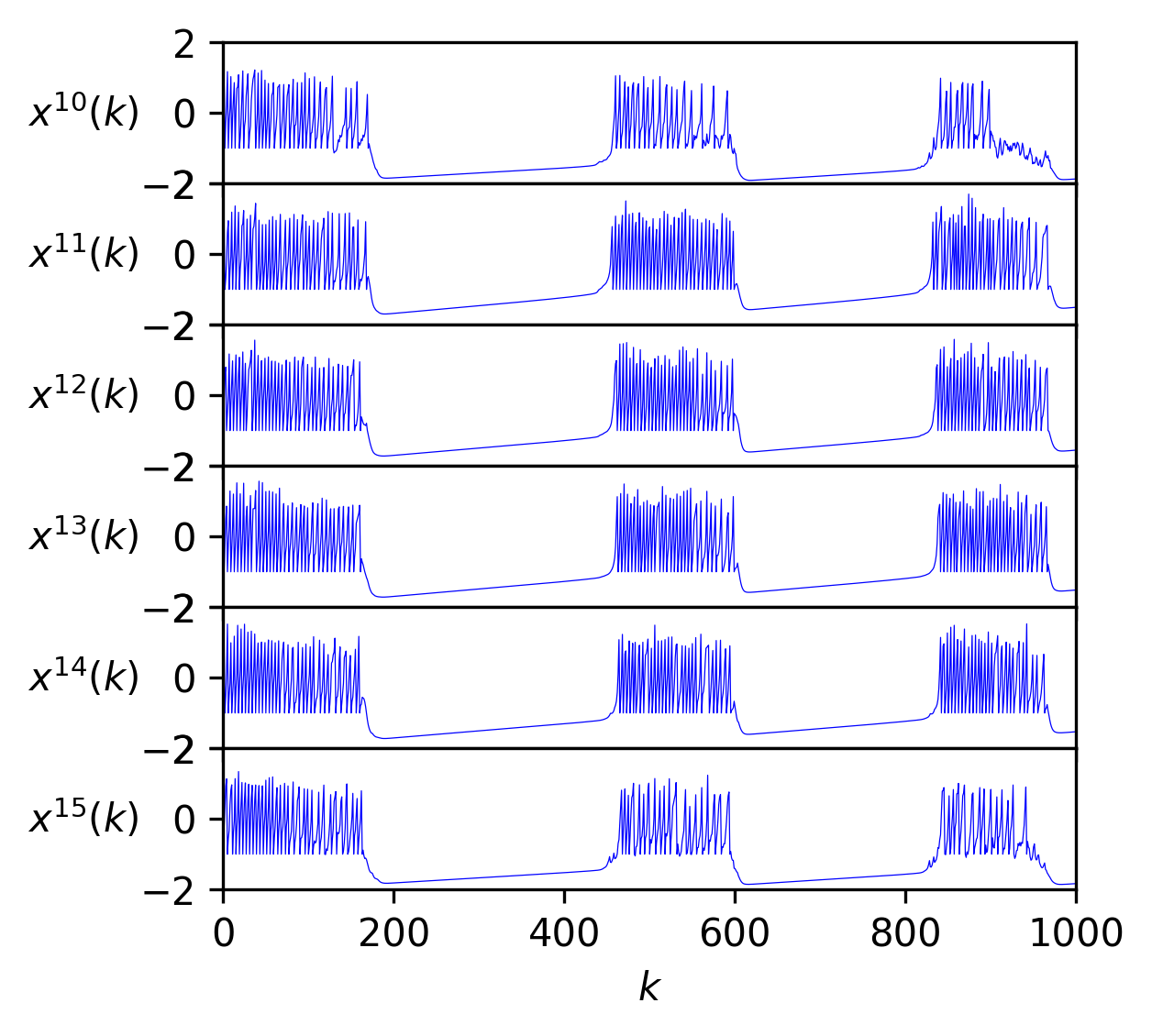}
        \caption{\centering $g=0.5$, $\lambda_1\approx 0.065$ (synchronized chaotic low-frequency bursting regime)}
        \label{fig:xvk_phetero_g0.5}
    \end{subfigure}
    \hfill
    \begin{subfigure}{0.32\textwidth}
        \centering
        \includegraphics[width=\textwidth]{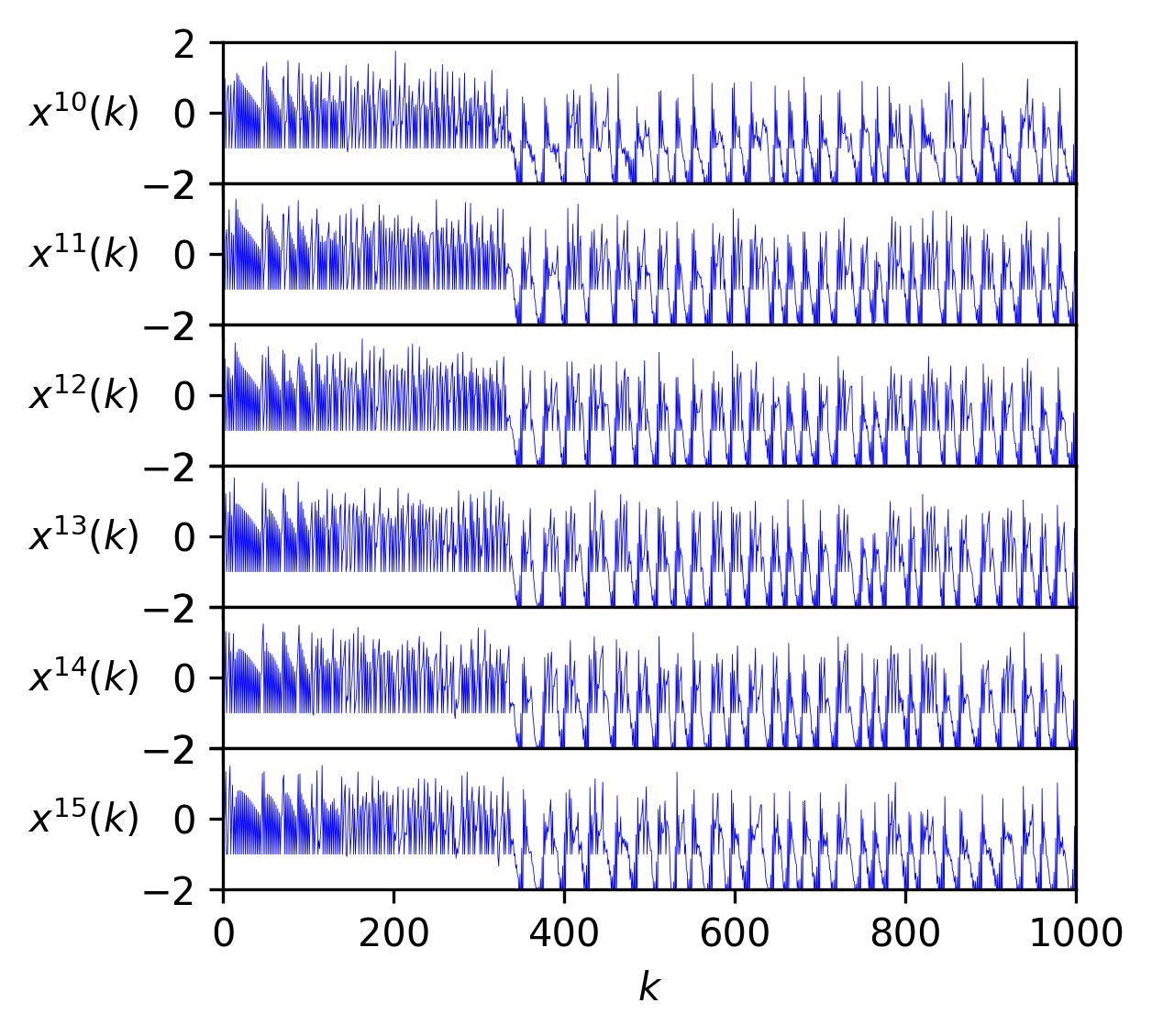}
        \caption{\centering $g=1$, $\lambda_1\approx 0.233$ (synchronized hyperchaotic regime)}
        \label{fig:xvk_phetero_g1}
    \end{subfigure}
    \caption{Graphs of the voltages $x(k)$ of six adjacent partially heterogeneous NN coupled neurons for five select values of $g$. The system is an $N=2$-dimensional lattice of NN electrically coupled partially heterogeneous Rulkov neurons with $\zeta=8$ neurons per side.}
    \label{fig:xvk_phetero_graphs}
\end{figure}

\begin{figure}[t!]
    \centering
    \begin{subfigure}{0.495\textwidth}
        \centering
        \includegraphics[width=0.8\textwidth]{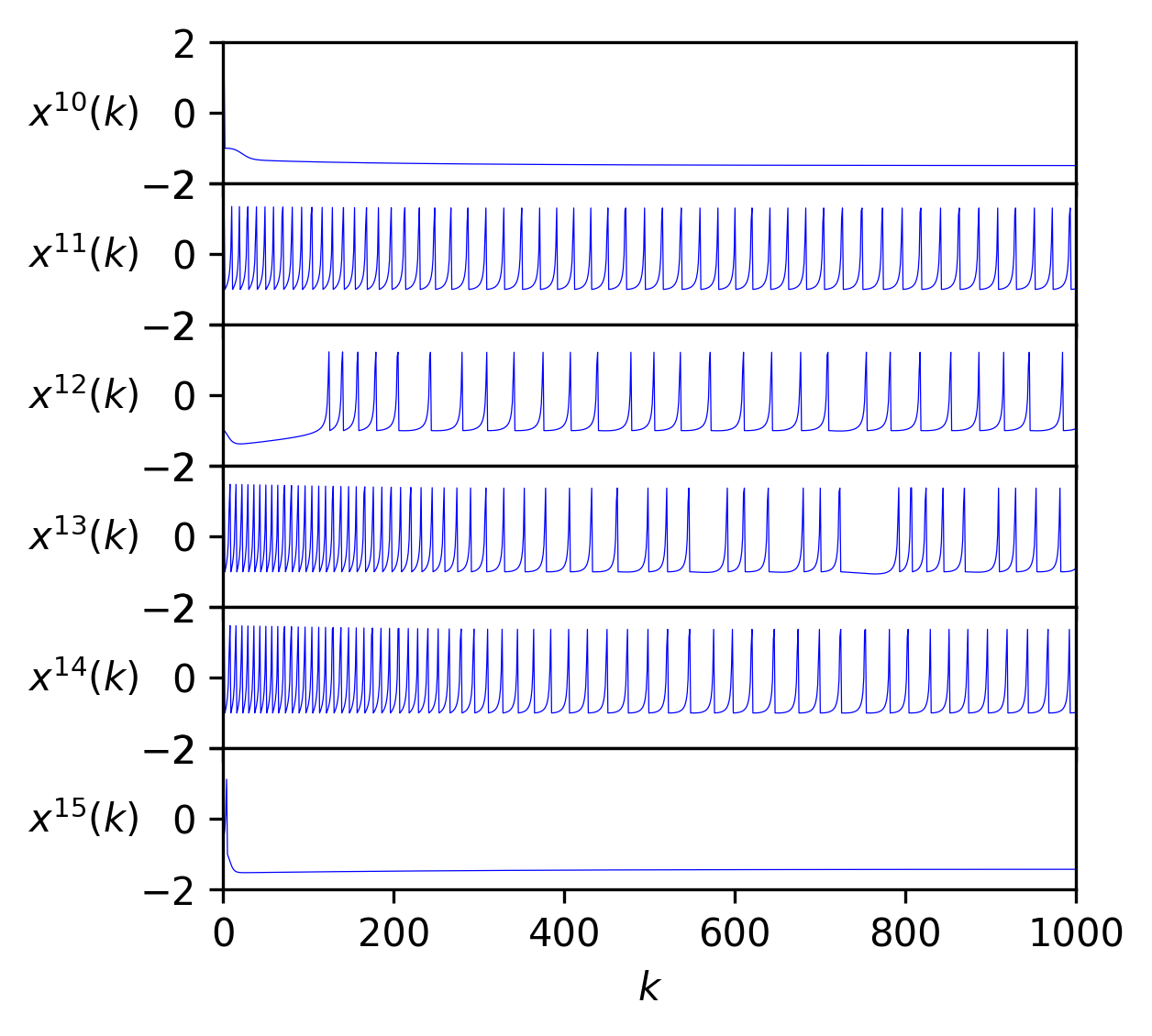}
        \caption{\centering $g=0$, $\lambda_1\approx 0.073$ (uncoupled regime)}
        \label{fig:xvk_fhetero_g0}
    \end{subfigure}
    \hfill
    \begin{subfigure}{0.495\textwidth}
        \centering
        \includegraphics[width=0.8\textwidth]{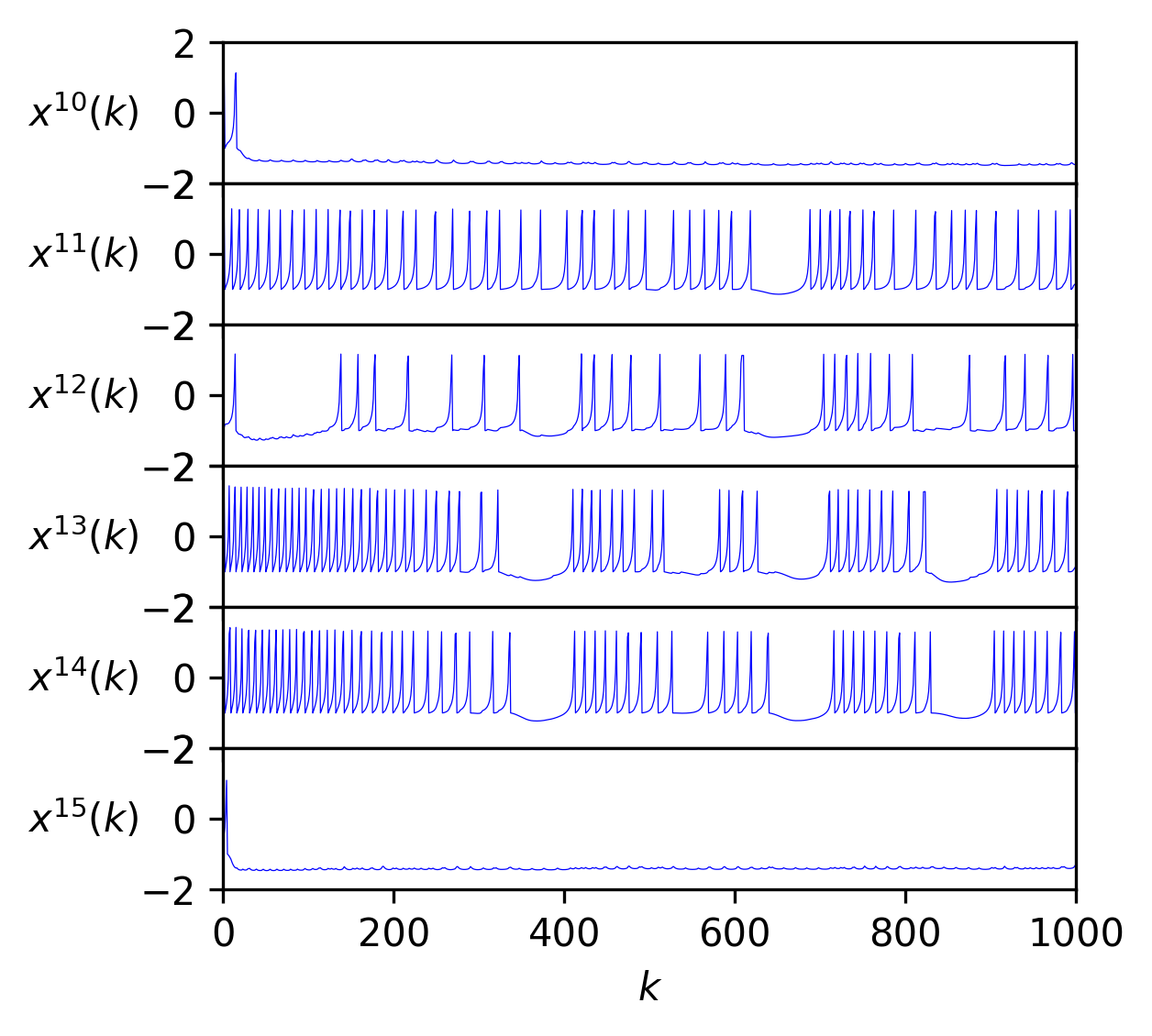}
        \caption{\centering $g=0.05$, $\lambda_1\approx 0.064$ (local quasi-bursting regime)}
        \label{fig:xvk_fhetero_g0.05}
    \end{subfigure} \\[0.5cm]
    \begin{subfigure}{0.32\textwidth}
        \centering
        \includegraphics[width=\textwidth]{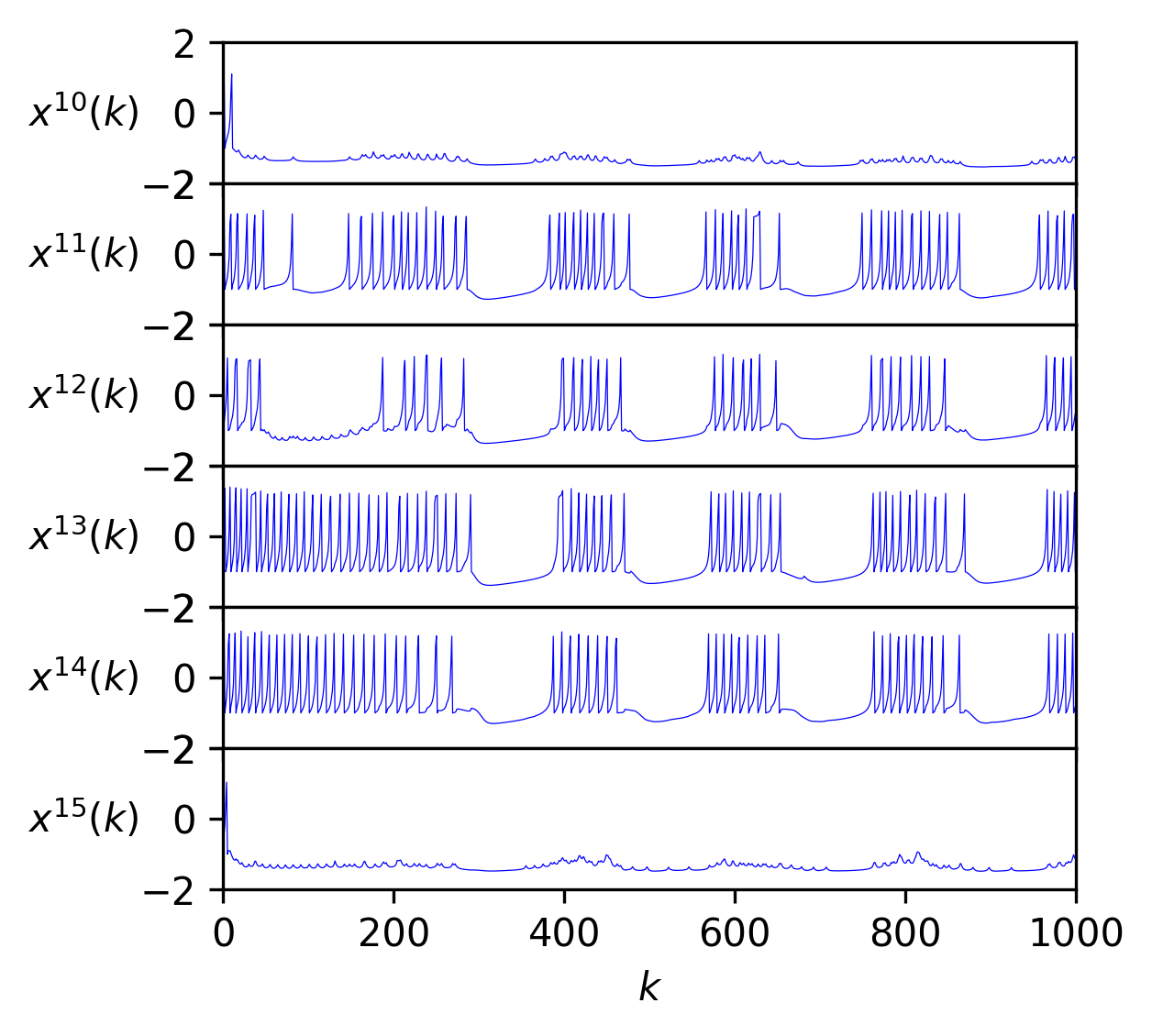}
        \caption{\centering $g=0.13$, $\lambda_1\approx 0.104$ (local chaotic high-frequency bursting regime)}
        \label{fig:xvk_fhetero_g0.13}
    \end{subfigure}
    \hfill
    \begin{subfigure}{0.32\textwidth}
        \centering
        \includegraphics[width=\textwidth]{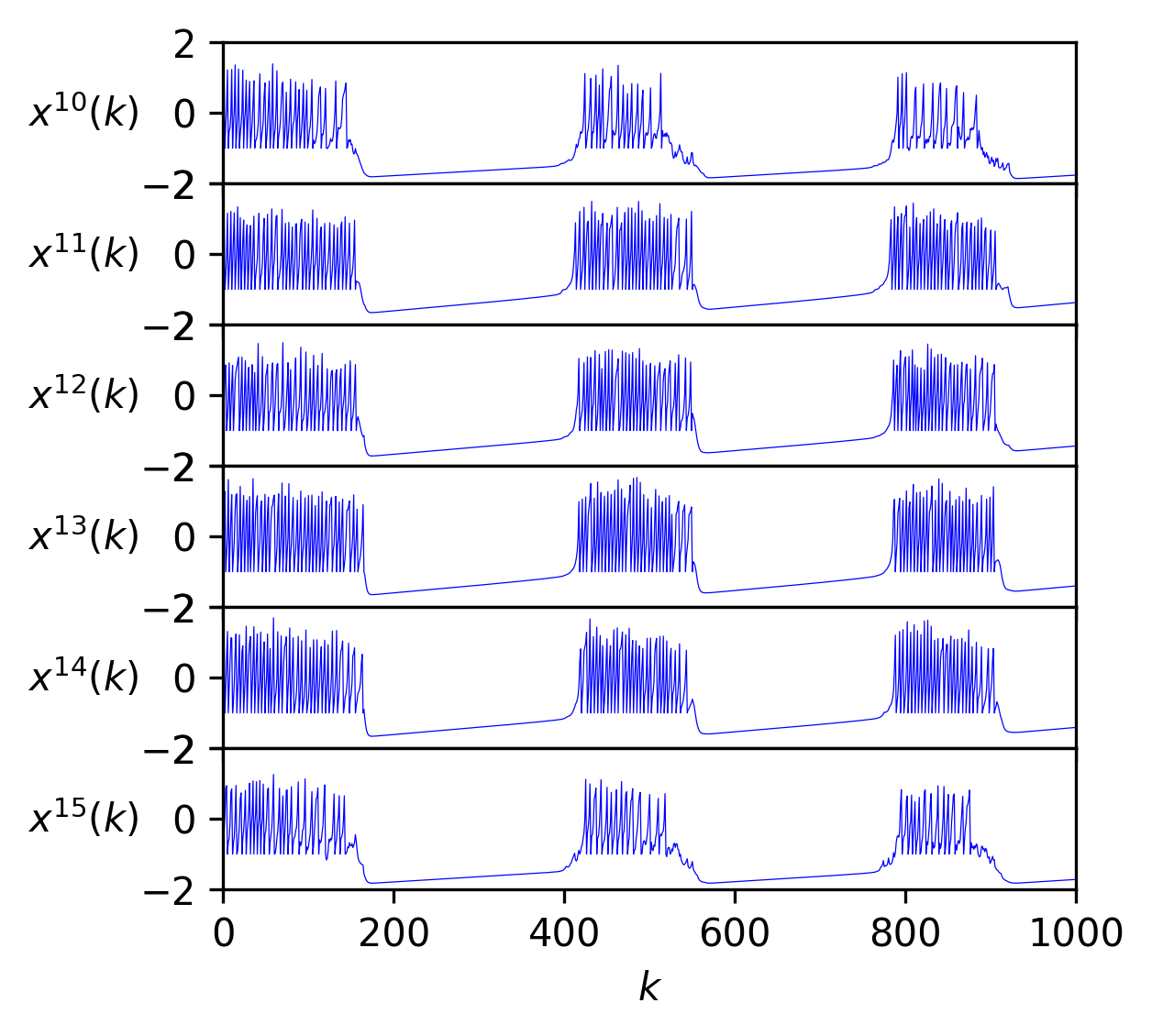}
        \caption{\centering $g=0.5$, $\lambda_1\approx 0.067$ (synchronized chaotic low-frequency bursting regime)}
        \label{fig:xvk_fhetero_g0.5}
    \end{subfigure}
    \hfill
    \begin{subfigure}{0.32\textwidth}
        \centering
        \includegraphics[width=\textwidth]{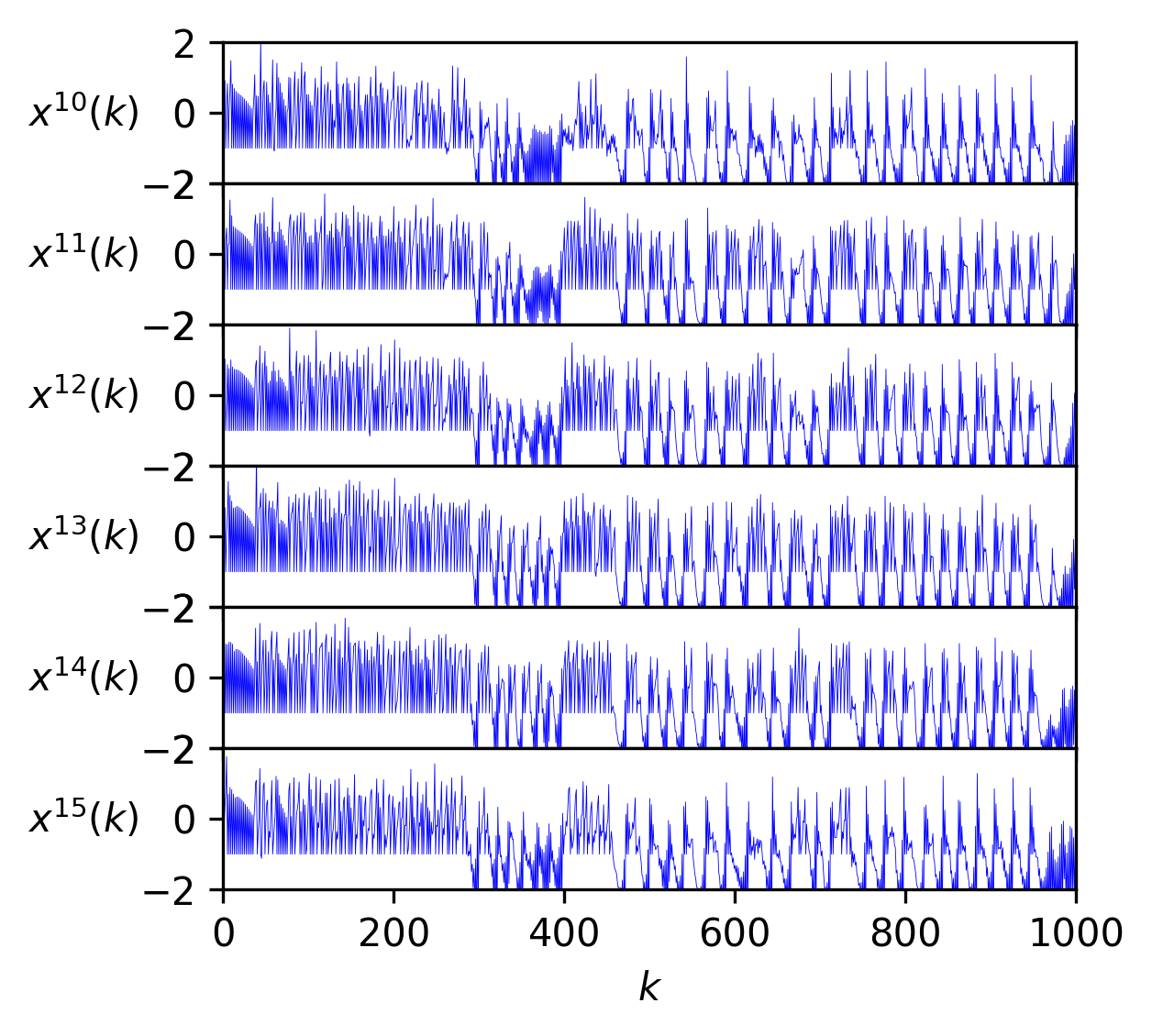}
        \caption{\centering $g=1$, $\lambda_1\approx 0.231$ (synchronized hyperchaotic regime)}
        \label{fig:xvk_fhetero_g1}
    \end{subfigure}
    \caption{Graphs of the voltages $x(k)$ of six adjacent fully heterogeneous NN coupled neurons for five select values of $g$ (the same neuron locations and $g$ values as Fig. \ref{fig:xvk_phetero_graphs}). The system is an $N=2$-dimensional lattice of NN electrically coupled fully heterogeneous Rulkov neurons with $\zeta=8$ neurons per side.}
    \label{fig:xvk_fhetero_graphs}
\end{figure}

We now turn our attention to the partially and fully heterogeneous cases of the two-dimensional neuron lattice. The plots of $\lambda_1$ versus $g$ for the heterogeneous cases (Figs. \ref{fig:lambda1_vs_g_phetero} and \ref{fig:lambda1_vs_g_fhetero}) reveal several similarities to the homogeneous case --- such as the gradual decline in $\lambda_1$ over the range $0.2 \lesssim g \lesssim 0.8$, and the sharp ascent for $0.9 \lesssim g \lesssim 1$ --- but also exhibit key differences. For example, $\lambda_1$ begins positive rather than negative, the left peaks of the graphs are much lower in the heterogeneous case as compared to the homogeneous case, and perhaps most strikingly, both graphs begin by sloping down when $g$ is close to 0. Although the chaotic dynamics of the partially and fully heterogeneous lattices follow the expected behavior in the electrically uncoupled regime ($g=0$) and the strongly coupled regime ($g\simeq 1$), their behavior in the intermediate values of $g$ features novel dynamics that arise as a result of neurons becoming electrically sensitive to their neighbors with different parameters $\sigma^\i$ and $\alpha^\i$. Specifically, we loosely categorize the heterogeneous dynamics into five dynamical regimes: (a) the uncoupled regime, (b) the local quasi-bursting regime, (c) the local chaotic high-frequency bursting regime, (d) the synchronized chaotic low-frequency bursting regime, and (e) the synchronized hyperchaotic regime. To fully understand the anomalous curves in Figs. \ref{fig:lambda1_vs_g_phetero} and \ref{fig:lambda1_vs_g_fhetero}, we will carefully analyze the dynamics in each of these regimes in turn.

Our first observation about the graphs of $\lambda_1$ vs. $g$ in the heterogeneous cases is that the dynamics of the lattice are chaotic in the uncoupled regime with $\lambda_1>0$. Specifically, for $g=0$, $\lambda_1\approx 0.075$ in the partially heterogeneous case and $\lambda_1\approx 0.073$ in the fully heterogeneous case. In the same manner as the behavior in the electrically uncoupled homogeneous lattice, the dynamics of each neuron is independent of that of any other neuron. Therefore, it is the inhomogeneous distribution of $\sigma^\i$ and $\alpha^\i$ that results in chaotic dynamics, which is expected because we set the parameters in the heterogeneous cases to include individually chaotic neurons. For example, in Figs. \ref{fig:xvk_phetero_g0} and \ref{fig:xvk_fhetero_g0}, we can identify neurons $\X^{12}$ and $\X^{13}$ as both individually chaotic because the frequency of their spikes varies with time. Furthermore, the figures clearly show two ways that the lattice heterogeneity manifests: low-frequency spiking neurons with different characteristic spiking frequencies ($\X^{11}$, $\X^{12}$, $\X^{13}$, and $\X^{14}$) and quiescent neurons ($\X^{10}$ and $\X^{15}$). The effect of heterogeneity is even more pronounced in the fully heterogeneous case (Fig. \ref{fig:xvk_fhetero_g0}) than the partially heterogeneous case (Fig. \ref{fig:xvk_phetero_g0}), with even higher frequency spikes and neurons with short periods of silence between small bursts of spikes. We will find that this dramatic heterogeneity is what leads to the different dynamical regimes when the neurons are coupled. We also note that the fact that $\lambda_1$ is lower for the fully heterogeneous case than the partially heterogeneous case is essentially a coincidence; the partially heterogeneous lattice simply contains the most individually chaotic neuron in the two lattices.

Now, unexpectedly, as the electrical coupling slowly increases away from $g=0$, the dynamics become less chaotic (see Figs. \ref{fig:lambda1_vs_g_phetero} and \ref{fig:lambda1_vs_g_fhetero}). This happens because individual low-frequency spiking neurons that are weakly connected to each other through their respective ranges of influence $\mathcal{N}^\i$ become locally sensitive to neighboring low-frequency spiking. This effect results in a new dynamical phase that is not seen in the homogeneous lattice, which we call ``local quasi-bursting.'' Local quasi-bursting is defined as a dynamical regime where clusters of individually spiking neurons exhibit oscillations between silence and spikes of similar frequency to the uncoupled neurons. This behavior is similar to the phenomenon of cluster synchronization \cite{cluster-sync-1, cluster-sync-2, cluster-sync-3}, which is characterized by complete synchronization in groups of coupled dynamical systems. However, local quasi-bursting is different from cluster synchronization because locally quasi-bursting neurons do not necessarily have to be completely synchronized due to the weakness of the electrical coupling. Low electrical coupling strength results in local quasi-bursting, because while the neurons are still individually chaotic, this minimal coupling allows neighborhoods of individually low-frequency spiking neurons to synchronize their low-frequency spikes, which lowers the slow variable, resulting in short periods of silence reminiscent of bursting (see Figs. \ref{fig:xvk_phetero_g0.05} and \ref{fig:xvk_fhetero_g0.05}). Because the spiking dynamics of the individual neurons don't significantly change, these short periods of silence are enough to lower the maximal Lyapunov exponent. We are careful to call this dynamical phase ``local quasi-bursting'' because it does not embody all the characteristics of synchronized chaotic bursting. Specifically, the quasi-bursting is ``local'' because the individually silent neurons remain silent; the low coupling strength isn't enough to synchronize them. The ``quasi-bursting'' isn't true bursting in the traditional way because the spikes within the ``bursts'' are essentially at the same frequency as the fastest individually spiking neurons, whereas true bursting has the characteristic of rapid spikes within the burst due to the fast variables' perception of a high slow variable (see Sec. \ref{subsec:rulkov-neuron}).

In the graphs of $\lambda_1$ vs. $g$, after the decrease through the local quasi-bursting regime, there is a sharp turn, then $\lambda_1$ starts to increase. This corresponds to the local chaotic high-frequency bursting regime, examples of which are displayed in Figs. \ref{fig:xvk_phetero_g0.13} and \ref{fig:xvk_fhetero_g0.13}. This regime of behavior is characterized by the individually spiking neurons undergoing locally synchronized high-frequency chaotic bursting, where ``high-frequency'' refers to the frequency of the slow oscillations between spiking and silence, and the individually quiescent neurons remaining silent. In this regime, the maximal Lyapunov exponent increases due to the fast unsynchronized chaotic spikes locking onto each other as the lattice is more strongly coupled. These interactions between spikes within the bursts are reminiscent of the unsynchronized chaotic spiking regime in the homogeneous system, as can be seen by comparing the bursts in Figs. \ref{fig:xvk_phetero_g0.13} and \ref{fig:xvk_fhetero_g0.13} with the spikes in Figs. \ref{fig:xvk_homo_g0.1}. 

The final two regimes of behavior, synchronized chaotic low-frequency bursting and synchronized hyperchaos, are similar to their counterparts for the homogeneous case with a few key differences. For $0.2\lesssim g\lesssim 0.8$, the coupling strength is strong enough to lift the individually quiescent neurons and synchronize the entire lattice in chaotic low-frequency bursting (see Fig. \ref{fig:xvk_phetero_g0.5} and \ref{fig:xvk_fhetero_g0.5}), where ``low-frequency'' refers to the frequency of the slow oscillations between spiking and silence. Comparing Figs. \ref{fig:xvk_phetero_g0.5} and \ref{fig:xvk_fhetero_g0.5} with Fig. \ref{fig:xvk_homo_g0.5}, a key qualitative, contrasting observation about the synchronized chaotic low-frequency bursting regime of the heterogeneous cases is that the periods of synchronized chaotic spiking are much shorter than their homogeneous counterparts, and the periods of quiescence are much longer than their homogeneous counterparts. This can be attributed to the contributions of individually quiescent neurons. In the neighborhood $g\simeq 1$, the neurons exhibit a synchronized hyperchaos similar to the homogeneous case, with the individually quiescent neurons contributing to more low voltages in the dynamics (compare Figs. \ref{fig:xvk_phetero_g1} and \ref{fig:xvk_fhetero_g1} with Fig. \ref{fig:xvk_homo_g0}).

Now that we have fully established the different dynamical regimes of behavior in the homogeneous, partially heterogeneous, and fully heterogeneous lattices, we will conclude this section with a brief comparison of the $\lambda_1$ vs. $g$ curves in Fig. \ref{fig:lambda1_vs_g_graphs}. The difference in the height of the left peak between the graphs of the homogeneous case and the heterogeneous cases is due to the fact that they embody different dynamical regimes: synchronized chaotic spiking and local chaotic high-frequency bursting. In the heterogeneous cases, the local chaotic high-frequency bursting has periods of silence, which lowers the maximal Lyapunov exponent. However, the similarity in the height of the right peak can be attributed to them all being associated with the same regime of behavior: synchronized hyperchaos, where the individual neuronal behaviors are dominated by the collective dynamics of the lattice.

\begin{figure}[t!]
    \centering
    \begin{subfigure}{0.6\textwidth}
        \centering
        \includegraphics[width=\textwidth]{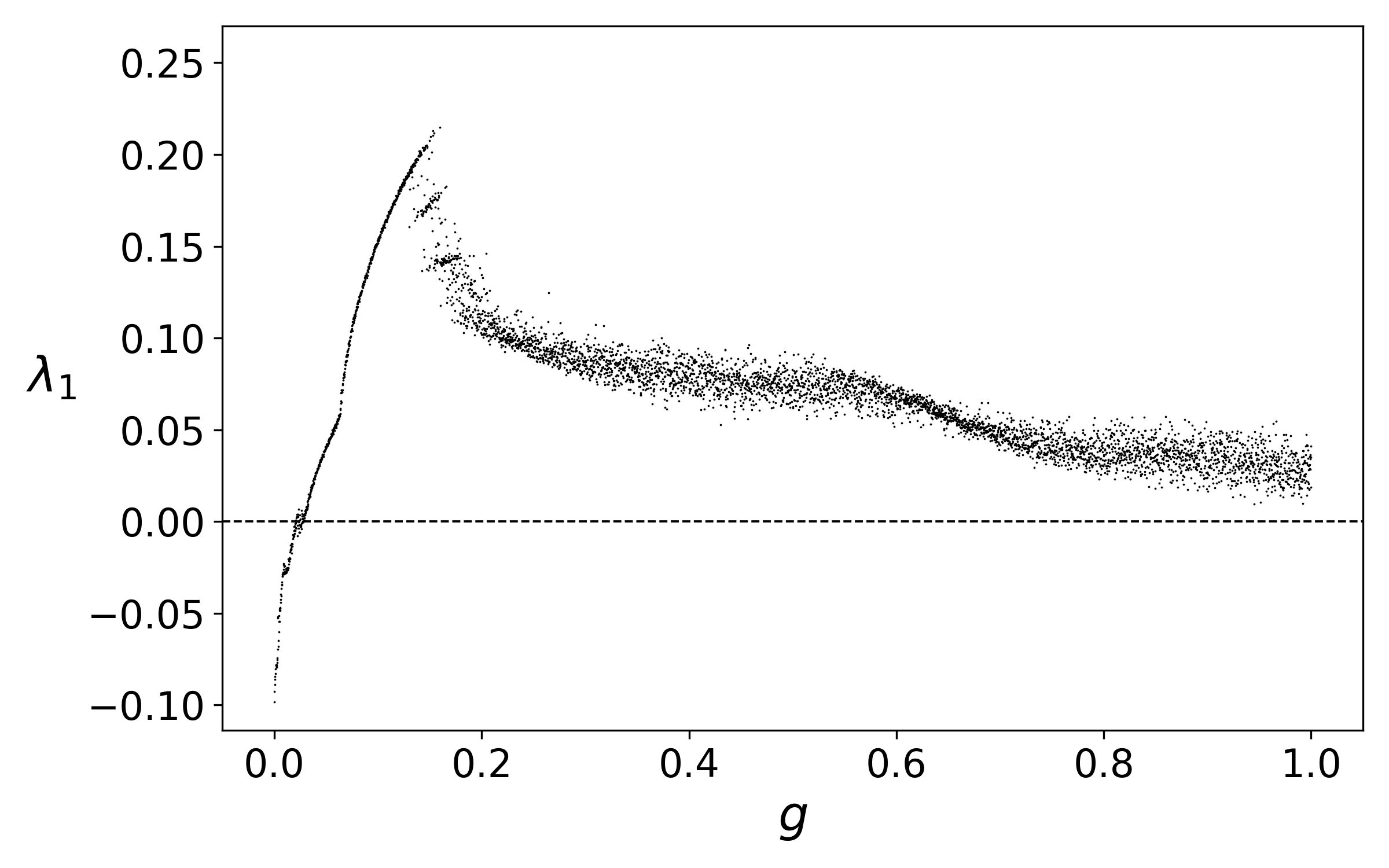}
        \caption{Homogeneous case}
        \label{fig:2d_nnn_lambda1_vs_g_homo}
    \end{subfigure} \\[0.5cm]
    \begin{subfigure}{0.475\textwidth}
        \centering
        \includegraphics[width=\textwidth]{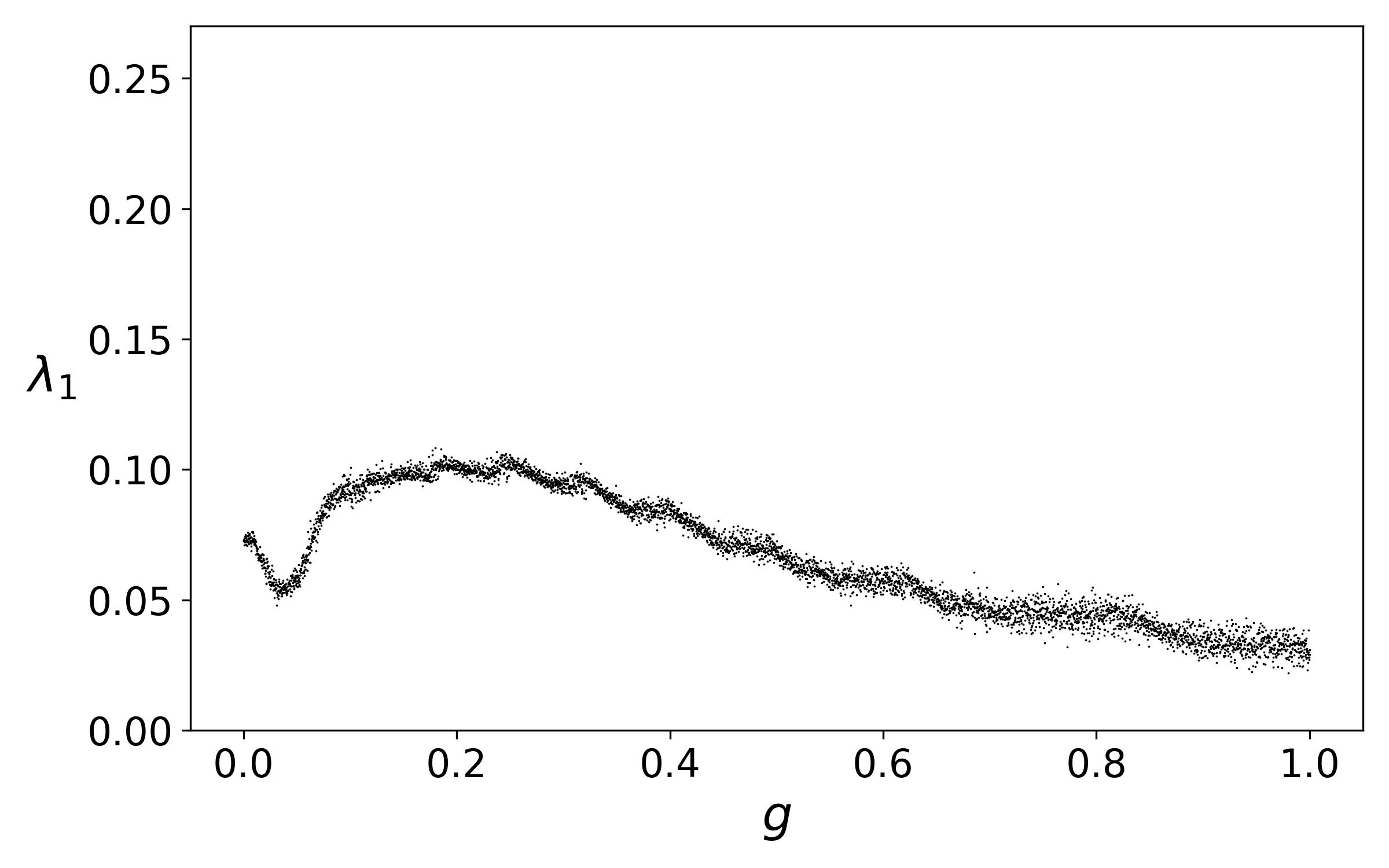}
        \caption{Partially heterogeneous case}
        \label{fig:2d_nnn_lambda1_vs_g_phetero}
    \end{subfigure}
    \hfill
    \begin{subfigure}{0.475\textwidth}
        \centering
        \includegraphics[width=\textwidth]{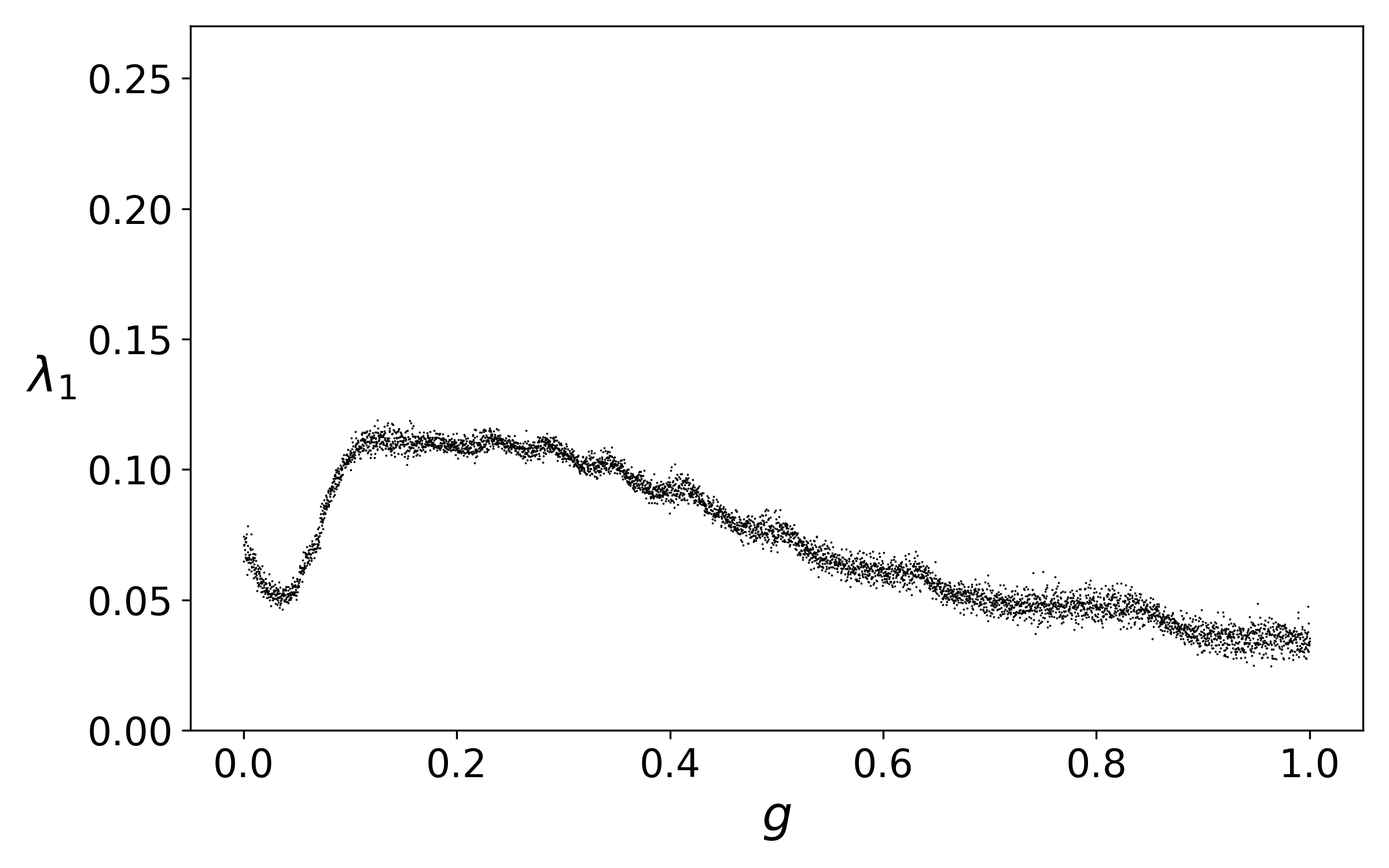}
        \caption{Fully heterogeneous case}
        \label{fig:2d_nnn_lambda1_vs_g_fhetero}
    \end{subfigure}
    \caption{Graphs of the maximal Lyapunov exponent $\lambda_1$ for 5000 electrical coupling strength values $g$ between 0 and 1 in the homogeneous, partially heterogeneous, and fully heterogeneous cases of an $N=2$-dimensional lattice of NNN electrically coupled Rulkov neurons with $\zeta=8$ neurons per side. The Lyapunov exponents are calculated using orbits of length $k=2000$.}
    \label{fig:2d_nnn_lambda1_vs_g}
\end{figure}

\begin{figure}[p!]
    \centering
    \begin{subfigure}{0.32\textwidth}
        \centering
        \includegraphics[width=\textwidth]{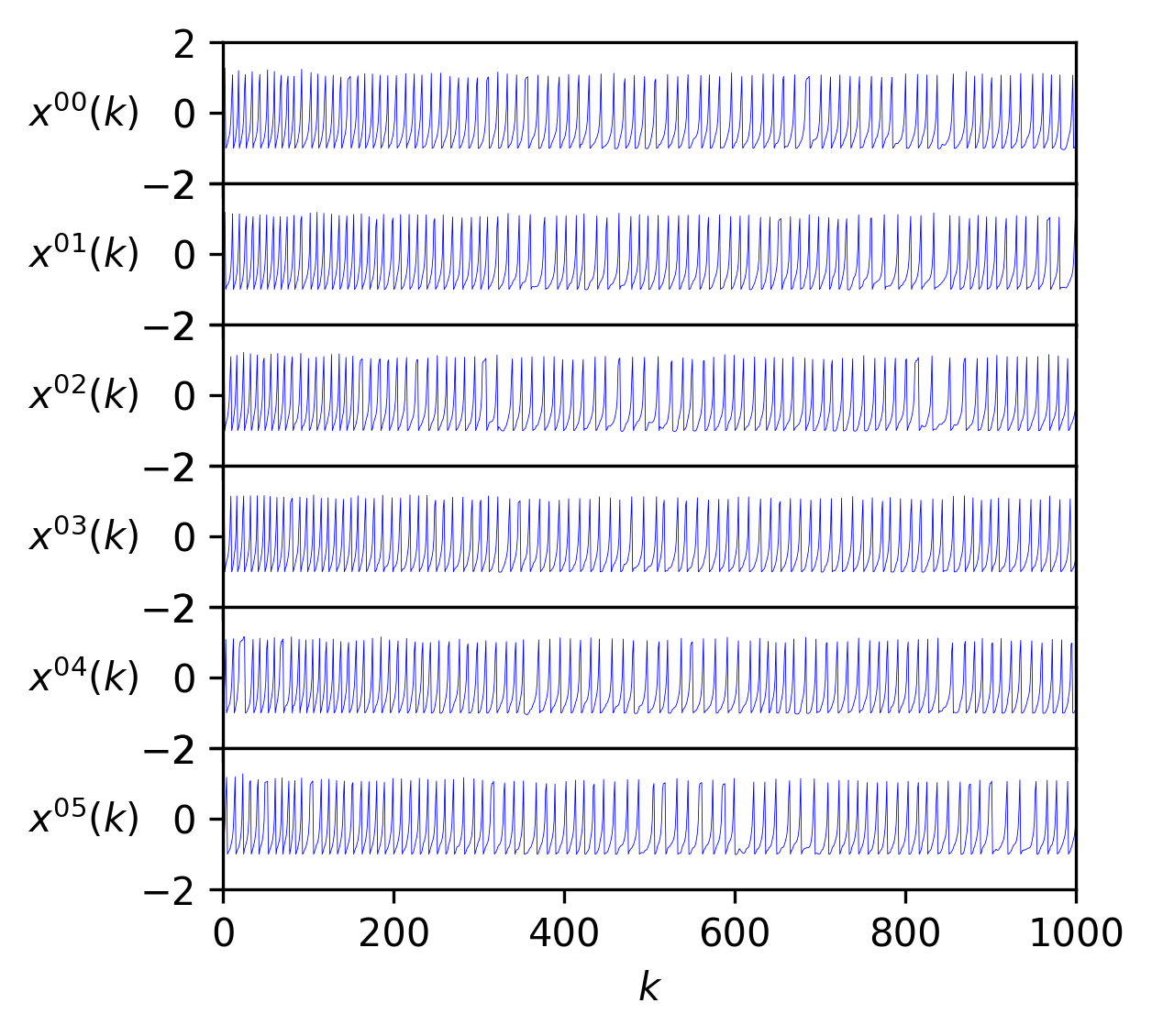}
        \caption{\centering $g=0.147$, $\lambda_1\approx 0.208$ (unsynchronized chaotic spiking regime)}
        \label{fig:nnn_xvk_homo_g0.147}
    \end{subfigure}
    \hfill
    \begin{subfigure}{0.32\textwidth}
        \centering
        \includegraphics[width=\textwidth]{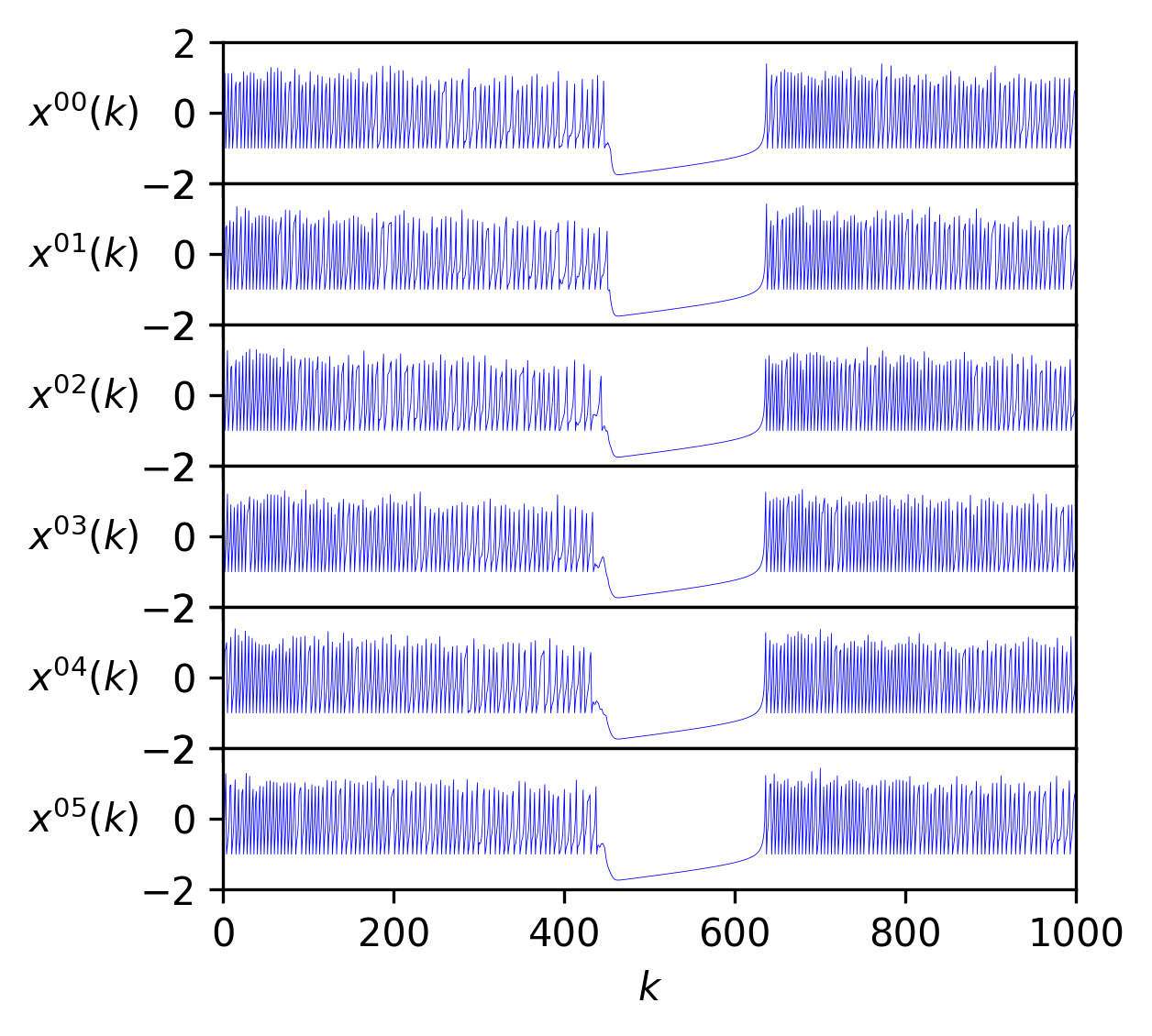}
        \caption{\centering $g=0.5$, $\lambda_1\approx 0.063$ (synchronized chaotic bursting regime)}
        \label{fig:nnn_xvk_homo_g0.5}
    \end{subfigure}
    \hfill
    \begin{subfigure}{0.32\textwidth}
        \centering
        \includegraphics[width=\textwidth]{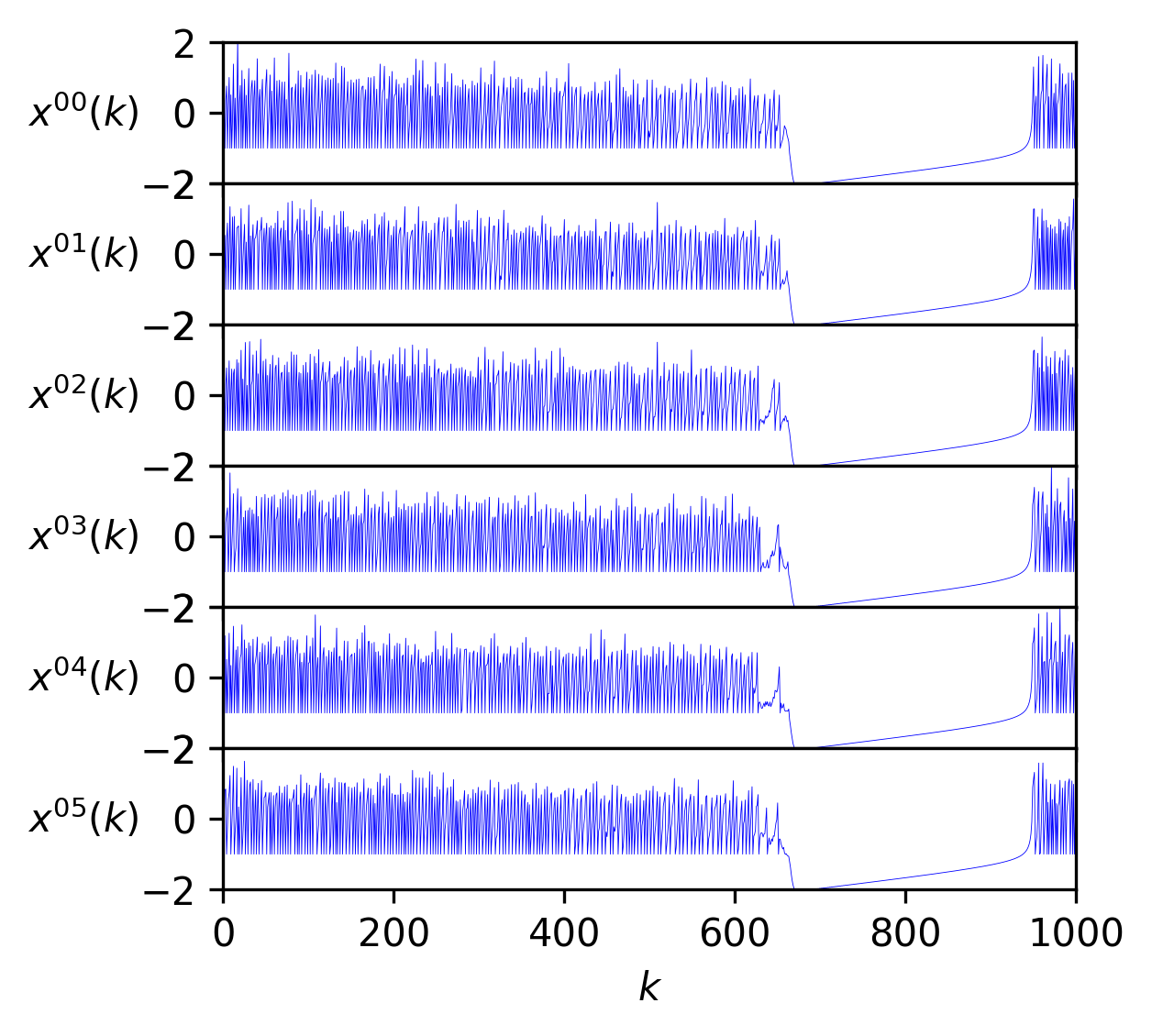}
        \caption{\centering $g=1$, $\lambda_1\approx 0.041$ (synchronized chaotic bursting regime)}
        \label{fig:nnn_xvk_homo_g1}
    \end{subfigure}
    \caption{Graphs of the voltages $x(k)$ of six adjacent homogeneous NNN coupled neurons for three select values of $g$. The system is an $N=2$-dimensional lattice of NNN electrically coupled homogeneous Rulkov neurons with $\zeta=8$ neurons per side.}
    \label{fig:NNN_xvk_homo_graphs}
\end{figure}

\begin{figure}[p!]
    \centering
    \begin{subfigure}{0.32\textwidth}
        \centering
        \includegraphics[width=\textwidth]{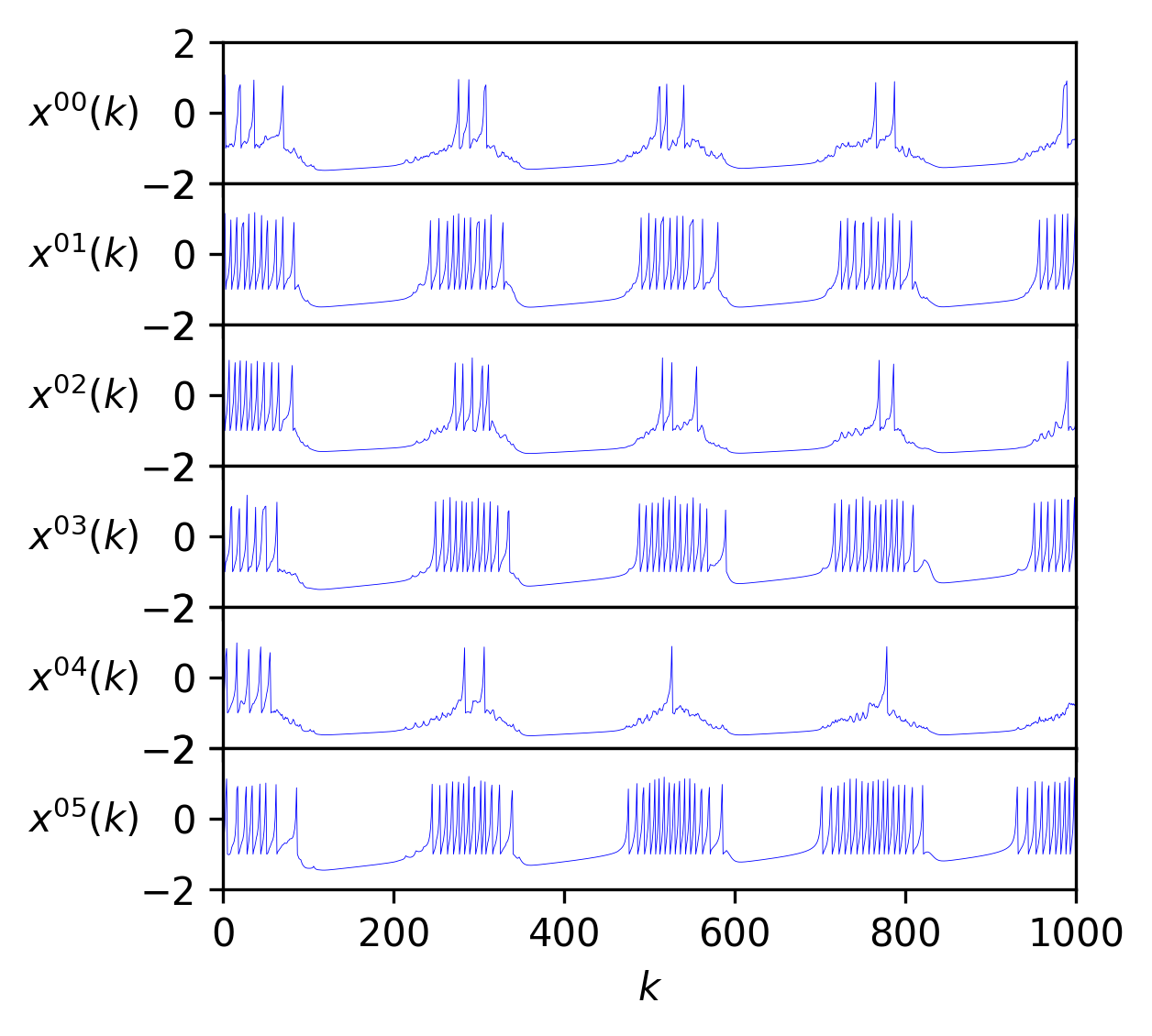}
        \caption{\centering $g=0.25$, $\lambda_1\approx 0.104$}
        \label{fig:nnn_xvk_fhetero_g0.25}
    \end{subfigure}
    \hfill
    \begin{subfigure}{0.32\textwidth}
        \centering
        \includegraphics[width=\textwidth]{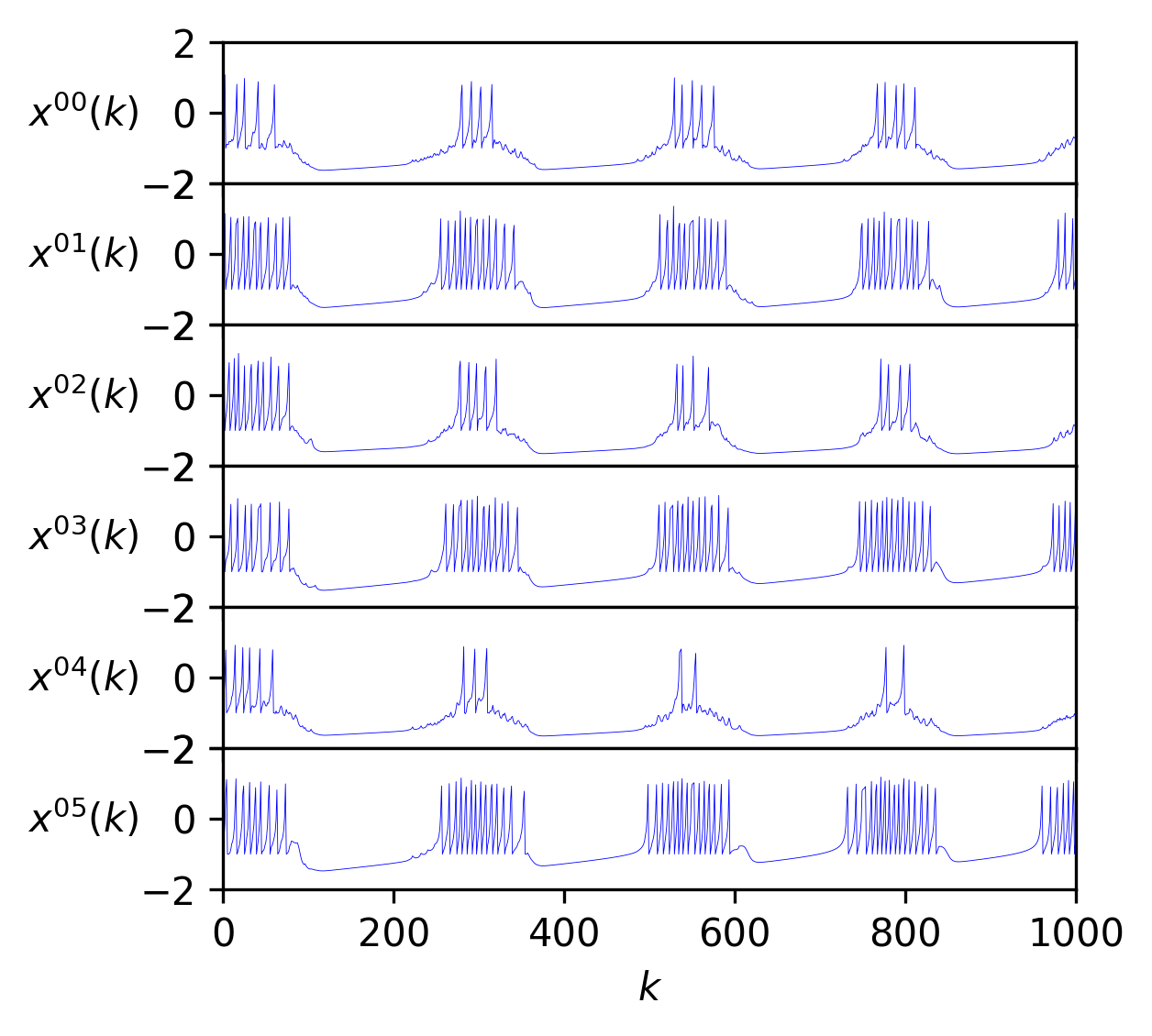}
        \caption{\centering $g=0.275$, $\lambda_1\approx 0.103$}
        \label{fig:nnn_xvk_fhetero_g0.275}
    \end{subfigure}
    \hfill
    \begin{subfigure}{0.32\textwidth}
        \centering
        \includegraphics[width=\textwidth]{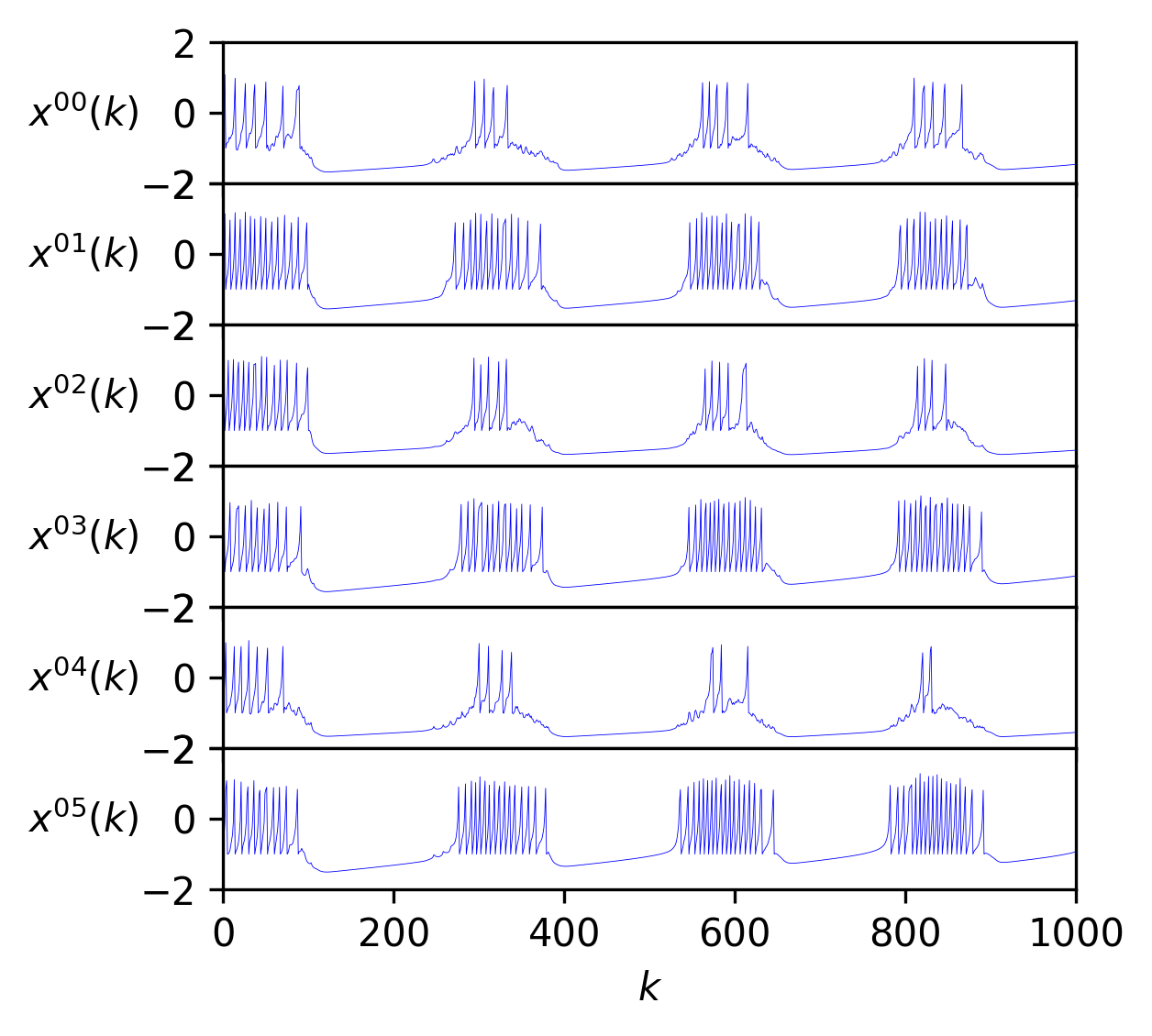}
        \caption{\centering $g=0.3$, $\lambda_1\approx 0.098$}
        \label{fig:nnn_xvk_fhetero_g0.3}
    \end{subfigure} \\[0.5cm]
    \begin{subfigure}{0.32\textwidth}
        \centering
        \includegraphics[width=\textwidth]{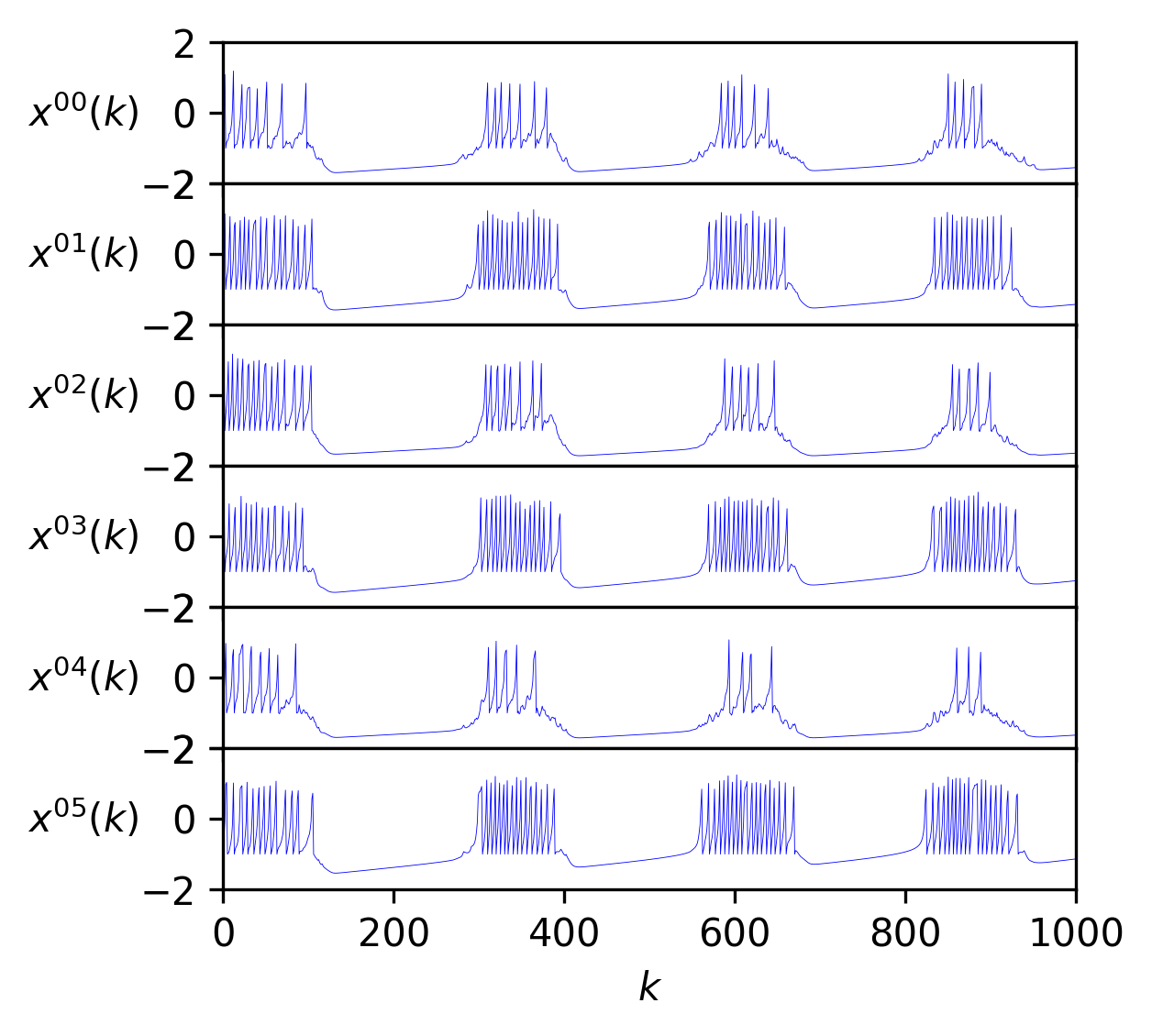}
        \caption{\centering $g=0.325$, $\lambda_1\approx 0.093$}
        \label{fig:nnn_xvk_fhetero_g0.325}
    \end{subfigure}
    \hfill
    \begin{subfigure}{0.32\textwidth}
        \centering
        \includegraphics[width=\textwidth]{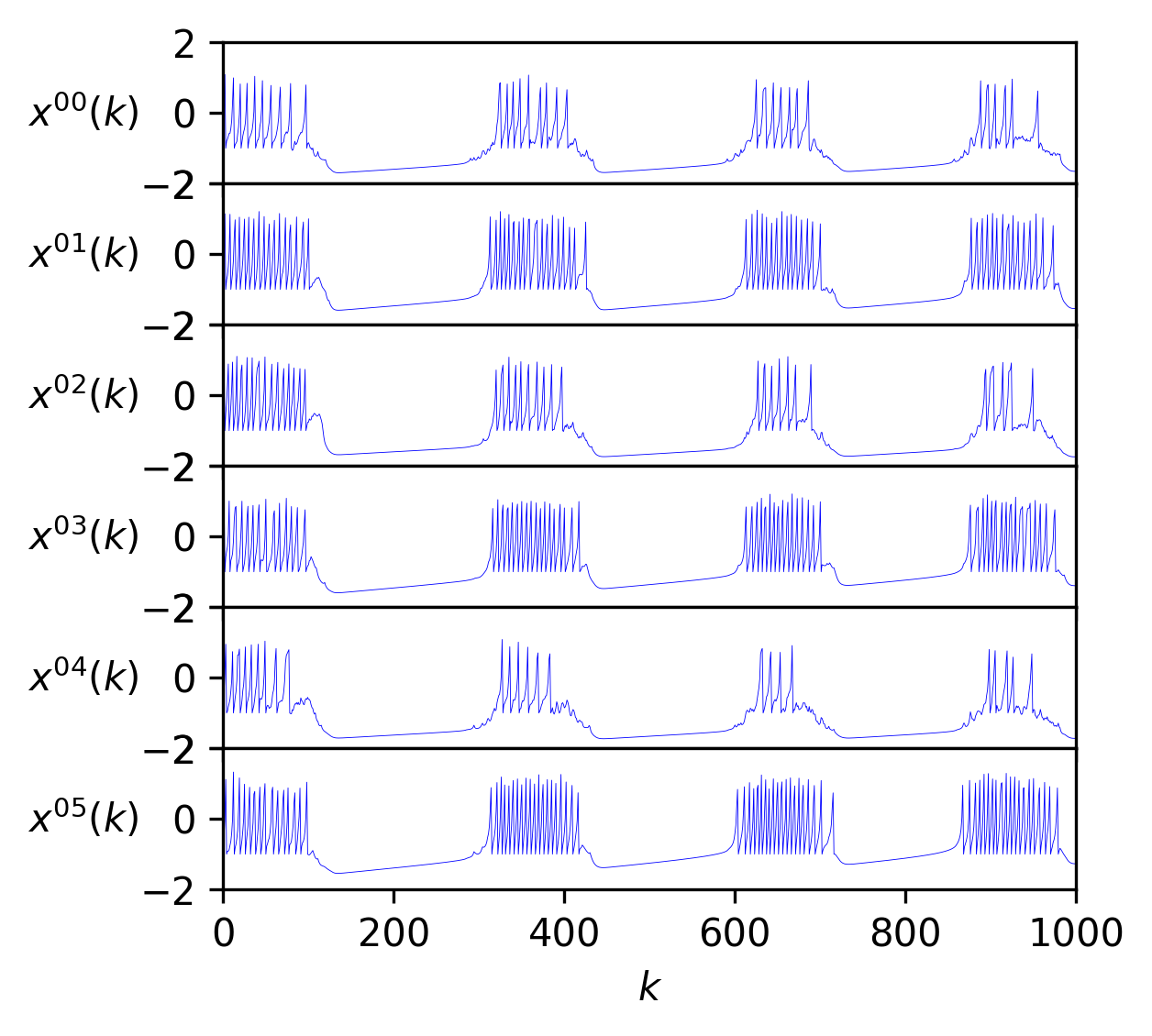}
        \caption{\centering $g=0.35$, $\lambda_1\approx 0.094$}
        \label{fig:nnn_xvk_fhetero_g0.35}
    \end{subfigure}
    \hfill
    \begin{subfigure}{0.32\textwidth}
        \centering
        \includegraphics[width=\textwidth]{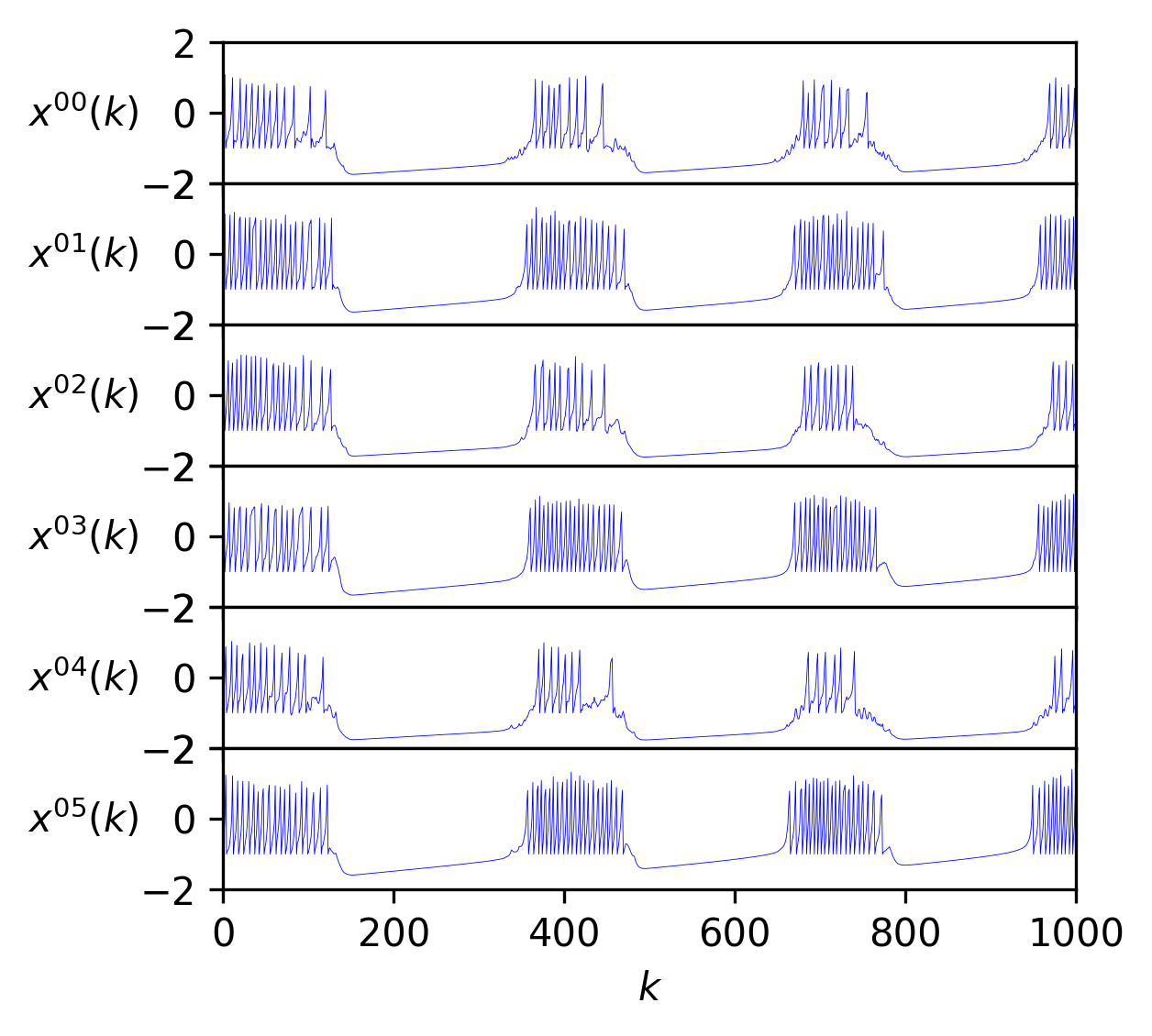}
        \caption{\centering $g=0.375$, $\lambda_1\approx 0.085$}
        \label{fig:nnn_xvk_fhetero_g0.375}
    \end{subfigure}
    \caption{Graphs of the voltages $x(k)$ of six adjacent fully heterogeneous NNN coupled neurons for three select values of $g$. The system is an $N=2$-dimensional lattice of NNN electrically coupled fully heterogeneous Rulkov neurons with $\zeta=8$ neurons per side.}
    \label{fig:NNN_xvk_fhetero_graphs}
\end{figure}

\subsection{NNN dynamics}
\label{subsec:2D-NNN}

We are now interested in how the dynamics of the two-dimensional lattice with $\zeta=8$ are affected if we change the electrical coupling connections to next-nearest-neighbor (NNN). As expounded in Sec. \ref{subsec:nd-model}, the NNN coupling in a two-dimensional lattice makes each neuron electrically coupled to $4N = 8$ other neurons [Eq. \eqref{eq:nnn-N}]. Just like the NN case, we examine the system's dynamics in the homogeneous, partially heterogeneous, and fully heterogeneous cases over a range of 5000 electrical coupling strength values between 0 and 1 through a computation of the system's Lyapunov exponents (Fig. \ref{fig:2d_nnn_lambda1_vs_g}). It is worth noting that for the rest of this paper, due to the presence of more complex Jacobian matrices [Eq. \eqref{eq:NNN-jacobian}] requiring more computational power, we compute Lyapunov exponents using orbits of length $k=2000$ instead of $k=10000$. By computer experiment, we confirm that this is long enough for the Lyapunov exponents to converge and that the dynamical trends are still clear, although this reduced orbit length introduces slightly more error into the Lyapunov spectrum (see Fig. \ref{fig:2d_nnn_lambda1_vs_g_homo} vs. Fig. \ref{fig:lambda1_vs_g_homo}).

As before, we begin by analyzing the homogeneous case. In Fig. \ref{fig:2d_nnn_lambda1_vs_g_homo}, we plot the maximal Lyapunov exponent $\lambda_1$ against the electrical coupling strength $g$. Comparing this to the NN coupled lattice (Fig. \ref{fig:lambda1_vs_g_homo}), we observe two major differences, both of which are relatively easy to understand. 

First, in Fig. \ref{fig:2d_nnn_lambda1_vs_g_homo}, the left peak of the graph, corresponding to the sharp transition between the unsynchronized chaotic spiking regime and the synchronized chaotic bursting regime, is both higher and shifted more to the right than the left peak in the graph for the NN coupled lattice. The reason behind this phenomenon lies in the weakness of NNN coupling. Because each neuron is coupled to eight other neurons as opposed to four, the NNN coupled lattice is more connected than the NN coupled lattice. However, this greater connectivity also reduces the individual contribution from a single neuron to one in its set of influence $\N^\i$ because that neuron is also influenced by seven others. Therefore, in this weak coupling domain, the eight unsynchronized neurons in a given neuron's set of influence $\N^\i$ ``destructively interfere'' to limit their overall effect. As a result, in order to reach the synchronized chaotic bursting regime, a higher electrical coupling strength is required to overcome the effects of ``destructive interference'' of the neuron voltages, leading to the more rightwards peak in Fig. \ref{fig:2d_nnn_lambda1_vs_g_homo}. Because the NNN coupled lattice remains in the unsynchronized chaotic spiking regime longer than the NN coupled lattice, this allows the neurons to interact more strongly with each other. What we mean with this observation is the following: In the NN coupled lattice, the maximal Lyapunov exponent of the system can only get so high before it is forced down by the silence that emerges in the synchronized chaotic bursting regime. On the other hand, in the NNN coupled lattice, the system can remain in the unsynchronized chaotic spiking regime even with the stronger interactions, which allows for the system to become more chaotic. We show an example of how the voltage of a neighborhood of NNN neurons behaves in this unsynchronized chaotic spiking regime in Fig. \ref{fig:nnn_xvk_homo_g0.147}.

Second, and perhaps more startling, is the disappearance of the characteristic ascent to synchronized hyperchaos that is present in all three cases of the NN coupled lattice (Fig. \ref{fig:lambda1_vs_g_graphs}). The observation indicates that the NNN coupled lattice has no synchronized hyperchaotic regime for $0\leq g\leq 1$. Indeed, in Figs. \ref{fig:nnn_xvk_homo_g0.5} and \ref{fig:nnn_xvk_homo_g1}, we can see that the lattice remains in the synchronized chaotic bursting all the way through the entire gradual downwards slope in Fig. \ref{fig:2d_nnn_lambda1_vs_g_homo} to the extreme value of $g=1$, where the system exhibits very slow oscillations between chaotic bursting and silence. This phenomenon can be similarly explained by the weakness of NNN coupling: because of the lessened influence of each individual neuron over its $\N^\i$, the extreme electrical coupling strength value of $g=1$ is not high enough to send the neurons into the synchronized hyperchaotic regime.

For the heterogeneous cases (Figs. \ref{fig:2d_nnn_lambda1_vs_g_phetero} and \ref{fig:2d_nnn_lambda1_vs_g_fhetero}), the rise to the synchronized hyperchaotic regime also vanishes for the exact same reason. Beyond this, the dynamical regimes of behavior are the same as the NN coupled lattice. However, there is still one interesting characteristic of these graphs that doesn't appear in the NN coupled system: the many small bumps in the graphs during the descent through the synchronized chaotic bursting regime. Because the sizes of these small oscillations are on the same order of magnitude as the variance in the Lyapunov exponent, it is difficult to present representative examples that show these bumps. Nonetheless, in Fig. \ref{fig:NNN_xvk_fhetero_graphs}, we display some standard visualizations of the voltages of six adjacent neurons in the fully heterogeneous lattice at six closely spaced values of $g$ ($\Delta g = 0.025$) in the synchronized chaotic bursting regime. In the figures, the neurons $\X^{00}$, $\X^{02}$, and $\X^{04}$ are all individually silent in the uncoupled regime.

As the coupling strength gradually increases, the figures show two main effects. First, as expected, the frequency of the slow oscillations decreases, with five visible bursts of spikes in Fig. \ref{fig:nnn_xvk_fhetero_g0.25} to four bursts in Fig. \ref{fig:nnn_xvk_fhetero_g0.375}. This decrease in the frequency of the slow oscillations is constant and always present, with the gradually increasing periods of silence causing the gradual slope down in the maximal Lyapunov exponent. The second effect is the increase in the number of spikes per burst of the individually quiescent neurons ($\X^{00}$, $\X^{02}$, $\X^{04}$). For example, focusing on the number of spikes in the first full burst of neuron $\X^{04}$, there are two spikes in Fig. \ref{fig:nnn_xvk_fhetero_g0.25}, followed by three spikes in Fig. \ref{fig:nnn_xvk_fhetero_g0.275}, four spikes in Fig. \ref{fig:nnn_xvk_fhetero_g0.3}, five spikes in Fig. \ref{fig:nnn_xvk_fhetero_g0.325}, six spikes in Fig. \ref{fig:nnn_xvk_fhetero_g0.35}, and finally eight spikes in Fig. \ref{fig:nnn_xvk_fhetero_g0.375}. We notice that the other silent neurons also go through simultaneous increases in the number of spikes per burst. The interplay between these two effects --- continuously decreasing slow oscillation frequency and increasing number of spikes per burst --- is the origin of the oscillations in the curves in Figs. \ref{fig:2d_nnn_lambda1_vs_g_phetero} and \ref{fig:2d_nnn_lambda1_vs_g_fhetero}. More precisely, the individually silent neurons in the highly connected NNN coupled lattice receive input from many individually spiking neurons in synchronized chaotic bursts. When the cumulative input passes a certain threshold, the individually silent neuron will add another spike. In the NN coupled lattice, the set of influence $\N^\i$ of each individually silent neuron is restricted to its neighbors, which causes them to add spikes out of sync with each other. This results in a steady decrease in $\lambda_1$ as $g$ increases. However, since the NNN coupled lattice is highly connected, the silent neurons tend to receive input synchronously, leading to coordinated threshold crossings reminiscent of a miniature ``phase transition'' of the lattice. This synchronized emergence of additional chaotic spikes throughout the synchronized chaotic bursting regime is what causes the small bumps observed in Figs. \ref{fig:2d_nnn_lambda1_vs_g_phetero} and \ref{fig:2d_nnn_lambda1_vs_g_fhetero}. These ``phase transitions'' emerging in the NNN coupled lattice are characteristic of more highly connected systems with individually silent neurons, and we will observe them again when we analyze $N$-dimensional lattices, which are also highly connected.

\begin{figure}[t!]
    \centering
    \begin{subfigure}{0.32\textwidth}
        \centering
        \includegraphics[width=\textwidth]{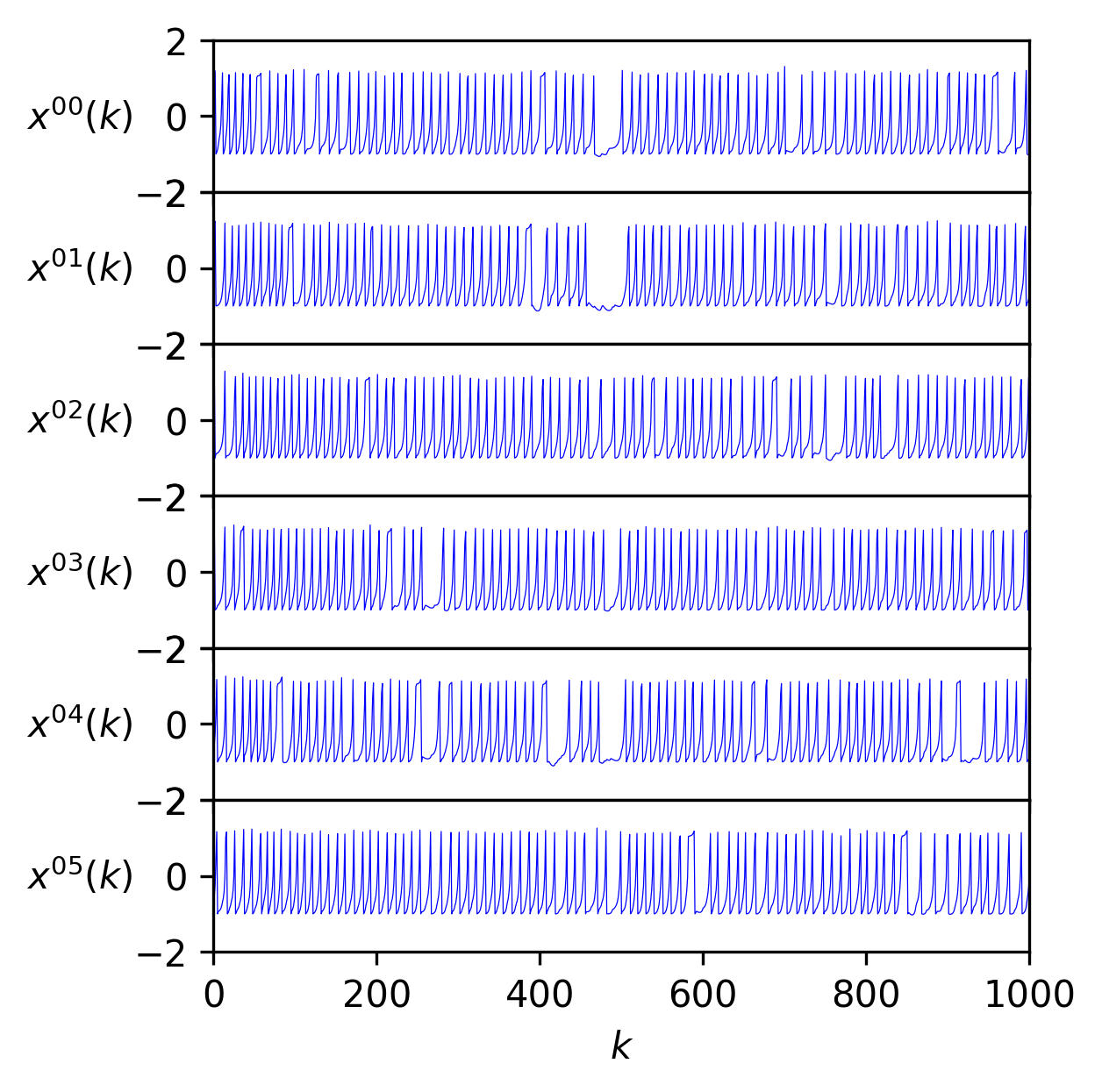}
        \caption{\centering $x(k)$ with $g=0.1$ (unsynchronized chaotic spiking regime)}
        \label{fig:2D-unsync-xvk}
    \end{subfigure}
    \hfill
    \begin{subfigure}{0.32\textwidth}
        \centering
        \includegraphics[width=\textwidth]{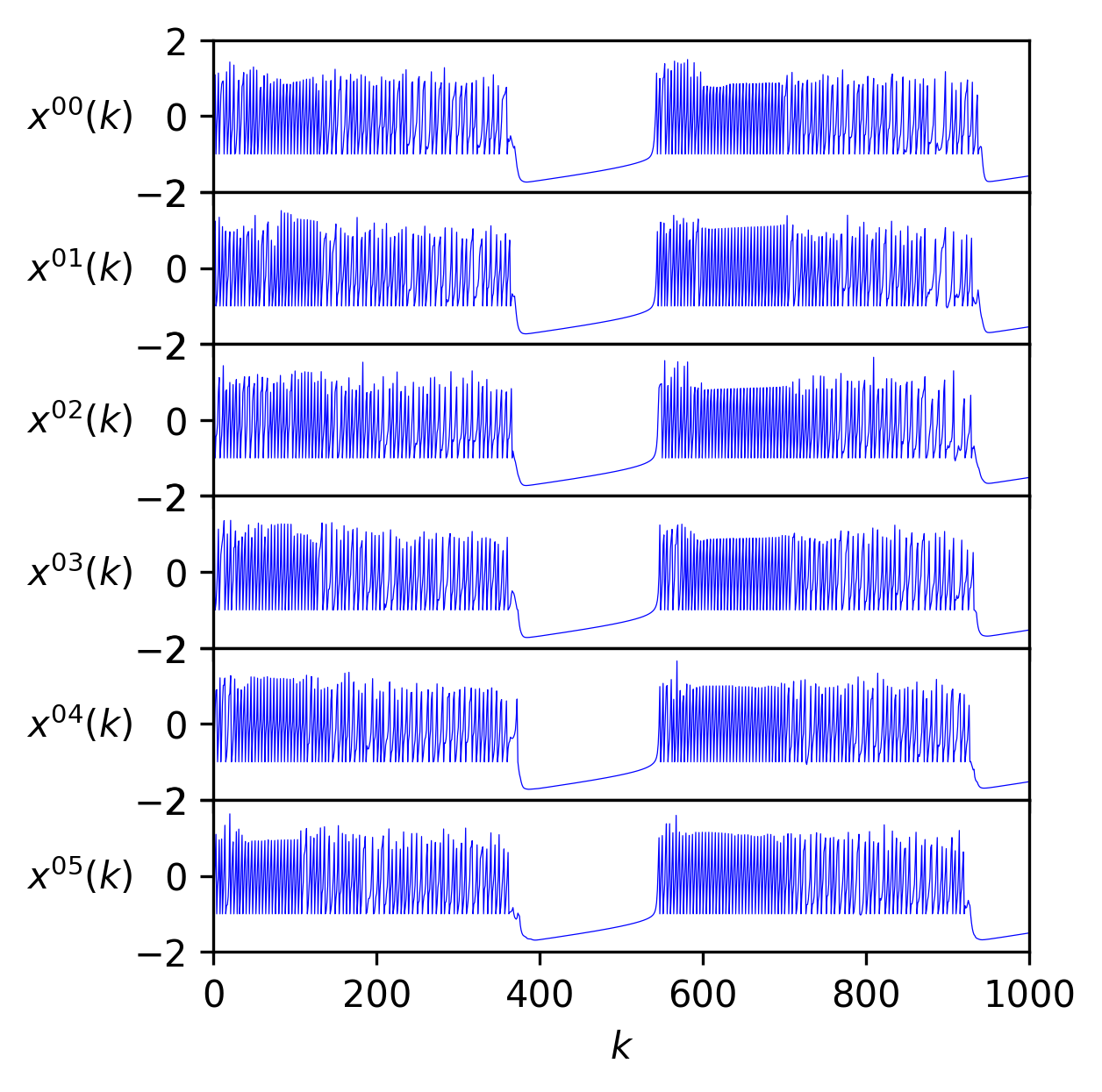}
        \caption{\centering $x(k)$ with $g=0.5$ (synchronized chaotic bursting regime)}
        \label{fig:2D-bursting-xvk}
    \end{subfigure}
    \hfill
    \begin{subfigure}{0.32\textwidth}
        \centering
        \includegraphics[width=\textwidth]{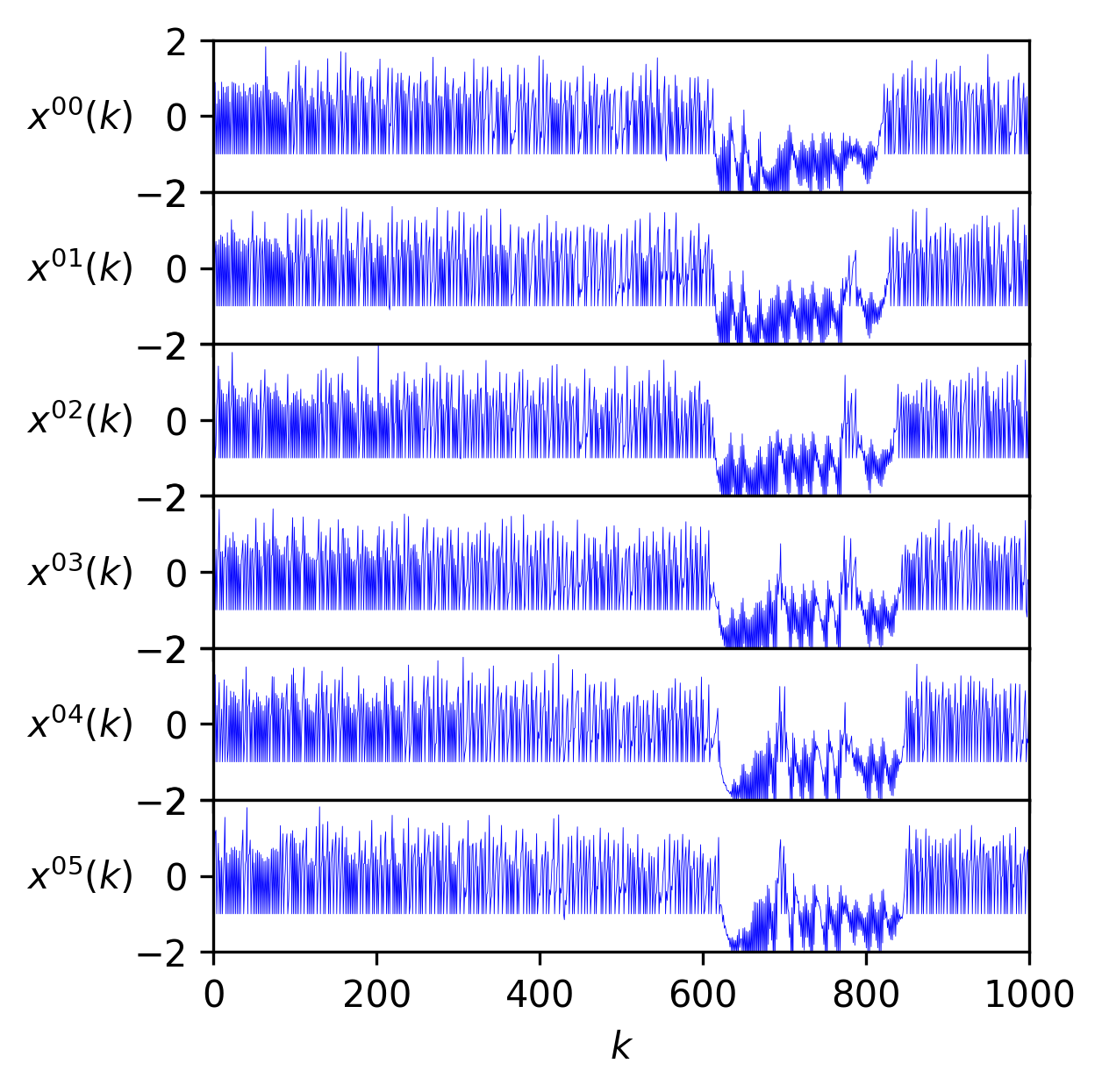}
        \caption{\centering $x(k)$ with $g=1$ (synchronized hyperchaotic regime)}
        \label{fig:2D-hyperchaos-xvk}
    \end{subfigure} \\[0.5cm]
    \begin{subfigure}{0.32\textwidth}
        \centering
        \includegraphics[width=\textwidth]{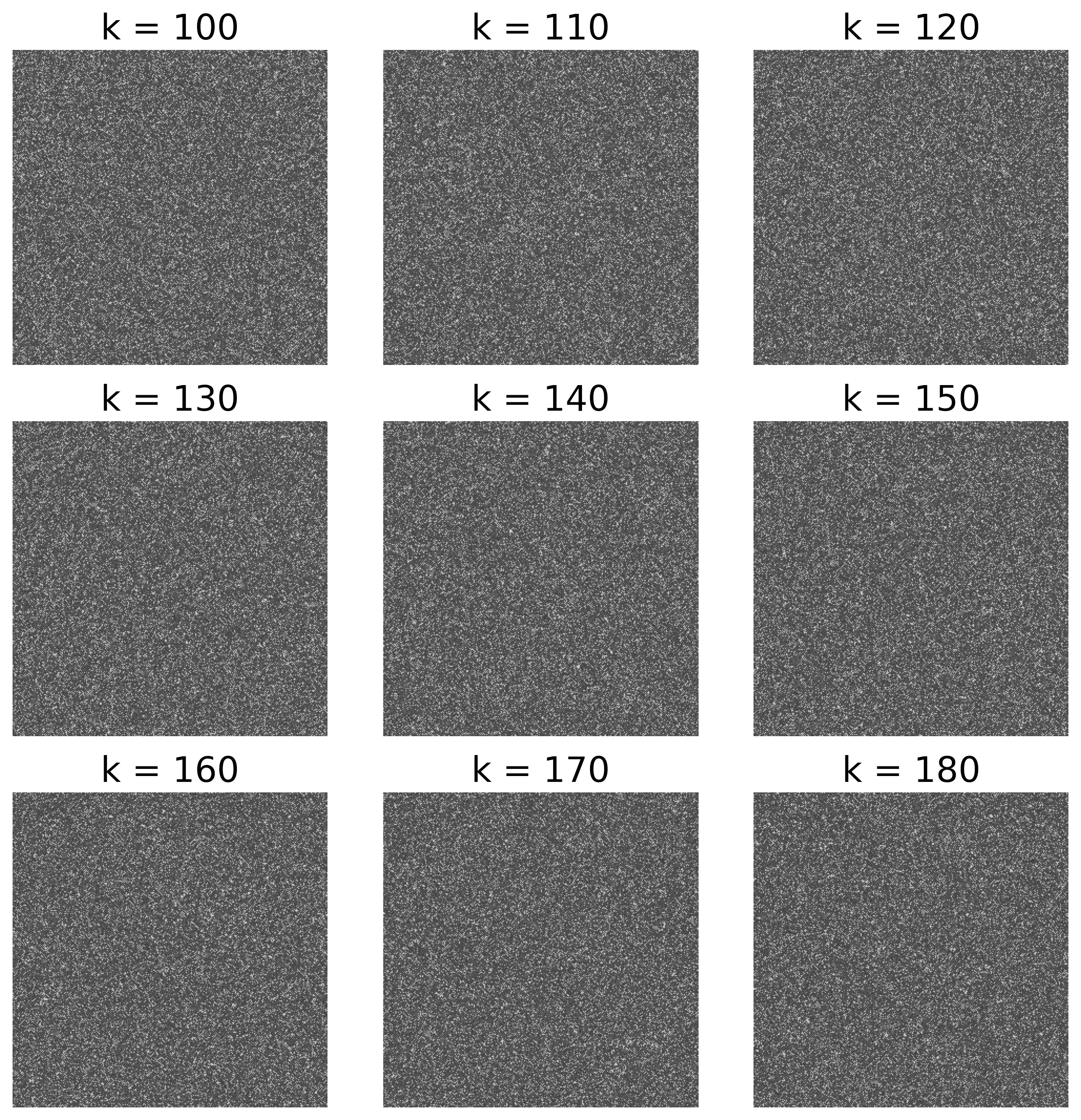}
        \caption{\centering Lattice snapshots with $g=0.1$ (unsynchronized chaotic spiking regime)}
        \label{fig:2D-unsync-snapshots}
    \end{subfigure}
    \hfill
    \begin{subfigure}{0.32\textwidth}
        \centering
        \includegraphics[width=\textwidth]{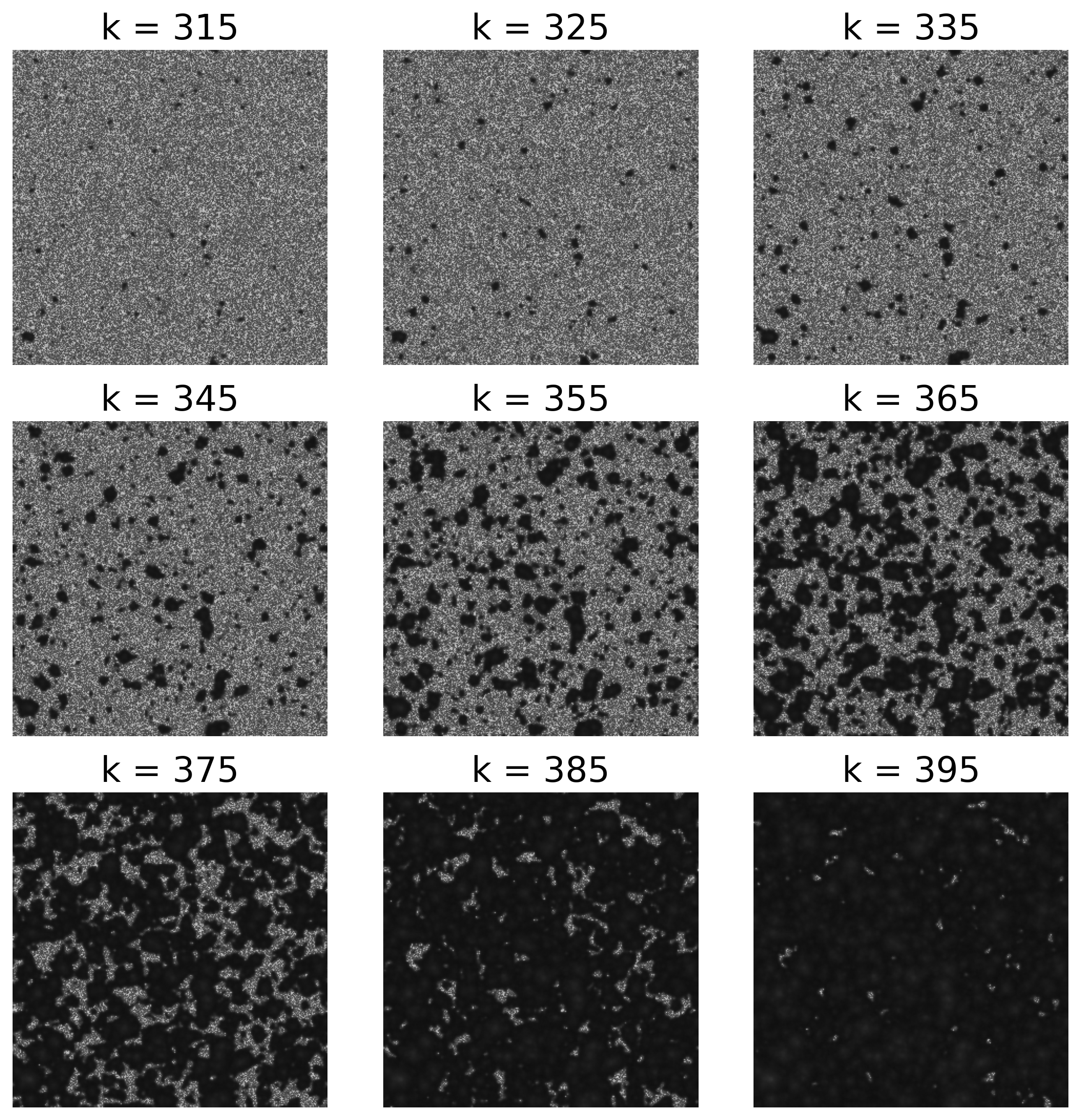}
        \caption{\centering Lattice snapshots with $g=0.5$ (synchronized chaotic bursting regime)}
        \label{fig:2D-bursting-snapshots}
    \end{subfigure}
    \hfill
    \begin{subfigure}{0.32\textwidth}
        \centering
        \includegraphics[width=\textwidth]{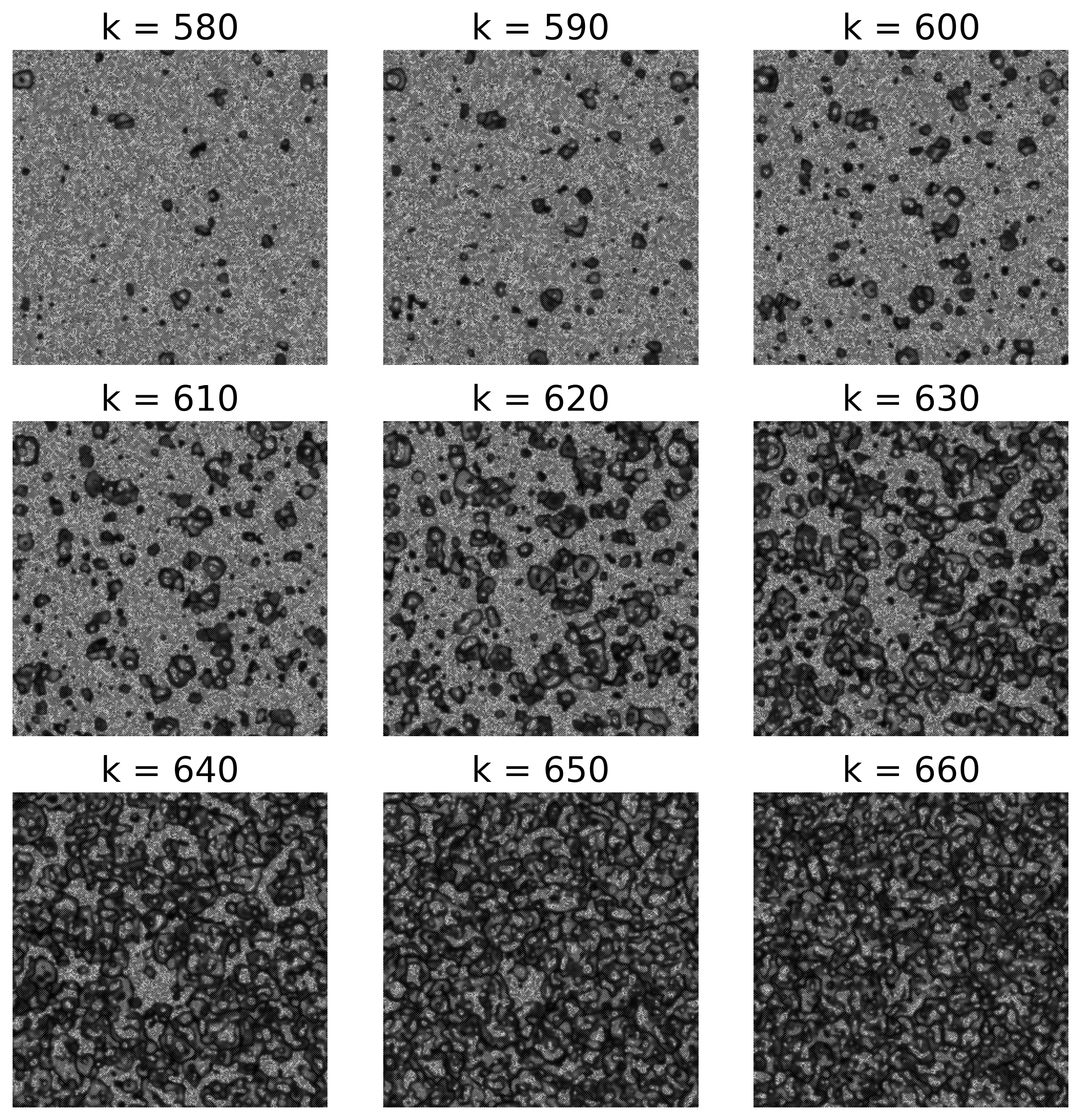}
        \caption{\centering Lattice snapshots with $g=1$ (synchronized hyperchaotic regime)}
        \label{fig:2D-hyperchaos-snapshots}
    \end{subfigure}
    \caption{Dynamics of an $N=2$-dimensional large lattice of NN electrically coupled homogeneous Rulkov neurons with $\zeta = 300$ neurons per side. The top graphs show the voltages $x(k)$ of six adjacent neurons in the lattice. The bottom graphs show snapshots of the voltages of all the neurons in the lattice.}
    \label{fig:large-lattice-graphs}
\end{figure}

\subsection{Collective dynamics of large lattices}
\label{subsec:2D-large}

Now that we have performed a quantitative analysis of the dynamics of an NN and NNN coupled two-dimensional Rulkov lattice with $\zeta=8$, we will qualitatively examine the dynamics of a larger two-dimensional neuron lattice. We now consider a two-dimensional NN coupled lattice of electrically coupled homogeneous Rulkov neurons with $\zeta=300$ neurons across both spatial dimensions. Since the state space of this system is $n=2\zeta^2 = 180000$-dimensional, it is computationally unrealistic to perform an accurate quantitative analysis of the dynamics of this system using Lyapunov exponents. However, we were able to establish an understanding of the dynamical regimes of the 2N lattice with $\zeta=8$ neurons in Sec. \ref{subsec:2D-NN}, and since the coupling connections of the lattice are local, we predict that the large lattice will exhibit similar dynamics.

In Sec. \ref{subsec:2D-NN}, we established that the two-dimensional lattice of NN coupled homogeneous neurons with $\zeta=8$ broadly has four regimes of dynamical behavior: the uncoupled regime, the unsynchronized chaotic spiking regime, the synchronized chaotic bursting regime, and the synchronized hyperchaotic regime. In the uncoupled regime, the dynamics of the large lattice ($\zeta=300$) will be the same as that of the small lattice ($\zeta=8$). Therefore, we turn our attention to the dynamics of the large lattice at electrical coupling strength values corresponding to the unsynchronized chaotic spiking, synchronized chaotic bursting, and synchronized hyperchaotic regimes of the small lattice. Accordingly, we choose the representative values of $g$ used for our analysis of the small lattice displayed in Fig. \ref{fig:xvk_homo_graphs}; namely, $g=0.1$ (unsynchronized chaotic spiking), $g=0.5$ (synchronized chaotic bursting), and $g=1$ (synchronized hyperchaos). 

In Figs. \ref{fig:2D-unsync-xvk}, \ref{fig:2D-bursting-xvk}, and \ref{fig:2D-hyperchaos-xvk}, we plot the voltages of six neurons in the large lattice as a function of the timestep $k$ for values of electrical coupling strength equal to $g=0.1, 0.5$, and $1$, respectively. Comparing the dynamics in these figures to those corresponding to the small lattice (Figs. \ref{fig:xvk_homo_g0.1}, \ref{fig:xvk_homo_g0.5}, and \ref{fig:xvk_homo_g1}), we find qualitatively identical behavior, supporting our prediction that, due to the local coupling of the neurons, increasing the size of the lattice does not significantly alter the familiar dynamical regimes of the lattice, nor does it introduce new dynamical regimes.

To gain more insight into the behavior of the larger lattice, with an emphasis on understanding the propagation of synchronized bursting, we constructed figures that display the voltages of all the neurons in the two-dimensional large lattice at a given time. In Figs. \ref{fig:2D-unsync-snapshots}, \ref{fig:2D-bursting-snapshots}, and \ref{fig:2D-hyperchaos-snapshots}, we present these visualizations of the three regimes of dynamical behavior in nine closely spaced timesteps (separated by $\Delta k=10$), where darker points represent lower voltage and lighter points represent a higher voltage. In Fig. \ref{fig:2D-unsync-snapshots} (unsynchronized chaotic spiking regime), the lattice resembles a random distribution of high and low voltages throughout the lattice, with each snapshot looking essentially the same. We expect such a distribution because the electrical coupling is not high enough to induce synchronization between the neurons. We also observe more dark points than light points in each snapshot, which is also expected: each individual neuron spends more time at a lower voltage, further enhanced by the occasional slow variable-induced quiescence, and only spends a few timesteps at the high voltages during the spike. Therefore, lower voltage is more prominent in the unsynchronized chaotic spiking regime.

In the regime of synchronized chaotic bursting, we have previously observed that although the moments of quiescence of groups of neurons are synchronized, the spiking behavior is not. Therefore, in the periods of spike bursting, we expect the voltages of the lattice to look like they do in the unsynchronized chaotic spiking regime (Fig. \ref{fig:2D-unsync-snapshots}). It is of more interest, then, to investigate how the lattice behaves during the transition between a burst of chaotic spikes and silence. For example, in Fig. \ref{fig:2D-bursting-xvk}, we see that this transition occurs around $k = 375$, so we collect snapshots of the two-dimensional voltage distribution around this timestep in Fig. \ref{fig:2D-bursting-snapshots}. In the figure, it is clear that the transition to silence does not happen simultaneously; rather, a single neuron transitions to silence, which influences its neighboring neurons around it to transition, and this effect cascades across the lattice. In other words, in the synchronized chaotic bursting regime for the large lattice, the transitions between bursts of spikes and silence are staggered and slowly spread radially outward through the lattice.

Let us now turn to the large-lattice dynamics in the synchronized hyperchaotic regime. Again, in Fig. \ref{fig:2D-hyperchaos-snapshots}, we visualize the interesting point where the neurons transition between higher voltages to lower voltages, which we can see in Fig. \ref{fig:2D-hyperchaos-xvk} happens around $k = 600$. In the figure, we can see that the transition down to lower voltage also occurs via a spreading of low-voltage neurons similar to the synchronized chaotic bursting case. By the last snapshot ($k=660$), almost all the neurons are at a lower voltage as compared to the first snapshot ($k=580$), indicating that the neurons can be more accurately described as quasi-synchronized. Quasi-synchronization with some error $\epsilon$ is characterized by a convergence of the difference between the states of two coupled systems to a value less than $\epsilon$ \cite{quasi-sync-1, quasi-sync-2}. Since all the neurons in the lattice follow the same general trend of high and low voltages, the lattice is quasi-synchronized up to an error of $\epsilon\simeq 2$. In Fig. \ref{fig:2D-hyperchaos-snapshots}, we can see that the dynamics of the lattice are described as locally synchronized \cite{local-sync-1, local-sync-2, local-sync-3}, exhibiting splotches of well-synchronized neurons and waves of propagating low voltage. This behavior suggests that, despite the hyperchaotic behavior, the lattice maintains a degree of spatial coherence, where local clusters of neurons evolve in step with one another. The coexistence of global quasi-synchronization and local synchronization highlights a rich interplay between order and chaos in this system, revealing a structural complexity characteristic of these strongly coupled nonlinear dynamics.

We conclude this section by commenting on the utility of small lattice simulations. Although the dynamics of larger neuron lattices reflect those of smaller lattices, we more easily identified emergent behavior in larger lattices resulting from the locality of the coupling connections. A quantitative analysis of smaller neuron lattices is valuable then because it translates well to understanding the overall dynamics of larger, more complex systems, i.e., it may provide a simpler model that may be understood before scaling up to larger models.

\section{\texorpdfstring{$N$}{N}-dimensional lattices}
\label{sec:nd-lattices}

In this section, we extend the spatial dimension of the neuron lattice from two to the general $N$ case. Although it is not particularly realistic to examine the dynamics of a Rulkov lattice in, say, $7$ spatial dimensions, because the number of nearest neighbors of a given neuron scales as $2N$, increasing the spatial dimension of the lattice effectively couples more neurons together. Through this lens, we can then study the trends in the dynamics as more neurons are coupled together in a more complex topology. We are especially interested in studying the ability of the Rulkov lattice to maintain synchronized dynamics and, if any, exhibit discontinuous transitions between dynamical regimes of behavior as we increase the electrical coupling strength $g$.

In Sec. \ref{subsec:small-nd}, we investigate the dynamics of an NN electrically coupled Rulkov lattice with $\zeta = 4$ across spatial dimensions $N = 1, 2, 3,$ and $4$. We systematically examine how increasing the spatial dimensionality affects the dynamical behavior in the homogeneous, partially heterogeneous, and fully heterogeneous cases. Special attention is given to features that emerge uniquely in certain dimensions. Building on this analysis of small $N$-dimensional lattices, Sec. \ref{subsec:large-3d} focuses specifically on the $N = 3$ case to explore the qualitative dynamics of a large-scale neuron lattice. This three-dimensional system serves as a toy model of a brain region and extends our earlier investigation of large two-dimensional lattices presented in Sec. \ref{subsec:2D-large}.

\begin{figure}[t!]
    \centering
    \begin{subfigure}{0.6\textwidth}
        \centering
        \includegraphics[width=\textwidth]{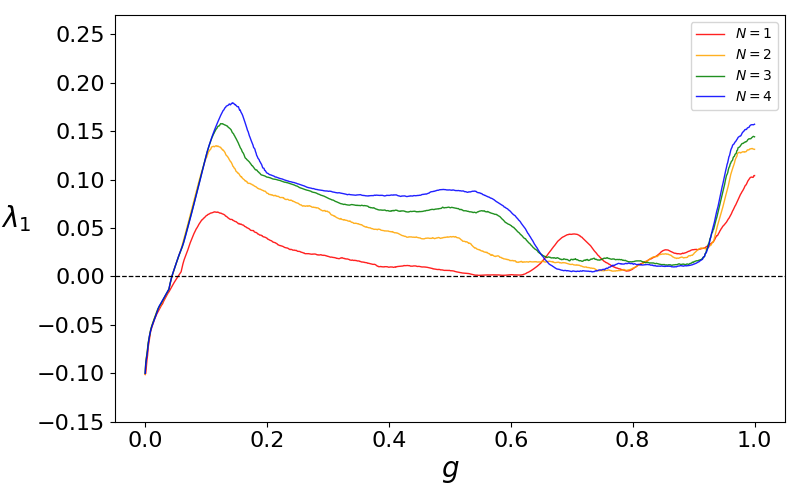}
        \caption{Homogeneous case}
        \label{fig:nd-neurons-homo}
    \end{subfigure} \\[0.5cm]
    \begin{subfigure}{0.475\textwidth}
        \centering
        \includegraphics[width=\textwidth]{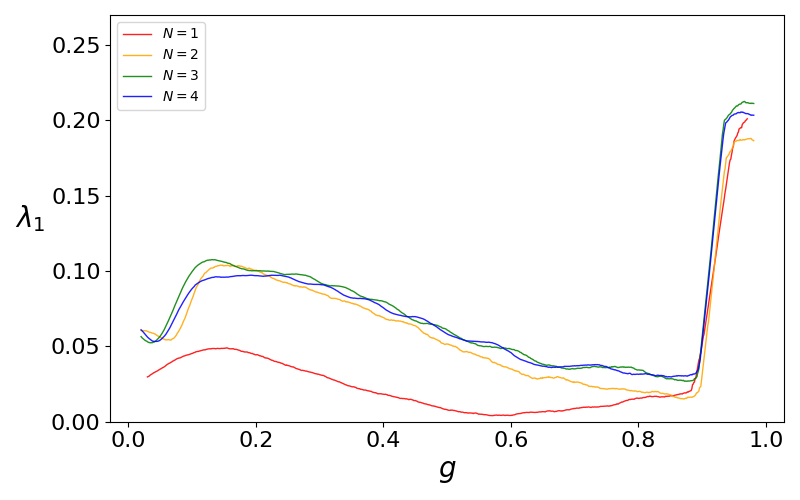}
        \caption{Partially heterogeneous case}
        \label{fig:nd-neurons-phetero}
    \end{subfigure}
    \hfill
    \begin{subfigure}{0.475\textwidth}
        \centering
        \includegraphics[width=\textwidth]{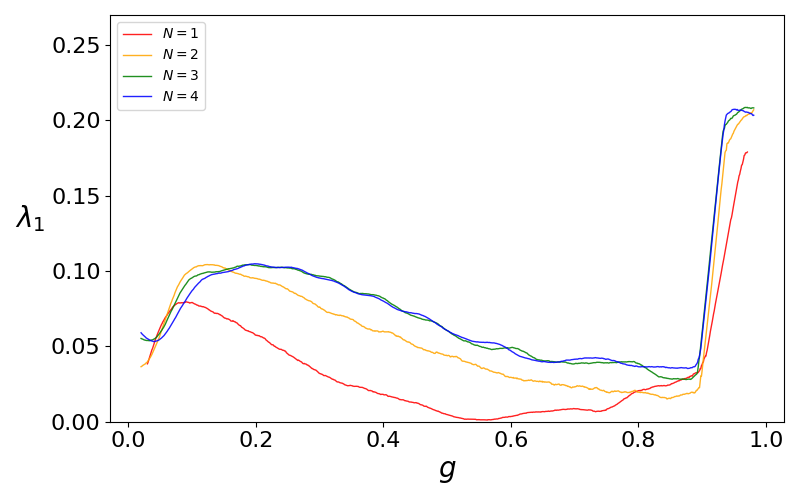}
        \caption{Fully heterogeneous case}
        \label{fig:nd-neurons-fhetero}
    \end{subfigure}
    \caption{Overlaid graphs of the maximal Lyapunov exponent $\lambda_1$ with 1000 electrical coupling strength values $g$ from $0$ to $1$ in the homogeneous, partially heterogeneous, and fully heterogeneous cases of a lattice in $N=1, 2, 3,$ and 4 spatial dimensions with $\zeta = 4$ neurons per dimension. To more clearly compare the trends, the $N = 1$ data has been smoothed with a moving window of 60 data points, and the $N=2,3,4$ data have been smoothed with a moving window of 40 data points. The Lyapunov exponents are calculated using orbits of length $k=2000$.}
    \label{fig:nd-neuron-lattice-results}
\end{figure}

\subsection{Small \texorpdfstring{$N$}{N}-dimensional lattice dynamics}
\label{subsec:small-nd}

We now begin the analysis of small lattice trends in various spatial dimensions using the established homogeneous and heterogeneous settings of the parameters $\sigma^\i$ and $\alpha^\i$. The simulations are performed with $\zeta = 4$ neurons per dimension for spatial dimensions $N = 1, 2, 3$, and $4$ using orbits of $2000$ timesteps and $1000$ electrical coupling strength values $g$ from $0$ to $1$. We will start with an overview of the qualitative details of our simulations, the results of which are displayed in Fig. \ref{fig:nd-neuron-lattice-results}. Then, we will discuss the dynamics of these lattices with reference to our corresponding figures and our dynamical characterization of the lattices in previous sections. Our objective is to describe the dynamics of small $N$-dimensional lattices by expounding their similarities and differences to the lattices we have investigated thus far.

We start with the striking observation that the initial nonchaotic ascent of $\lambda_{1}$ in the homogeneous case (Fig. \ref{fig:nd-neurons-homo}) is nearly identical across all spatial dimensions. For $N = 1$, the curve diverges slightly near $\lambda_{1} \lesssim 0$, but in dimensions $N = 2$, $3$, and $4$, the rise is almost indistinguishable. In fact, it is an artifact of the moving average that this slope appears to diverge before $\lambda_{1}$ becomes positive; if we use a smaller window, the trend for the $N = 1$ lattice remains colinear with that of the other lattices. This similarity is not limited to the initial ascent: the location of the first peak in $\lambda_{1}$, signaling the transition from unsynchronized chaotic bursting to a more ordered bursting phase, shifts to the right in a consistent, dimension-dependent fashion. The regularity of this shift is astounding. If one were to simulate these small lattices extending into more spatial dimensions, it would be reasonable to predict that the first chaotic peak in $\lambda_{1}$ grows slightly in height and shifts further to the right. Since this trend is clear and continuous across several spatial dimensions, we might suspect that if it is violated discontinuously at higher dimensions, it could signal an exotic dynamical phase transition. We found in Sec. \ref{subsec:2D-NNN} that the prolonged onset of synchronized bursting from the unsynchronized chaotic spiking regime to the synchronized chaotic bursting regime in the NNN coupled lattice was due to the ``destructive interference'' of a larger number of nearest-neighbors. This effect is also responsible for the shift to the right in the left peak as $N$ increases.

There is another consistent pattern in the homogeneous lattice: increasing the spatial dimension pushes $\lambda_{1}$ upward in the chaotic regimes. In particular, around $g \approx 0.2$ and again for $g \gtrsim 0.9$, we observe that $\lambda_{1}$ increases monotonically with dimension, reflecting more chaotic dynamics in both the unsynchronized chaotic spiking and synchronized hyperchaotic regimes. This suggests that higher spatial connectivity not only delays the onset of synchrony but also amplifies the overall dynamical instability in both ends of the coupling spectrum.

Zooming in on intermediate values of $g$ in Fig. \ref{fig:nd-neurons-homo}, especially near $g \simeq 0.55$, there is an additional, subtle detail we can glean upon inspection. For $N = 2$ a slight bump appears in $\lambda_{1}$ --- which is not particularly noticeable in $N = 1$ --- and this structure becomes more pronounced as the dimension increases. The bump rounds out into a broader plateau, flattening on its left flank and descending more steeply on its right. This sharpening of features is, in fact, a more general one: increasing spatial dimension accentuates the curvature of $\lambda_{1}$ as a function of coupling in the regions of dynamical phase transitions. This observation similarly allows us to make predictions about this trend into higher spatial dimensions, i.e., if we continue to increase the spatial dimension, we might expect that the first peak around $g \simeq 0.2$ and the descent into synchronized bursting around $g \simeq 0.6$ become sharper and approach a discontinuous transition. These extrapolations are important because simulating higher-dimensional lattices can be computationally prohibitive. 

\begin{figure}[t!]
    \centering
    \begin{subfigure}{0.32\textwidth}
        \centering
        \includegraphics[width=\textwidth]{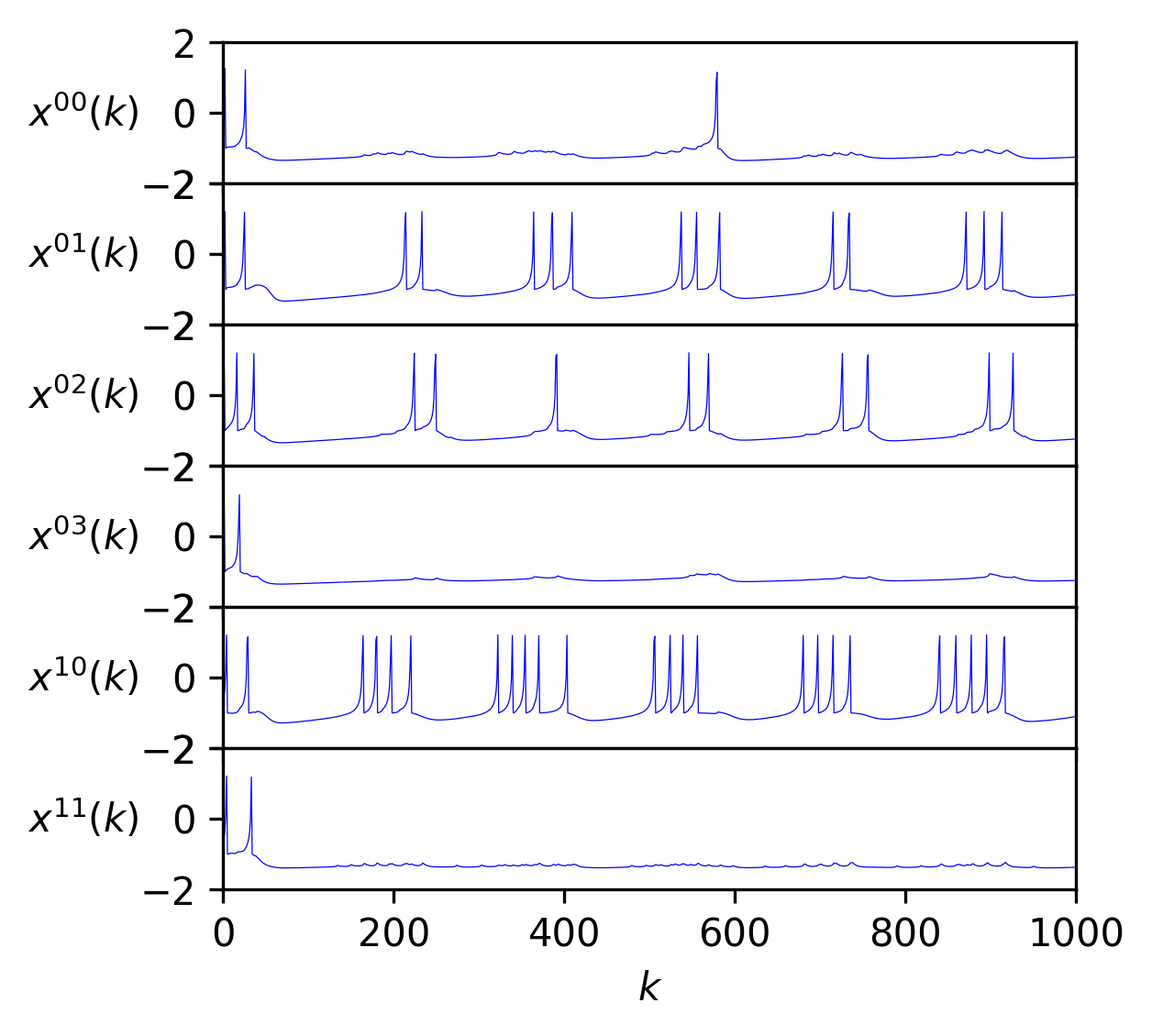}
        \caption{\centering $N = 2$, $\lambda_1 \approx 0.044$}
        \label{fig:xvk_2d_phetero_g0.045}
    \end{subfigure}
    \hfill
    \begin{subfigure}{0.32\textwidth}
        \centering
        \includegraphics[width=\textwidth]{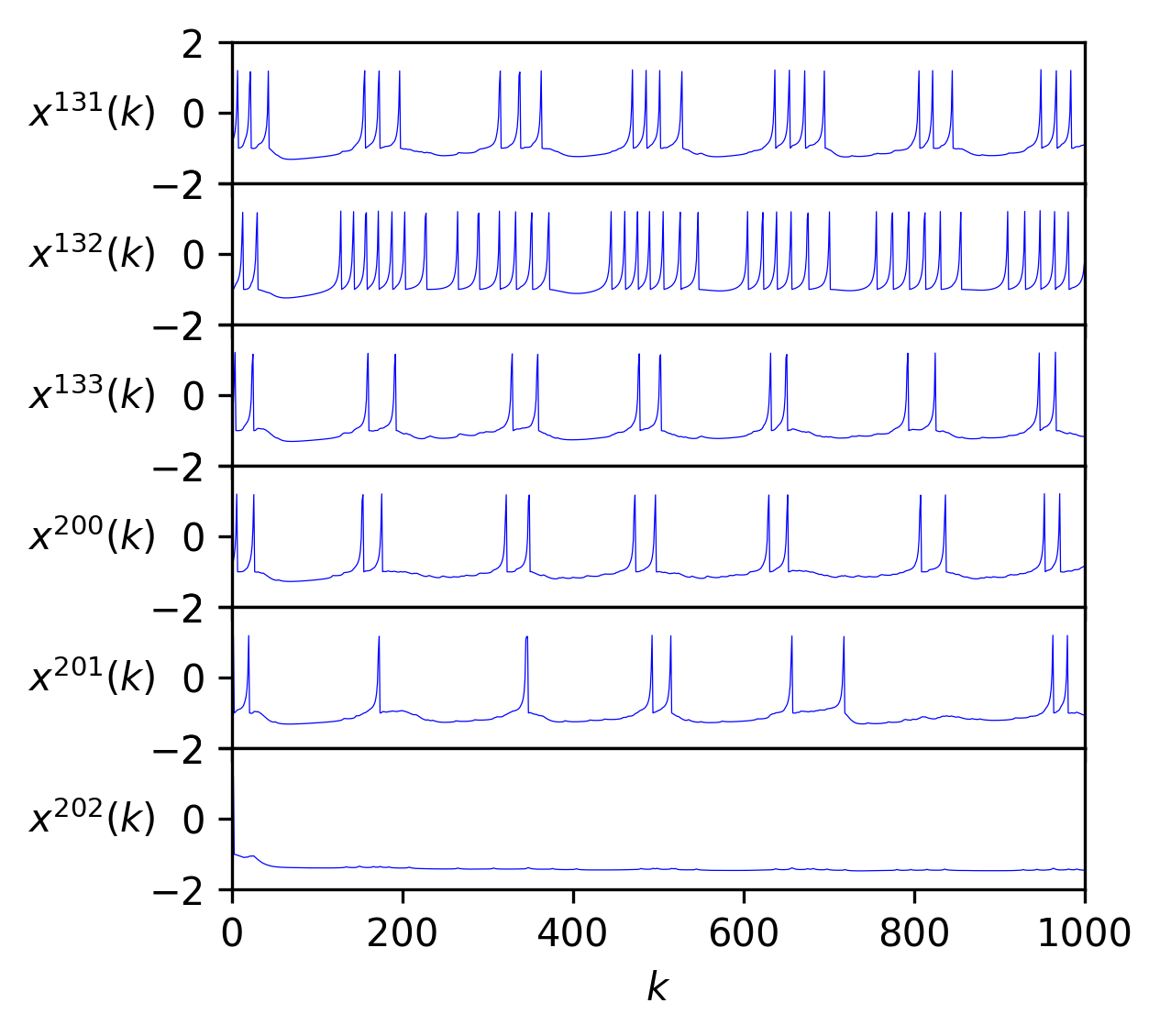}
        \caption{\centering $N = 3$, $\lambda_1 \approx 0.057$}
        \label{fig:xvk_3d_phetero_g0.045}
    \end{subfigure}
    \hfill
    \begin{subfigure}{0.32\textwidth}
        \centering
        \includegraphics[width=\textwidth]{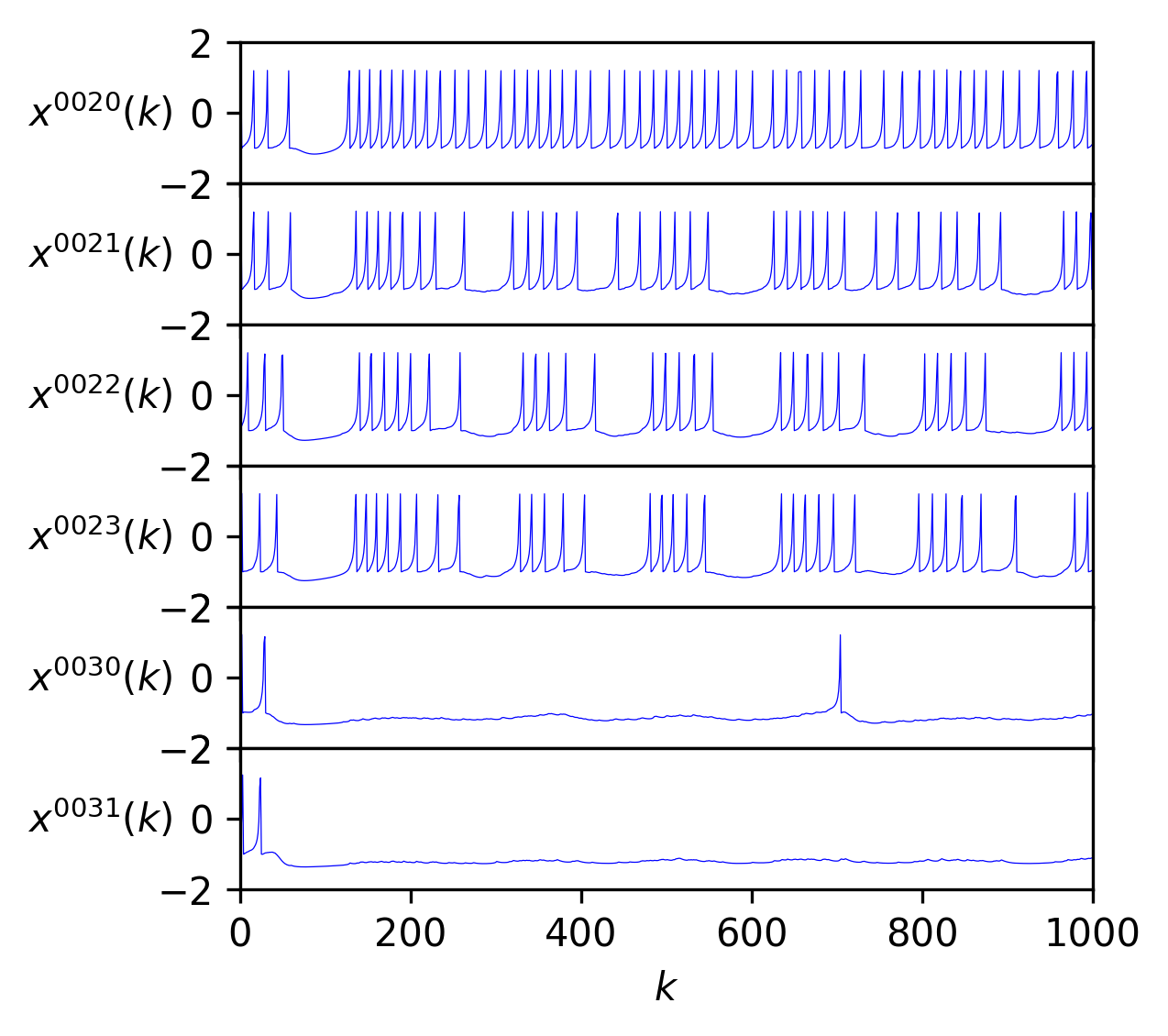}
        \caption{\centering $N = 4$, $\lambda_1\approx 0.057$}
        \label{fig:xvk_4d_phetero_g0.045}
    \end{subfigure}
    \caption{Graphs of the voltages $x(k)$ of six adjacent partially heterogeneous NN coupled neurons with $g = 0.045$. These neurons are in the local quasi-bursting regime, which exists in this lattice with $\zeta = 4$ in $N=2, 3$, and $4$ spatial dimensions.}
    \label{fig:nd-lattice-examples}
\end{figure}

Let us now turn to the heterogeneous lattices. As in the homogeneous case, we find that the trend of $\lambda_{1}$ as we vary $g$ is quite similar in spatial dimensions $N=2,3,$ and 4. Consider first Fig. \ref{fig:nd-neurons-phetero}: we find that in the $0 \lesssim g$ region, these small lattices are able to exhibit local quasi-bursting (see Fig. \ref{fig:nd-lattice-examples}), which explains the early decrease of $\lambda_{1}$ as discussed in Sec. \ref{subsec:2D-NN}. In contrast, the $N = 1$ lattice does not exhibit this dynamical phase. As the spatial dimension of the lattice increases, the location of this valley moves closer to $g = 0$. If this trend continues, the location of this valley will eventually approach $g = 0$, fully extinguishing this dynamical phase. It is thus conjectured that this dynamical phase may only be realized in local lattice topologies where the cooperative effect of ``destructive interference'' (see Sec. \ref{subsec:2D-NNN}) does not significantly hinder this quasi-bursting phase at low conductance.

We should also comment on the overlap of the slopes governing the descent of $\lambda_{1}$ after the first chaotic peak in Fig. \ref{fig:nd-neurons-phetero}. Recall that, physically, this descent in $\lambda_{1}$ corresponds to the onset of synchronized bursting phases in the lattices. What is revealed to us in these simulations is that the manner in which these partially heterogeneous lattices achieve this dynamical phase is similar.

We will finally comment on the fully heterogeneous lattices. In Fig. \ref{fig:nd-neurons-fhetero}, we find that the local quasi-bursting phase in the $N = 2$ lattice has been extinguished. In contrast, the $N = 3$ and $N = 4$ lattices retain this dynamical phase. While the $N = 1$ and $N = 2$ do not possess this dynamical phase, all four lattices then perform the familiar ascent to the first chaotic peak as $g$ continues to increase from $0$. Interestingly, the manner by which the first chaotic peak is reached differs slightly with increasing dimension, namely, with a decreasing slope of the curve. Given that we have not performed an analysis of the trend beyond $N = 4$ dimensions and that we have calculated a moving average to produce the figures in Fig. \ref{fig:nd-neuron-lattice-results}, it is unclear whether this trend is more general and if in higher dimensions we can expect the slope to effectively flatten out.

All four lattices then exhibit a synchronized chaotic bursting phase. As we have discussed previously (Sec. \ref{subsec:2D-NN}), synchronized chaotic bursting occurs when the conductance reaches a value just high enough for the slow variable map to induce alternating periods of chaotic unsynchronized spikes and quiescence in the fast variable map. We remark, then, with that discussion in mind, that it appears that the slope of $\lambda_{1}$ as a function of $g$ is roughly the same in the region of $0.27 < g < 0.52$. That is, the manner in which the bursting synchronization phase is achieved is the same. Furthermore, this trend in the $N = 3$ and $N = 4$ lattices almost mirrors each other up to the underlying oscillatory fluctuation previously discussed in Sec. \ref{subsec:2D-NNN}. We note that these oscillations are expected in these higher-dimensional lattices because they are the result of miniature ``phase transitions'' arising from the highly connected nature of the lattice.

As we have now come to expect, the trend of the maximal Lyapunov exponent in each lattice is to rise rapidly to its highest value when the conductance is raised all the way to $g = 1$. The rapid ascent to chaos is not unexpected; as $g$ approaches $1$, each neuron’s influence on its neighbors is strong enough to induce near-perfect phase locking of their chaotic spiking. However, we again note that the $N = 1$ lattice stands out. Its ascent to hyperchaos differs in both shape and steepness from that of the $N = 2,3,$ and 4 lattices. This feature differs from its partially heterogeneous counterpart, whose slope to hyperchaos is similar to the other lattices (Fig. \ref{fig:nd-neurons-phetero}). Nevertheless, in these higher-dimensional systems, the appearance of a rapid increase to synchronized hyperchaos has come to be expected, and the slope of this increase is likely to be roughly similar in different spatial dimensions, perhaps hinting at a kind of \quotesaround{universal saturation} dynamic for highly coupled heterogeneous lattices in dimensions $N \geq 2$. This universality in the strongly coupled regime across dimensions is further supported by the fact that the dynamics of these $N$-dimensional lattices in the hyperchaotic regime matches that of the $N=1$ system with $\zeta=30$ neurons studied in Ref. \cite{ring}, suggesting that the anomalous $N=1$ case we observe here is a result of the small number of neurons ($\zeta^1=4$) in the one-dimensional lattice compared to the higher-dimensional lattices.

\begin{figure}[p]
    \centering
    \includegraphics[width=0.75\textwidth]{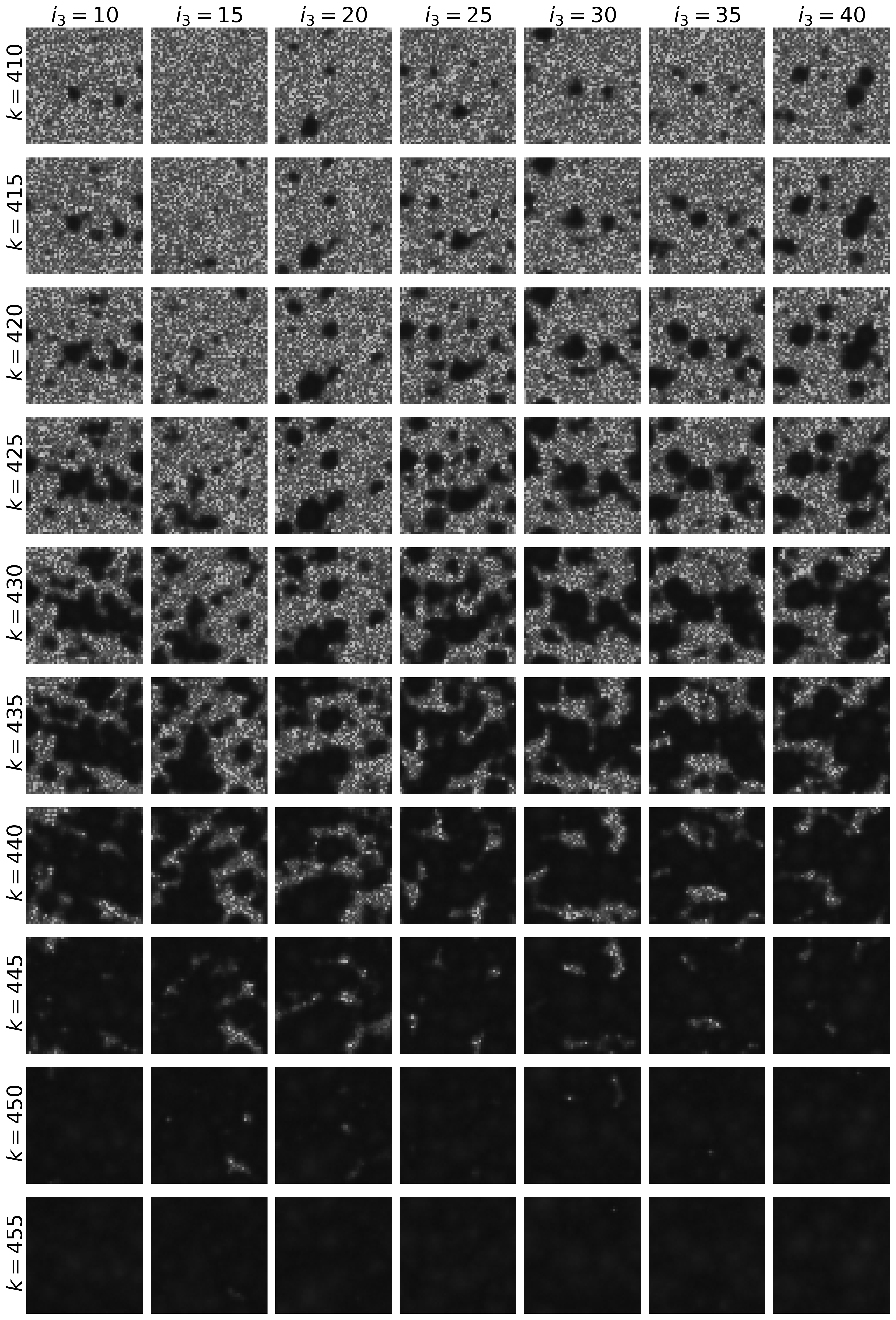}
    \caption{Dynamics of an $N=3$-dimensional large lattice of NN electrically coupled homogeneous Rulkov neurons with $\zeta=50$ neurons per side. The neurons are in the synchronized chaotic bursting regime with electrical coupling strength $g=0.5$. Each row displays a snapshot of the voltages of the three-dimensional lattice at a certain timestep $k$, and each column displays a certain two-dimensional $(i_1,i_2)$ slice of the lattice. The darker points indicate lower voltage, and the lighter points indicate higher voltage. The snapshots are taken at the transition between chaotic spiking and silence.}
    \label{fig:3D-large-g0.5}
\end{figure}
\begin{figure}[p]
    \centering
    \includegraphics[width=0.75\textwidth]{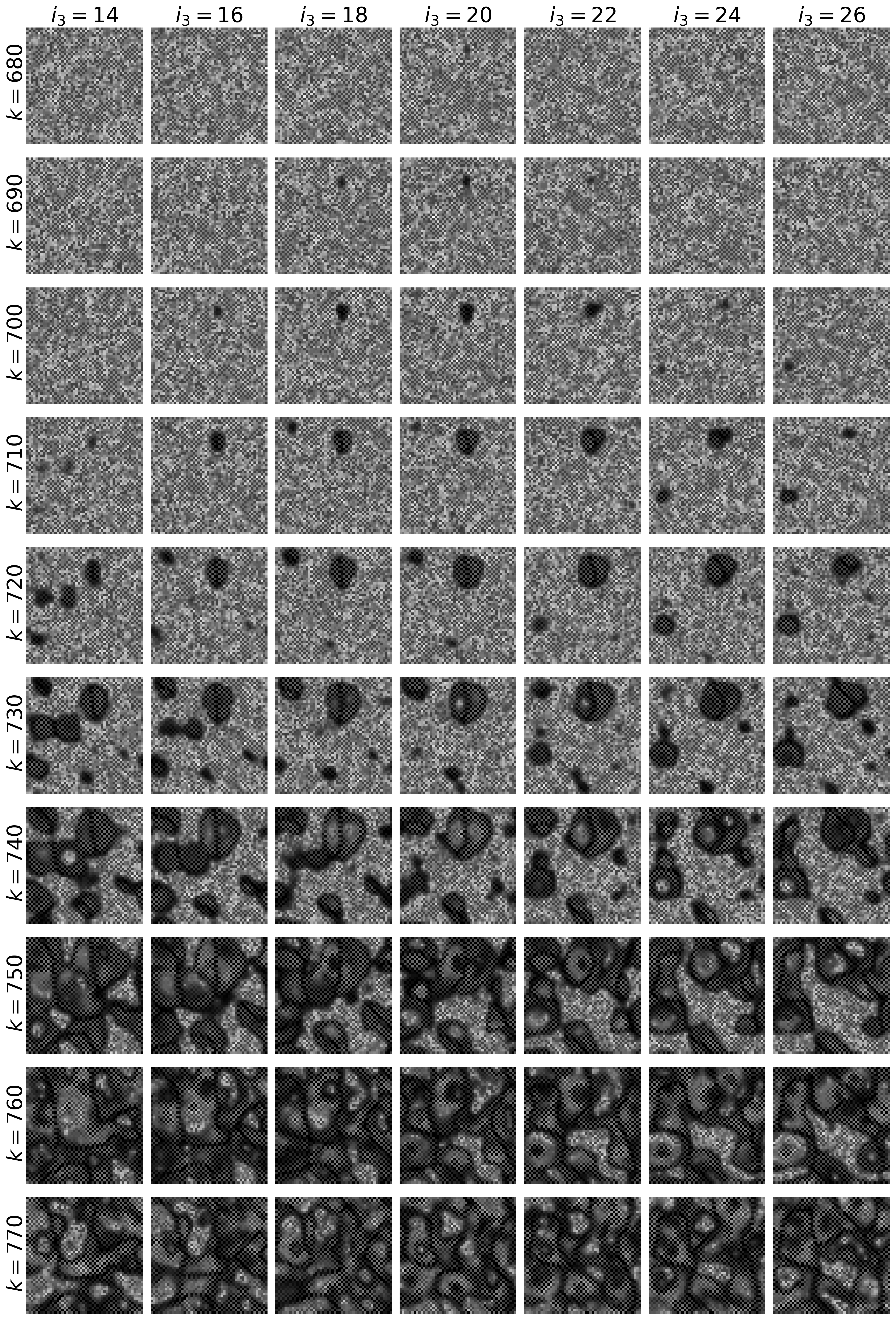}
    \caption{Dynamics of an $N=3$-dimensional large lattice of NN electrically coupled homogeneous Rulkov neurons with $\zeta=50$ neurons per side. The neurons are in the (quasi-)synchronized hyperchaotic regime with electrical coupling strength $g=1$. Each row displays a snapshot of the voltages of the three-dimensional lattice at a certain timestep $k$, and each column displays a certain two-dimensional $(i_1,i_2)$ slice of the lattice. The darker points indicate lower voltage, and the lighter points indicate higher voltage. The snapshots are taken at the transition between a period of higher voltage and a period of lower voltage.}
    \label{fig:3D-large-g1}
\end{figure}

\begin{figure}[t]
    \centering
    \begin{subfigure}{0.9\textwidth}
        \centering
        \includegraphics[width=\textwidth]{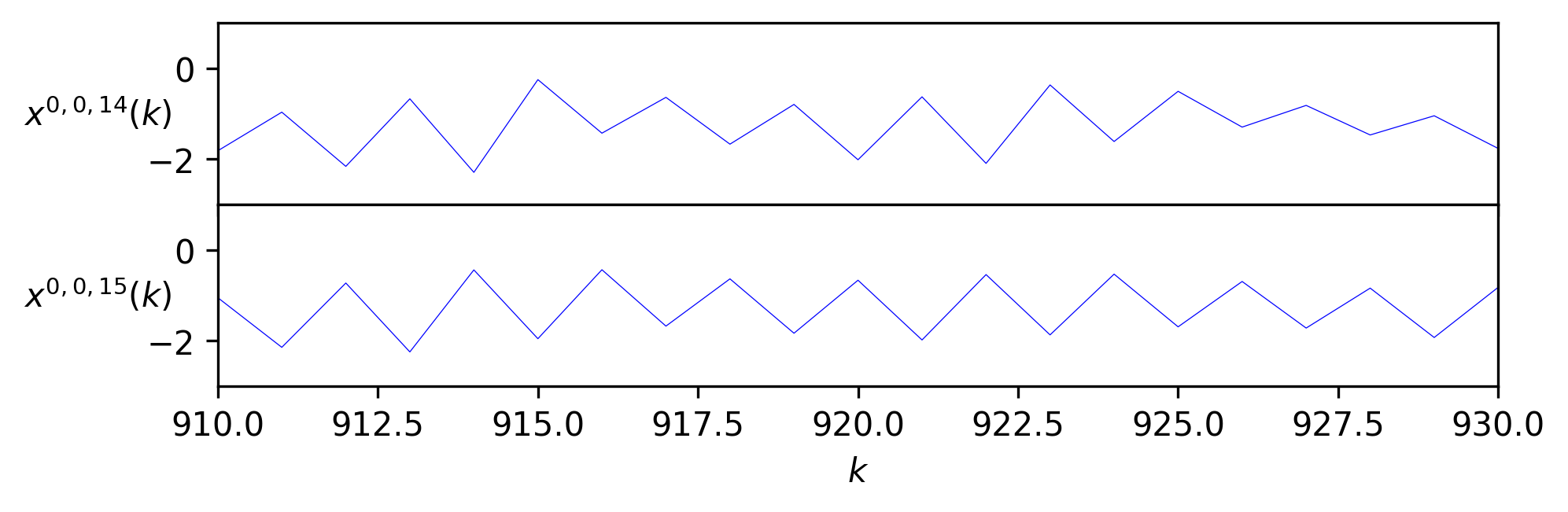}
        \vspace{-0.5cm}
        \caption{Voltages $x(k)$ of two adjacent neurons.}
        \label{fig:3D_lag_xvk}
        \vspace{0.5cm}
    \end{subfigure}
    \begin{subfigure}{0.9\textwidth}
        \centering
        \includegraphics[width=\textwidth]{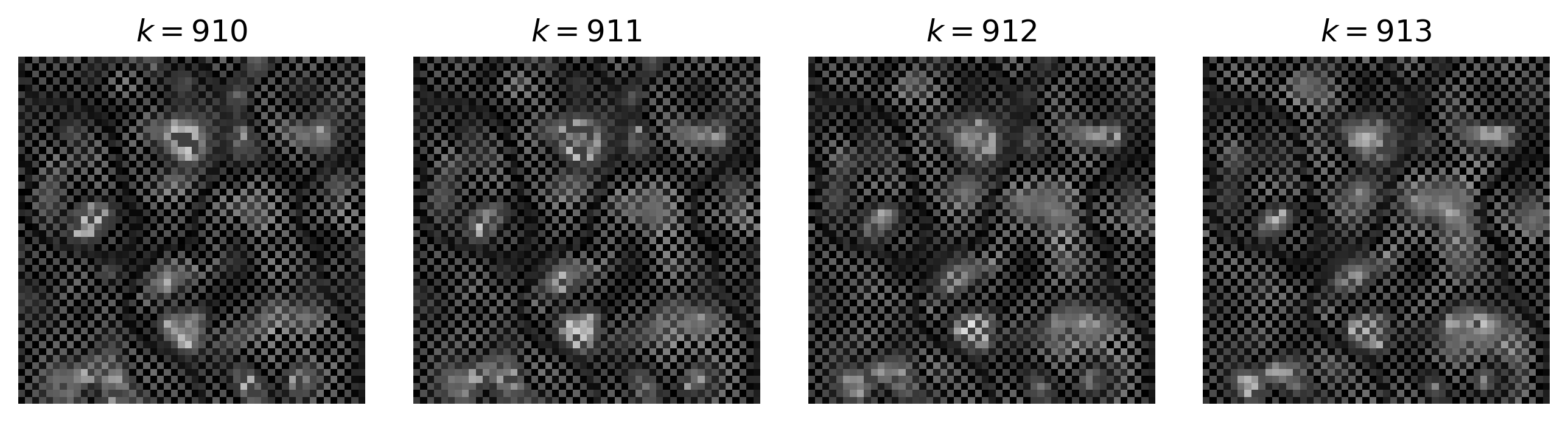}
        \caption{Four consecutive snapshots of a two-dimensional $(i_1,i_2)$ slice of the three-dimensional lattice. The slices are taken at constant $i_3=20$. The darker points indicate lower voltage, and the lighter points indicate higher voltage.}
        \label{fig:3D_lag_snapshots}
    \end{subfigure}
    \caption{Dynamics of an $N=3$-dimensional large lattice of NN electrically coupled homogeneous Rulkov neurons with $\zeta=50$ neurons per side and electrical coupling strength $g=1$. These graphs show the existence of extreme lag synchronization in the (quasi-)synchronized hyperchaotic regime.}
    \label{fig:3D_lag}
\end{figure}

\subsection{Large three-dimensional lattice dynamics}
\label{subsec:large-3d}

Similar to the two-dimensional case (Sec. \ref{subsec:2D-large}), we are now interested in the qualitative dynamics of a large Rulkov neuron lattice in the $N=3$ spatial dimensions, which is a rough simulation of a brain region. Specifically, we consider a three-dimensional NN coupled lattice of electrically coupled homogeneous Rulkov neurons with a side length of $\zeta=50$ neurons. This results in a state space of dimension $n=2\zeta^3 = 250000$. In this section, we examine the behavior of the large three-dimensional lattice in the synchronized chaotic bursting and synchronized hyperchaotic regimes. We do not consider the uncoupled or unsynchronized chaotic spiking regimes because it is easy to see that these are not qualitatively different from the two-dimensional case: seemingly random distributions of high and low voltages across the lattice (see Fig. \ref{fig:2D-unsync-snapshots}).

In Fig. \ref{fig:3D-large-g0.5}, we present a visualization of the three-dimensional lattice in the synchronized chaotic bursting regime with electrical coupling strength $g=0.5$. Similar to the two-dimensional case (Fig. \ref{fig:2D-bursting-snapshots}), we show the lattice at the transition between chaotic spiking and silence because the voltages in the chaotic spiking phase look qualitatively identical to unsynchronized spikes. To visualize the evolution of the three-dimensional lattice, Fig. \ref{fig:3D-large-g0.5} shows snapshots of select two-dimensional $(i_1,i_2)$ slices of the lattice, with slices taken at seven constant $i_3$ values spaced by $\Delta i_3 = 5$. The time axis points downwards in the figure, with each row showing evenly spaced snapshots ($\Delta k=5$) of the two-dimensional $(i_1,i_2)$ slices of the lattice at a given timestep $k$. At the top row, when $k=410$, most of the neurons are still in the chaotic spiking phase. There are also a few small black spots in most of the slices, representing neurons that have already transitioned. In a similar manner to the two-dimensional case (Fig. \ref{fig:2D-bursting-snapshots}), the quiescent neurons pull their neighboring chaotic spiking neurons down to the silence phase, causing the low voltage black regions to spread radially throughout the lattice until the whole lattice is in the silence phase. We note that this transition between chaotic spiking and silence happens faster in this three-dimensional large lattice than the associated two-dimensional large lattice in Sec. \ref{subsec:2D-large} (Notice that the time between snapshots in Fig. \ref{fig:2D-bursting-snapshots} is $\Delta k = 10$, while the time between snapshots in Fig. \ref{fig:3D-large-g0.5} is $\Delta k = 5$.).
This occurs because although the number of neurons in the three-dimensional lattice is larger, its effective size is smaller due to the smaller side length ($\zeta = 50$ in the three-dimensional lattice vs. $\zeta=300$ in the two-dimensional lattice) and greater number of coupling connections, which makes neurons that are already in the quiescent phase able to pull its many adjacent neurons down more quickly.

In Fig. \ref{fig:3D-large-g1}, we present a similar visualization of the three-dimensional lattice in the synchronized hyperchaotic regime with electrical coupling strength $g=1$. Here, we display the dynamics of the lattice at the transition between a period of higher voltage and a period of lower voltage and take evenly spaced snapshots with $\Delta k = 10$. As we discovered in Sec. \ref{subsec:2D-large}, this regime for the two-dimensional large lattice case is better described as ``quasi-synchronized'' or ``locally synchronized.'' This also holds true in the three-dimensional case, where the entire lattice follows the general trend of alternation between higher voltages and lower voltages, and in the lower voltage phase, the lattice exhibits splotches of locally quasi-synchronized neurons. Figure \ref{fig:3D-large-g1} also clearly shows the radial spreading of low voltages at the displayed transition: a small splotch of low voltage neurons appears in the $i_3=20$ slice at $k=680$, and as time progresses, the low voltages spread both radially in the $i_1$ and $i_2$ directions within the same slice and radially in the $i_3$ direction to adjacent slices. In the lower voltage phase, similar to the two-dimensional case, we can see clear splotches of locally quasi-synchronized neurons and waves of low voltage propagating throughout the three-dimensional lattice.

Something that does appear in the two-dimensional large lattice in the synchronized hyperchaotic regime (Fig. \ref{fig:2D-hyperchaos-snapshots}) but that is abundantly clear in the three-dimensional large lattice (Fig. \ref{fig:3D-large-g1}) is the appearance of a checkerboard pattern, with alternating high and low voltage neurons. This is present at all of the timesteps in Fig. \ref{fig:3D-large-g1}, but it is most clear in the lower voltage phase ($k=750, 760, 770$). This checkerboard pattern does not appear in the less strongly coupled regimes, and it is a bit counterintuitive since we know that the positive electrical coupling strength $g$ leads to a synchronization of adjacent neurons, not an anti-synchronization. Indeed, rather than anti-synchronization, this observation is a manifestation of an extreme form of lag synchronization \cite{lag-sync-1, lag-sync-2, lag-sync-3, lag-sync-4, lag-sync-model}, which often appears in hyperchaotic systems. Since the Rulkov map is a discrete-time system, it is to be expected that there is a very small amount of lag with regard to the injection of current into a neuron, since the neuron's voltage will only reflect the injection one timestep after injection begins (see Sec. \ref{sec:the-model}). For more weakly coupled lattices, this effect does not manifest since the amount of flowing current is relatively small and slow to change. However, when $g=1$, the current flow is equal to the difference in the voltages between adjacent neurons [Eq. \eqref{eq:coup-param}], so the magnitudes of the flowing current are large, and their direction can change instantaneously. Therefore, the one-timestep lag synchronization can manifest itself in the synchronized hyperchaotic regime, which is made especially clear in Fig. \ref{fig:3D_lag}. In Fig. \ref{fig:3D_lag_xvk}, we plot the voltages $x(k)$ of two adjacent neurons in the three-dimensional large lattice with $g=1$ over a short time interval. As we can see, the two neurons are spiking in antiphase, which is equivalent to the one-timestep lag synchronization. Here, the strong current between the two neurons switches at every timestep: the current flows out of the higher voltage neuron into the lower voltage neuron, reversing their voltages in the next timestep, which causes the current to flow in the opposite direction, and the process repeats. To see this lag synchronization on a larger scale, we plot four consecutive snapshots of a whole two-dimensional $(i_1,i_2)$ slice of the lattice at $i_3=20$. The checkerboard pattern is evident in the lattice, and looking closely, we can see that the pattern alternates between each timestep. Specifically, a dark neuron turns light in the next timestep, and vice versa, which is the defining characteristic of this extreme lag synchronization.

\section{Conclusions}
\label{sec:conclusions}

In this article, we explored the dynamics that arise when electrically coupling Rulkov neurons in $N$-dimensional lattice configurations, with an emphasis on understanding the emergent cooperative behavior. Our primary objective was to extend previous studies on Rulkov neurons arranged in a ring and to determine whether similar dynamical phases emerge in higher-dimensional lattice structures or if novel collective behavior is enabled by the more complex geometries and highly coupled lattices. Our analysis relied on comprehensive computer simulations performed across a wide range of coupling strengths and configurations.

We began with a comprehensive study of two-dimensional lattices with nearest-neighbor and next-nearest-neighbor coupling across 5000 different electrical coupling strength values. By computing the Lyapunov spectra, examining the trend of the maximal Lyapunov exponent, and producing representative orbits of groups of neurons, we identified and classified unique dynamical regimes according to the systems' tendencies to exhibit chaos and synchronization. In homogeneous lattices, we consistently observed four dynamical regimes: nonchaotic spiking, unsynchronized chaotic spiking, synchronized chaotic bursting, and synchronized hyperchaos. We also discovered that the degree of heterogeneity in a Rulkov system significantly affects the possible cooperative behaviors of the lattice; for example, in heterogeneous lattices, a novel phase was realized that we termed local quasi-bursting --- a state of reduced chaos where local neighborhoods of neurons exhibit intermittent bursting activity against other quiescent groups at low conductance values. In the next-nearest-neighbor coupled lattices, we found that the more highly coupled system led to ``destructive interference'' effects that resulted in a delay of the transition to dynamical regimes of higher electrical coupling strength. Additionally, we observed miniature ``phase transitions'' that emerged from the coordinated addition of spikes to individually quiescent neurons. Finally, we found that the locality of neuron interactions is responsible for the maintenance of neuron dynamics in large lattices. This allowed us to apply the results from small lattice simulations to many more neurons and predict large lattice emergent behavior such as local synchronization, quasi-synchronization, and radial low-voltage spreading.

We then analyzed lattices in $N=1,2,3,$ and 4 spatial dimensions to better understand how spatial dimensionality affects the lattice dynamics. We discovered several unexpected trends in the approach to chaos and settling into synchrony. Although many features of the two-dimensional lattice persisted in higher dimensions, including the structured sequence of dynamical phases and miniature phase transitions, there were also important dimensional trends. For instance, the peak of the maximal Lyapunov exponent $\lambda_1$ corresponding to the transition between unsynchronized chaotic spiking and synchronized chaotic bursting exhibits a regular rightward and upward shift as $N$ is increased. Additionally, the slope of $\lambda_{1}$ in the synchronized bursting regime becomes increasingly uniform as $N$ increases, suggesting a dimensional convergence in the route to synchrony. The final transition to synchronized hyperchaos, characterized by steep rises in $\lambda_{1}$ near the extreme electrical coupling strength values, remains present in all dimensions, though the sharpness of this rise depends on spatial connectivity. We also specialized to the $N=3$ case to examine large three-dimensional lattices, a simplified model of a brain region, and observed similar dynamics to large two-dimensional lattices. The large three-dimensional lattice made the emergence of an extreme form of lag synchronization in the strongly coupled case clear, which manifested as an anti-synchronization of adjacent neurons.

These findings suggest that dimensionality (topology of connectivity) plays a dual role: it both stabilizes certain dynamical behaviors (e.g., synchrony) and enables a richer set of phase transitions by increasing the pathways through which local dynamics can propagate and interact. However, our results also indicate that dimensional increases may eventually smooth out or suppress lower-dimensional features such as local quasi-bursting, raising questions about universality and scaling, a subject of future work. Including questions about universality, we hope to characterize the fractal geometry of the attractors associated with these lattice systems, extending the chaotic attractor analysis from the ring lattice studied in Ref. \cite{ring} to higher dimensions. Such investigations may further illuminate the structural complexity underlying emergent collective dynamics in discrete-time neuronal maps. 

The results of this paper have several important implications. First, they demonstrate that even minimal neuron models like Rulkov maps, when embedded in higher-dimensional, biologically relevant topologies, can generate an unexpectedly rich spectrum of spatiotemporal dynamics. Second, the clear shifts in transition thresholds and the emergence of dimension-dependent synchrony patterns suggest potential mechanisms by which biological neural tissue could leverage spatial structure to tune or regulate collective activity. Lastly, our work offers a scalable and computationally efficient framework for studying complex neuronal synchronization phenomena, with possible applications to understanding brain disorders \cite{brain-disorders}, seizure dynamics \cite{seizure-dynamics}, and neuromorphic engineering \cite{neuromorphic}. Specifically, our findings suggest that increasing neural connectivity --- like that seen in cortical networks --- can both stabilize and destabilize neural activity depending on coupling strength, potentially modeling transitions seen in epileptic seizures or other disorders of synchronization. The work also highlights how localized chaotic dynamics can emerge in heterogeneous networks, offering insight into how pathological activity might originate and spread in real neural tissue. \\

\noindent\textbf{Acknowledgements} B.B.L. thanks Isaac Assink for beneficial discussions. \\

\noindent\textbf{Data availability} The data that support the findings of this article are openly available \cite{data}. \\

\noindent\textbf{Conflict of interest} The authors have no competing interests to declare.

\appendix

\section{Jacobian derivation}
\label{appdx:jacobian_deriv}

Let us first consider NN coupling, where $\N^{\i}$ is given by Eq. \eqref{eq:nn-N}. We introduce the economical notation $\i^*_{m\pm} = \{i_1,\hdots,i_{m-1},i_m\pm 1,i_{m+1},\hdots,i_N\}$ for some $1\leq m\leq N$, which allows us to easily refer to the NN neuron along the $m$th spatial dimension. (Recall that there is $\bmod\, N$ on every index $i$.) Then, by Eq. \eqref{eq:coup-param}, the NN coupling parameter $\C^\i$ is
\begin{equation}
    \C^\i = \frac{g}{2N}(\X^{\i^*_{1+}0} + \X^{\i^*_{1-}0} + \cdots + \X^{\i^*_{N+}0} + \X^{\i^*_{N-}0} - 2N\X^{\i0}),
    \label{eq:nn-coup-param}
\end{equation}
where the summed terms include all the elements of Eq. \eqref{eq:nn-N}. With this coupling parameter, we want to find the associated Jacobian tensor $\tensor{J}{^{\i'a'}_{\i a}} = \partial\F^{\i'a'}/\partial\X^{\i a}$. For $a' = 1$, substituting Eq. \eqref{eq:nn-coup-param} into Eq. \eqref{eq:rulkov-iter-lattice} yields
\begin{equation}
    \F^{\i'1} = \X^{\i'1} - \mu \X^{\i'0} + \mu\left[\sigma^{\i'} + \frac{g}{2N}(\X^{\i'^*_{1+}0} + \cdots + \X^{\i'^*_{N-}0} - 2N\X^{\i'0})\right].
\end{equation}
We want to differentiate this function with respect to every variable $\X^{\i a}$, but the only variables present in the function are: $\X^{\i'1}$ ($\i = \i'$, $a=1$), $\X^{\i'0}$ ($\i = \i'$, $a=0$), $\X^{\i'^*_{1+}0}$ ($\i = \i'^*_{1+}$, $a=0$), \ldots, and $\X^{\i'^*_{N-}0}$ ($\i = \i'^*_{N-}$, $a=0$). Since differentiating with respect to any other variables will yield zero, we only need to evaluate the derivatives with respect to these variables:
\begin{equation}
    \frac{\partial\F^{\i'1}}{\partial\X^{\i'1}} = 1, \quad \frac{\partial\F^{\i'1}}{\partial\X^{\i'0}} = -\mu(1+g),\quad \frac{\partial\F^{\i'1}}{\partial\X^{\i'^*_{1+}0}} = \cdots = \frac{\partial\F^{\i'1}}{\partial\X^{\i'^*_{N-}0}} = \frac{\mu g}{2N}.
\end{equation}
Putting this together, we have
\begin{equation}
    \tensor{\J}{^{\i'1}_{\i a}} = \frac{\partial\F^{\i'1}}{\partial\X^{\i a}} = \begin{cases}
        1, & \text{for $\i = \i'$ and $a=1$} \\
        -\mu(1+g), & \text{for $\i = \i'$ and $a=0$} \\
        \mu g/2N, & \text{for $\i = \i'^*_{m\pm}$ and $a=0$} \\
        0, & \text{otherwise}
    \end{cases}.
\end{equation}
For $a' = 0$, we will introduce the notation $\D^{\i'} = \alpha^{\i'} + \X^{\i'1} + \C^{\i'}$.
Then, substituting Eq. \eqref{eq:nn-coup-param} into Eqs. \eqref{eq:rulkov-iter-lattice} and \eqref{eq:rulkov_fast_equation} yields
\begin{equation}
    \F^{\i'0} = \begin{cases}
        \displaystyle \frac{\alpha^{\i'}}{1-\X^{\i'0}} + \X^{\i'1} + \frac{g}{2N}(\X^{\i'^*_{1+}0} + \cdots + \X^{\i'^*_{N-}0} - 2N\X^{\i'0}), & \X^{\i'0}\leq 0 \\
        \displaystyle \alpha^{\i'} + \X^{\i'1} + \frac{g}{2N}(\X^{\i'^*_{1+}0} + \cdots + \X^{\i'^*_{N-}0} - 2N\X^{\i'0}), & 0 < \X^{\i'0} < \D^{\i'} \\
        -1, & \X^{\i'0}\geq \D^{\i'}
    \end{cases}.
\end{equation}
In the first piece ($\X^{\i'0}\leq 0$), we have the same variables ($\X^{\i'1},\X^{\i'0},\X^{\i'^*_{1+}0},\hdots,\X^{\i'^*_{N-}0}$) yielding nonzero derivatives:
\begin{equation}
    \frac{\partial\F^{\i'0}}{\partial\X^{\i'1}} = 1, \quad \frac{\partial\F^{\i'0}}{\partial\X^{\i'0}} = \frac{\alpha^{\i'}}{(1-\X^{\i'0})^2} - g,\quad \frac{\partial\F^{\i'0}}{\partial\X^{\i'^*_{1+}0}} = \cdots = \frac{\partial\F^{\i'0}}{\partial\X^{\i'^*_{N-}0}} = \frac{g}{2N}.
\end{equation}
The same is true for the second piece ($0 < \X^{\i'0} < \D^{\i'}$), which yields the following partial derivatives:
\begin{equation}
    \frac{\partial\F^{\i'0}}{\partial\X^{\i'1}} = 1, \quad \frac{\partial\F^{\i'0}}{\partial\X^{\i'0}} = - g,\quad \frac{\partial\F^{\i'0}}{\partial\X^{\i'^*_{1+}0}} = \cdots = \frac{\partial\F^{\i'0}}{\partial\X^{\i'^*_{N-}0}} = \frac{g}{2N}.
\end{equation}
Of course, all the derivatives of the third piece ($\X^{\i'0}\geq \D^{\i'}$) are zero. Now that we have considered all the cases, combining yields the full Jacobian tensor of the Rulkov lattice system in the NN coupling case:
\begin{equation}
    \tensor{\J}{^{\i'a'}_{\i a}} = 
    \begin{cases}
        \begin{cases}
            0, & \hspace{2.78cm}\text{if $a' = 0$ and $\X^{\i'0}\geq \D^{\i'}$,} \\
            1, & \hspace{2.78cm}\text{otherwise},
        \end{cases} & \text{for $\i = \i'$ and $a=1$} \\
        \\
        \begin{cases}
            -\mu(1+g), & \text{if $a'=1$,} \\
            \alpha^{\i'}(1-\X^{\i'0})^{-2} - g, & \text{if $a'=0$ and $\X^{\i'0}\leq 0$,} \\
            -g, & \text{if $a'=0$ and $0 < \X^{\i'0} < \D^{\i'}$,} \\
            0, & \text{otherwise,}
        \end{cases} & \text{for $\i = \i'$ and $a=0$} \\
        \\
        \begin{cases}
            \mu g/2N, & \hspace{1.92cm}\text{if $a'=1$,} \\
            0, & \hspace{1.92cm}\text{if $a' = 0$ and $\X^{\i'0}\geq \D^{\i'}$,} \\
            g/2N, & \hspace{1.92cm}\text{otherwise}
        \end{cases} & \text{for $\i = \i'^*_{m\pm}$ and $a=0$} \\
        \\
        0, & \text{otherwise}
    \end{cases}.
    \label{eq:NN-jacobian-appdx}
\end{equation}
It is worth noting that in the $N=1$ case, this is identical to the Jacobian of the ring lattice system studied in Ref. \cite{ring} (as it should be). It is also worth noting that when implementing this Jacobian into code, it is more efficient to mirror this derivation by systematically filling the Jacobian with nonzero derivatives rather than using the compact form of Eq. \eqref{eq:NN-jacobian-appdx} (see Ref. \cite{data}).

The Jacobian for the NNN coupling case follows easily from this derivation. Introducing the notation $\i^{**}_{m\pm} = \{i_1,\hdots,i_{m-1},i_m\pm 2,i_{m+1},\hdots,i_N'\}$ for some $1\leq m\leq N$, Eq. \eqref{eq:coup-param} indicates that the NNN coupling parameter is
\begin{equation}
    \C^\i = \frac{g}{4N}(\X^{\i^*_{1+}0} + \X^{\i^*_{1-}0} + \X^{\i^{**}_{1+}0} + \X^{\i^{**}_{1-}0} \cdots + \X^{\i^{**}_{N-}0} - 4N\X^{\i0}),
    \label{eq:nnn-coup-param}
\end{equation}
where the summed terms include all the elements of Eq. \eqref{eq:nnn-N}. Following similar lines as the NN coupling, we can conclude that the Jacobian tensor in the NNN coupling case is
\begin{equation}
    \tensor{\J}{^{\i'a'}_{\i a}} = 
    \begin{cases}
        \begin{cases}
            0, & \hspace{2.78cm}\text{if $a' = 0$ and $\X^{\i'0}\geq \D^{\i'}$,} \\
            1, & \hspace{2.78cm}\text{otherwise},
        \end{cases} & \text{for $\i = \i'$ and $a=1$} \\
        \\
        \begin{cases}
            -\mu(1+g), & \text{if $a'=1$,} \\
            \alpha^{\i'}(1-\X^{\i'0})^{-2} - g, & \text{if $a'=0$ and $\X^{\i'0}\leq 0$,} \\
            -g, & \text{if $a'=0$ and $0 < \X^{\i'0} < \D^{\i'}$,} \\
            0, & \text{otherwise,}
        \end{cases} & \text{for $\i = \i'$ and $a=0$} \\
        \\
        \begin{cases}
            \mu g/4N, & \hspace{1.92cm}\text{if $a'=1$,} \\
            0, & \hspace{1.92cm}\text{if $a' = 0$ and $\X^{\i'0}\geq \D^{\i'}$,} \\
            g/4N, & \hspace{1.92cm}\text{otherwise}
        \end{cases} & \text{for $\i = \i'^*_{m\pm}$ or $\i'^{**}_{m\pm}$ and $a=0$} \\
        \\
        0, & \text{otherwise}
    \end{cases}.
    \label{eq:NNN-jacobian}
\end{equation}

\section{Tensorial implementation}
\label{appdx:tensor-matrix-scheme}

Our tensorial formulation of the Rulkov lattice model results in clean equations because it respects the model's topology. However, the type $(N+1, 0)$ state tensor $\X$ is fundamentally a state vector of dimensionality $n=2\zeta^N$, and the type $(N+1, N+1)$ Jacobian tensor $\J$ is fundamentally a $2\zeta^N\times 2\zeta^N$ Jacobian matrix. In this section, we explain how to transform from the intuitive tensorial formulation to the fundamental vectorial formulation, which is essential for a computational implementation of the system.

We consider the state vector 
\begin{equation}
    \begin{split}
        \mb{X} &= (\X^{0\cdots000,0}, \X^{0\cdots000,1}, \X^{0\cdots001,0}, \X^{0\cdots001,1}, \hdots, \X^{0\cdots00(\zeta-1),0}, \X^{0\cdots00(\zeta-1),1},\hdots \\
        &\phantom{\mathrel{=(}}\,\X^{0\cdots010,0}, \X^{0\cdots010,1}, \X^{0\cdots011,0}, \X^{0\cdots011,1}, \hdots, \X^{0\cdots01(\zeta-1),0}, \X^{0\cdots01(\zeta-1),1},\hdots\hdots \\
        &\phantom{\mathrel{=(}}\,\X^{(\zeta-1)\cdots(\zeta-1)(\zeta-1)0,0}, \X^{(\zeta-1)\cdots(\zeta-1)(\zeta-1) 0,1}, \hdots, \X^{(\zeta-1)\cdots(\zeta-1)(\zeta-1)(\zeta-1),1})^\top,
    \end{split}
    \label{eq:explicit-x-vector}
\end{equation}
where we write $\X^{\i a} = \X^{i_1\cdots i_{N-2}i_{N-1}i_N,a}$ for clarity. Let us index $\mb{X}$ with $\nu$, which can take integer values between 0 and $2\zeta^N-1$. Then, we want to construct a function $\I$ mapping $(\i,a)\mapsto\nu$ such that $\X^{\i a} = \mb{X}^{\I(\i,a)}$. An intuitive way to construct $\I$ is to first think of the set of indices $\i = i_1i_2\cdots i_N$ as an $N$-digit number in base-$\zeta$. This converts the $N$ indices specifying coordinate locations in the lattice to a single number between 0 and $\zeta^N-1$ that runs over all the neurons in the lattice. To get the final function $\I$, we simply multiply this location index by 2 and add $a$ to get the index $\nu$ in the state vector $\mb{X}^\nu$. Explicitly,
\begin{equation}
    \I(i_1,i_2,\hdots,i_N,a) = 2\sum_{k=1}^N i_k\zeta^{N-k} + a.
    \label{eq:I}
\end{equation}
Inverting this function to get $(i_1,i_2,\hdots,i_N,a) = \I^{-1}(\nu)$ is straightforward, provided that $\zeta$ and $N$ are known. We can also use Eq. \eqref{eq:I} to convert the Jacobian tensor $\tensor{\J}{^{\i'a'}_{\i a}}$ into the Jacobian matrix $\tensor{\mb{J}}{^{\nu'}_{\nu}}$:
\begin{equation}
    \tensor{\J}{^{\i'a'}_{\i a}} = \tensor{\mb{J}}{^{\I(\i',a')}_{\I(\i, a)}}.
\end{equation}
When implementing this model, we store the state as $\mb{X}$ and the Jacobian as $\mb{J}$, and we use the function $\I$ to translate $\X$ and $\J$ into the computational data structures. This allows us to take advantage of tools such as efficient matrix multiplication and QR factorization.

\bibliographystyle{ieeetr.bst}
\bibliography{refs}

\end{document}